\documentclass[10pt]{article}
\usepackage{graphicx}
\usepackage{amsmath,amsfonts,amssymb}
\usepackage{array}
\usepackage{authblk}
\usepackage{caption}
\usepackage{tikz}
\usetikzlibrary{snakes}
\usetikzlibrary{plotmarks}
\newcommand{\be}{\begin{equation}}
\newcommand{\ee}{\end{equation}}
\newcommand{\bea}{\begin{eqnarray}}
\newcommand{\eea}{\end{eqnarray}}
\usepackage[british]{babel}
\usepackage[dvips]{geometry}

\usepackage{epstopdf}
\usepackage{psfrag}

\usepackage{tikz}
\usetikzlibrary{decorations.markings}

\usepackage{color}

\newcommand{\Rii}{
\begin{tikzpicture}[scale=0.5,baseline={([yshift=-.5ex]current bounding box.center)}]
\draw [black, line width=0.2pt] (0,-1)  -- (1,0); 
\draw [black, line width=0.2pt] (0,1) -- (1,0); 
\draw [black, line width=0.2pt] (-1,0) -- (0,-1); 
\draw [black, line width=0.2pt] (-1,0) -- (0,1);  
\draw[blue, line width=0.3mm, rounded corners=7pt] (-0.5,-0.5) -- (-0.2,0.0) -- (-0.5,0.5);
\draw[blue, line width=0.3mm, rounded corners=7pt] (0.5,-0.5) -- (0.2,0.0) -- (0.5,0.5);
\draw[red, line width=0.3mm, rounded corners=7pt] (-0.4,-0.6) -- (0,0) -- (-0.4,0.6);
\draw[red, line width=0.3mm, rounded corners=7pt] (0.4,-0.6) -- (0,0) -- (0.4,0.6);
\end{tikzpicture}}
\newcommand{\Ref}{
\begin{tikzpicture}[scale=0.5,baseline={([yshift=-.5ex]current bounding box.center)}]
\draw [black, line width=0.2pt] (0,-1)  -- (1,0); 
\draw [black, line width=0.2pt] (0,1) -- (1,0); 
\draw [black, line width=0.2pt] (-1,0) -- (0,-1); 
\draw [black, line width=0.2pt] (-1,0) -- (0,1);  
\draw[blue, line width=0.3mm, rounded corners=7pt] (-0.5,-0.5) -- (0.,0.0) -- (0.5,-0.5);
\draw[blue, line width=0.3mm, rounded corners=7pt] (-0.5,0.5) -- (0.,0.0) -- (0.5,0.5);
\draw[red, line width=0.3mm, rounded corners=7pt] (-0.4,-0.6) -- (-0,-0.2) -- (0.4,-0.6);
\draw[red, line width=0.3mm, rounded corners=7pt] (-0.4,0.6) -- (0,0.2) -- (0.4,0.6);
\end{tikzpicture}}
\newcommand{\Rei}{
\begin{tikzpicture}[scale=0.5,baseline={([yshift=-.5ex]current bounding box.center)}]
\draw [black, line width=0.2pt] (0,-1)  -- (1,0); 
\draw [black, line width=0.2pt] (0,1) -- (1,0); 
\draw [black, line width=0.2pt] (-1,0) -- (0,-1); 
\draw [black, line width=0.2pt] (-1,0) -- (0,1);  
\draw[blue, line width=0.3mm, rounded corners=7pt] (-0.5,-0.5) -- (0.,0.0) -- (-0.5,0.5);
\draw[blue, line width=0.3mm, rounded corners=7pt] (0.5,-0.5) -- (0.,0.0) -- (0.5,0.5);
\draw[red, line width=0.3mm, rounded corners=7pt] (-0.4,-0.6) -- (-0,-0.2) -- (0.4,-0.6);
\draw[red, line width=0.3mm, rounded corners=7pt] (-0.4,0.6) -- (0,0.2) -- (0.4,0.6);
\end{tikzpicture}}
\newcommand{\Rif}{
\begin{tikzpicture}[scale=0.5,baseline={([yshift=-.5ex]current bounding box.center)}]
\draw [black, line width=0.2pt] (0,-1)  -- (1,0); 
\draw [black, line width=0.2pt] (0,1) -- (1,0); 
\draw [black, line width=0.2pt] (-1,0) -- (0,-1); 
\draw [black, line width=0.2pt] (-1,0) -- (0,1);  
\draw[red, line width=0.3mm, rounded corners=7pt] (-0.5,-0.5) -- (0.,0.0) -- (-0.5,0.5);
\draw[red, line width=0.3mm, rounded corners=7pt] (0.5,-0.5) -- (0.,0.0) -- (0.5,0.5);
\draw[blue, line width=0.3mm, rounded corners=7pt] (-0.4,-0.6) -- (-0,-0.2) -- (0.4,-0.6);
\draw[blue, line width=0.3mm, rounded corners=7pt] (-0.4,0.6) -- (0,0.2) -- (0.4,0.6);
\end{tikzpicture}}
\newcommand{\Rul}{
\begin{tikzpicture}[scale=0.5,baseline={([yshift=-.5ex]current bounding box.center)}]
\draw [black, line width=0.2pt] (0,-1)  -- (1,0); 
\draw [black, line width=0.2pt] (0,1) -- (1,0); 
\draw [black, line width=0.2pt] (-1,0) -- (0,-1); 
\draw [black, line width=0.2pt] (-1,0) -- (0,1);  
\draw[blue, line width=0.3mm, rounded corners=7pt] (-0.5,-0.5) -- (-0.2,0.0) -- (-0.5,0.5);
\draw[red, line width=0.3mm, rounded corners=7pt] (-0.4,-0.6) -- (0,0) -- (-0.4,0.6);
\end{tikzpicture}}
\newcommand{\Rur}{
\begin{tikzpicture}[scale=0.5,baseline={([yshift=-.5ex]current bounding box.center)}]
\draw [black, line width=0.2pt] (0,-1)  -- (1,0); 
\draw [black, line width=0.2pt] (0,1) -- (1,0); 
\draw [black, line width=0.2pt] (-1,0) -- (0,-1); 
\draw [black, line width=0.2pt] (-1,0) -- (0,1);  
\draw[blue, line width=0.3mm, rounded corners=7pt] (0.5,-0.5) -- (0.2,0.0) -- (0.5,0.5);
\draw[red, line width=0.3mm, rounded corners=7pt] (0.4,-0.6) -- (0,0) -- (0.4,0.6);
\end{tikzpicture}}
\newcommand{\Rud}{
\begin{tikzpicture}[scale=0.5,baseline={([yshift=-.5ex]current bounding box.center)}]
\draw [black, line width=0.2pt] (0,-1)  -- (1,0); 
\draw [black, line width=0.2pt] (0,1) -- (1,0); 
\draw [black, line width=0.2pt] (-1,0) -- (0,-1); 
\draw [black, line width=0.2pt] (-1,0) -- (0,1);  
\draw[blue, line width=0.3mm, rounded corners=7pt] (-0.5,-0.5) -- (0.,0.0) -- (0.5,-0.5);
\draw[red, line width=0.3mm, rounded corners=7pt] (-0.4,-0.6) -- (-0,-0.2) -- (0.4,-0.6);
\end{tikzpicture}}
\newcommand{\Ruu}{
\begin{tikzpicture}[scale=0.5,baseline={([yshift=-.5ex]current bounding box.center)}]
\draw [black, line width=0.2pt] (0,-1)  -- (1,0); 
\draw [black, line width=0.2pt] (0,1) -- (1,0); 
\draw [black, line width=0.2pt] (-1,0) -- (0,-1); 
\draw [black, line width=0.2pt] (-1,0) -- (0,1);  
\draw[blue, line width=0.3mm, rounded corners=7pt] (-0.5,0.5) -- (0.,0.0) -- (0.5,0.5);
\draw[red, line width=0.3mm, rounded corners=7pt] (-0.4,0.6) -- (0,0.2) -- (0.4,0.6);
\end{tikzpicture}}
\newcommand{\Rvlr}{
\begin{tikzpicture}[scale=0.5,baseline={([yshift=-.5ex]current bounding box.center)}]
\draw [black, line width=0.2pt] (0,-1)  -- (1,0); 
\draw [black, line width=0.2pt] (0,1) -- (1,0); 
\draw [black, line width=0.2pt] (-1,0) -- (0,-1); 
\draw [black, line width=0.2pt] (-1,0) -- (0,1);  
\draw[blue, line width=0.3mm] (-0.5,-0.5) -- (0.5,0.5);
\draw[red, line width=0.3mm] (-0.4,-0.6) -- (0.6,0.4);
\end{tikzpicture}}
\newcommand{\Rvrl}{
\begin{tikzpicture}[scale=0.5,baseline={([yshift=-.5ex]current bounding box.center)}]
\draw [black, line width=0.2pt] (0,-1)  -- (1,0); 
\draw [black, line width=0.2pt] (0,1) -- (1,0); 
\draw [black, line width=0.2pt] (-1,0) -- (0,-1); 
\draw [black, line width=0.2pt] (-1,0) -- (0,1);  
\draw[blue, line width=0.3mm] (0.5,-0.5)-- (-0.5,0.5);
\draw[red, line width=0.3mm] (0.4,-0.6) -- (-0.6,0.4);
\end{tikzpicture}}
\newcommand{\Rt}{
\begin{tikzpicture}[scale=0.5,baseline={([yshift=-.5ex]current bounding box.center)}]
\draw [black, line width=0.2pt] (0,-1)  -- (1,0); 
\draw [black, line width=0.2pt] (0,1) -- (1,0); 
\draw [black, line width=0.2pt] (-1,0) -- (0,-1); 
\draw [black, line width=0.2pt] (-1,0) -- (0,1);  
\end{tikzpicture}}
\newcommand{\RR}{\check{R}}

\newcommand{\ket}[1]{|{#1}\rangle}
\newcommand*\diff{\mathop{}\!\mathrm{d}}
\newcommand{\arrowNE}[2]{
\draw[black,thick] (#1,#2) -- (#1-0.2,#2);
\draw[black,thick] (#1,#2) -- (#1,#2-0.2);
}
\newcommand{\arrowNW}[2]{
\draw[black,thick] (#1,#2) -- (#1+0.2,#2);
\draw[black,thick] (#1,#2) -- (#1,#2-0.2);
}
\newcommand{\arrowSE}[2]{
\draw[black,thick] (#1,#2) -- (#1-0.2,#2);
\draw[black,thick] (#1,#2) -- (#1,#2+0.2);
}
\newcommand{\arrowSW}[2]{
\draw[black,thick] (#1,#2) -- (#1+0.2,#2);
\draw[black,thick] (#1,#2) -- (#1,#2+0.2);
}

\title{On truncations of the Chalker-Coddington model}

\author[1,2,3]{Romain Couvreur}
\author[4]{Eric Vernier}
\author[1,2,3]{Jesper Lykke Jacobsen}
\author[3,5]{Hubert Saleur}

\affil[1]{Laboratoire de Physique Th\'eorique, D\'epartement de Physique de l'ENS, \'Ecole Normale Sup\'erieure, Sorbonne Universit\'e, CNRS, PSL Research University, 75005 Paris, France}
\affil[2]{Sorbonne Universit\'e, \'Ecole Normale Sup\'erieure, CNRS, \newline Laboratoire de Physique Th\'eorique (LPT ENS), 75005 Paris, France} 
\affil[3]{Institut de Physique Th\'eorique, Universit\'e Paris Saclay, CEA, CNRS, F-91191 Gif-sur-Yvette, France}
\affil[4]{The Rudolf Peierls Centre for Theoretical Physics, University of Oxford, Oxford OX1 3NP, UK}
\affil[5]{USC Physics Department, Los Angeles CA 90089, USA}

\begin{document}
\maketitle

\begin{abstract}
The supersymmetric  reformulation of physical observables in the  Chalker-Coddington model (CC) for the plateau transition in the integer quantum Hall effect leads to a reformulation of its critical properties in terms of a 2D non-compact loop model or a 1D non-compact $gl(2|2)$ spin chain. Following a proposal by Ikhlef, Fendley and Cardy \cite{Ikhlef}, we define and study a series of truncations of these loop models and spin chains, involving a finite and growing number of degrees of freedom per site. The case of the first truncation is solved analytically using the Bethe-ansatz. It is shown to exhibit many of the qualitative features expected for the untruncated theory, including a quadratic spectrum of exponents with a continuous component, and a normalizable ground state below that continuum. Quantitative properties are however at odds with the results of simulations on the CC model. Higher truncations are studied only numerically. While their properties are found to get closer to those of the CC model, it is not clear whether this is a genuine effect, or the result of strong finite-size corrections.

 \end{abstract}

\section{Introduction}

The transition between plateaux of the integer quantum Hall effect remains one of the most fascinating open problems in the field of  quantum criticality: it is simple enough to admit very accessible formulations, is well documented experimentally---and yet, analytical predictions or any kind of ``exact solution'' remain essentially out of reach (despite a lot of recent progress, see below).

The universality class of this problem is believed to be well captured by the Chalker-Coddington network model \cite{ChalkerCoddington}.
The non-interacting nature of this model, together with supersymmetry techniques, allow a reformulation of most questions of interest  in terms of a  well
defined two-dimensional vertex model of statistical mechanics.%
\footnote{Note that the initial problem takes place in $2+1$ dimensions (see below).}
The tools of integrability and conformal field theory, usually so powerful in this context,  should then apply. But they have failed, so far, in providing any kind of  analytical solution of this model: an exact calculation of say the correlation length exponent, or the multi-fractal spectrum of the wave function, still awaits. 

There are well identified reasons why the problem remains unsolved. First, the vertex model (obtained by disorder average in the second quantized formulation of the network model) involves infinite-dimensional representations of the super algebra $gl(2|2)$ on the links. There is  so far little experience in deriving the continuum limit of such models, even in the integrable case---and moreover the $gl(2|2)$ model does not seem to be integrable \cite{Zirnbauer,Gade98}. Second, the presence of a superalgebra---inherited from the supersymmetry
approach---means that the ``Hamiltonian'' is in fact non-hermitian. The corresponding conformal field theory (CFT) can thus be expected to be non-unitary, and exhibit logarithmic features. In contrast with the unitary case, where simple arguments (such as the presence of a Lie algebra symmetry in the lattice model) allow one to narrow down the possible CFTs describing the continuum limit of a given model, the non-unitarity leaves open a very large number of  candidates, and a ``educated guess'' approach is  hard to implement. 

Based on numerical results and a string of subtle arguments, the most reasonable such educated guess \cite{Zirnbauer,Tsveliketal} is a Wess-Zumino-Novikov-Witten (WZNW) model on the supergroup $PSL(2|2)$. In contrast with ordinary WZNW models, this model is conformally invariant for arbitrary values of the coupling constant. We write the action as
\begin{equation}
S={1\over g}\int {\rm d}^2x \, \hbox{Str}(\partial_\mu G^{-1}\partial_\mu G)+i \kappa \Gamma[G] \,, \label{wzwact}
\end{equation}
where the field $G$ lives on the supergroup manifold, $\Gamma[G]$ is the usual topological term, and ${\rm Str}$ denotes the supertrace. For topological reasons, the parameter $\kappa$ has to be integer, but the coupling constant $g$ can in general be different from $1/\kappa$ and the theory still be conformal.  Only when $g={1\over \kappa}$ does the model have left and right symmetries $PSL(2|2)_L\times PSL(2|2)_R$. In this case---which is advocated in
\cite{Tsvelik}---the conformal weights appearing in the space of states of the theory are of the form \cite{Gotzetal}
\begin{equation}
h={1\over \kappa}[ j_1(j_1+1)-j_2(j_2+1)] \,. \label{Tsvset}
\end{equation}
Here, $j_1=0,1/2,\ldots, {\kappa-2\over 2}$ is an  $su(2)$ spin, whereas  $j_2$ is an $sl(2,\mathbb{R})$ spin: either $j_2$ is  a negative real number with $-{\kappa+1\over 2}<j_2<-{1\over 2}$ (discrete series), or $j_2$ takes the form  $j_2=-{1\over 2}+is$ with $s \in \mathbb{R}$ (principal continuous series). 

As usual for models with non-compact targets, one has to be careful with issues of normalizability, and state operator correspondence. For instance, arbitrary values of $j_2\equiv -q$ are usually considered in the literature---and associated with the 
$q$-th power of the density of states---but do not appear, strictly speaking, in the spectrum (\ref{Tsvset}). Note that the field with vanishing conformal weight (formally $j_1=j_2=0$) does not appear in the set (\ref{Tsvset}) either, indicating that the vacuum in this $PSL(2|2)$ theory is not normalizable. Also, it is important to notice that, because of the supergroup nature of the target, observables can appear in various types of representations (typical, atypical etc) of the algebra \cite{Gotzetal}. Finally, we emphasize that the action (\ref{wzwact}) would require considerable more care with  the definition of the target space  \cite{Zirnbauer}. Despite all these caveats, numerical predictions based on (\ref{Tsvset}) are somewhat reasonable if one chooses $\kappa=8$. No reason is known that would justify a priori such a peculiar choice.

The model with $g\neq {1\over \kappa}$ has also been discussed as a potential candidate \cite{Zirnbauer}. In this case however, the presence of two couplings constants ($g$ and $k$) having to be adjusted independently makes the educated guess approach even trickier. Moreover, it is not entirely clear whether the exponents would keep the simple form (\ref{Tsvset}) under this perturbation \cite{Quellaetal}.

More recently, a considerable effort has been devoted to better understand the nature of the lattice observables in relation with their continuum limit, making heavy use of symmetry considerations and the simple but very fruitful idea that wave intensities obey an operator product expansion of Abelian form. Abandoning speculation on the sigma model description of the full critical theory, the authors in \cite{BWZ,BWZII} have strongly advocated a Gaussian free field description of critical wave functions at the transition, with, in particular, a parabolic spectrum exactly of the form (\ref{Tsvset})---even though the identification of the theory as a ``simple''  $PSL(2|2)$ WZNW model is unlikely. Thorough numerical evidence seems to support this claim. 

\bigskip

Our approach in this paper is more elementary, and deals  with the  question of the spectrum of the transfer matrix of the $gl(2|2)$ vertex model \cite{Zirnbauer}. 

To avoid the difficulty of having to deal with infinite-dimensional representations, it has been suggested in the past \cite{Ikhlef,MarstonTsai} to consider {\sl truncations} of this model with a finite number of states per link, hoping that by increasing the number of states the properties of the exact theory might be approached. Decent estimates  of the thermal exponent $\nu$ have for instance been obtained in this way  \cite{MarstonTsai}. But it is not clear whether  the idea has a good chance of working. The essence of the problem is that the properties of an untruncated model might be profoundly different from those of any truncation, however many states this truncation might allow. An example in point is the two-dimensional Brownian walk. Any truncation limiting the number of times the walk can go over an edge (or through a vertex) can be interpreted in terms of a ``weak'' on-site or on-edge repulsion. It is well known that such repulsion takes the model in the universality class of self-avoiding walks \cite{DombJoyce,Clisby}, with completely different critical exponents.%
\footnote{In terms of RG terms, any truncation introduces a relevant $\phi^4$ coupling, which drives the system to a different fixed point (the same is true in all dimensions $d<4$).}
Although it can be expected that crossovers might occur slowly enough to allow higher truncations to approach the Brownian exponents at intermediate lengths, any naive model of ``truncated Brownian motion" can probably be expected to fail at providing good insights on the Brownian motion itself. It is not clear whether the situation is so bad for the Chalker-Coddington model itself---in part, because we know so little about the universality class we are after (see \cite{CJS} for a discussion of this problem). 

One might, in contrast,  be extremely lucky, and have the  opposite  situation where the properties of the truncated model are the same as those of the untruncated one.  Since the untruncated, original model is expected  to exhibit a continuous spectrum of critical exponents,   this would require  the  target of the continuum theory for the truncated model to be non-compact. Unexpectedly, this is in fact possible.  
It has indeed  been realized over the last decade that  a CFT with non-compact target can
well be obtained as the continuum limit of a statistical mechanics model with only a finite number of degrees of freedom per link, provided this model does not have a hermitian ``Hamiltonian''.  A well-known example of this  is the staggered six-vertex model studied in \cite{IJS12}, which turns out to be described by the $SL(2,\mathbb{R})/U(1)$ ``black hole'' sigma model in the continuum limit. It is thus not so clear that truncations of the $gl(2|2)$ vertex model---which involve indeed a non-hermitian Hamiltonian---will  have a continuum limit that is so  different from the one of the untruncated model. 

It is with this idea in mind that we have revisited the model proposed by \cite{Ikhlef}, henceforth referred to as the {\em first truncation} of the $gl(2|2)$ model. While the authors of this paper studied the main exponents numerically, we are able to exploit the exact solvability (already noticed in \cite{Ikhlef}) to extract the full continuum limit. Remarkably, we find indeed that the target of the corresponding CFT is non-compact, with a continuum spectrum of critical exponents.  It is then possible to compare in detail the results for this truncated model with what is expected or known for the untruncated one.

Spurred by this success, one can then turn to a study of the higher truncations. While this generalization was only briefly suggested in \cite{Ikhlef}, we here carry this program further (in section~\ref{sec:higher-truncations}) by defining precisely the corresponding set of configurations and Boltzmann weights, and carrying out numerical calculations of the exponents. In particular we determine the correlation length exponent, and give numerical evidence that its value approaches that of the untruncated model upon increasing the order of the truncation. The question of whether this is a crossover effect (and all truncations are in  the same universality class) or whether the higher truncations are truly in different universality classes approaching the one of the  untruncated model, remains however open.

Our paper is organized as follows. 

In section 2, we discuss the Chalker-Coddington model and  its geometrical reformulation. We then define, following \cite{Ikhlef},  the first truncation and its integrable modification. Symmetries are discussed in great detail. 
In section 3, general properties of the integrable modified truncated model are discussed. The vertex model associated with the geometrical Boltzmann weights is introduced, and the corresponding $\check{R}$ matrix is identified with the fundamental $\check{R}$ matrix for the $b_2^{(1)}$ algebra. Bethe-ansatz equations are given, and the various regimes are identified. 
Section 4 is a detour where we remind the reader of earlier results on two models which turn out to have very similar continuum limits to the integrable modified truncated model: the $a_2^{(2)}$ and the $a_3^{(2)}$ models. The $b_2^{(1)}$ model per se is then discussed in section 5. The associated loop model---i.e., the geometrical version of the integrable modified truncated model---is discussed in section 6. Physical properties relevant to the integer Hall transition are discussed in section 7. Some results for higher truncations are discussed in section 8, together with the exponent $\nu$, and how it is related with conformal weights of the CFTs. Conclusions are presented in section 9. In two appendices we discuss the interpretation of the continuum limit of the integrable modified truncated model as a sigma model, as well as the possible geometrical interpretation of scaling observables defined in the recent work \cite{BWZ,BWZII}. 

Finally, a note about notations. We denote by $(h,\bar{h})$ or $(\Delta,\bar{\Delta})$---depending on the context---the values of the conformal weights, and by $x=h+\bar{h}$ (or $x=\Delta+\bar{\Delta}$) the scaling dimensions of the operators (often referred to simply as dimension). Unfortunately, in some  parts of the literature, $x$ happens to be denoted by ``$\Delta$'': in this case, extra care has to be exercised  in comparing results. 

\section{Modifying the model and the question of universality class}

In this section we review (and expand on) the construction of geometrical models for the transition between plateaux, refering to the original article \cite{Ikhlef} for full details.

\subsection{The Chalker-Coddington model as a supersymmetric path integral}
\label{sec:pathintegral}

The Chalker-Coddington model \cite{ChalkerCoddington} is a network representation of the discrete time evolution of single-electronic wavefunctions in a disordered potential subject to a transverse magnetic field (for a review see \cite{NetworkReview}). 
The equipotential lines, along which the magnetic field imposes a directed motion, are modeled by the edges of an oriented square lattice, while the lattice nodes represent the saddle points where tunneling between different equipotential lines may occur. The presence of disorder is modeled by random phases accompanying the propagation on each edge.  
In this problem of non-interacting electrons the energy appears as a parameter whose tuning allows to observe the plateau transition, so it appears as if the dimension can be reduced from 2+1 to 2, as will become manifest in the following reformulation.

Following \cite{Ikhlef}, observables in the Chalker-Coddington model such as the point-contact conductance can be reformulated as correlation functions of a supersymmetric path integral. Fermions are introduced to average over the $U(1)$ disorder instead of using  a method such as the replica trick. The model is built at every site from advanced/retarded ($+/-$) bosons and advanced/retarded ($+/-$) fermions. Averaging over the $U(1)$ random phase at an edge projects the Fock space on the subspace where there are an equal number of advanced particles and retarded particles.

To each oriented edge $e$ we associate four pairs of variables, $\left(b_{+,L}(e),f_{+,L}(e)\right)$, $\left(b_{-,L}(e),f_{-,L}(e)\right)$, \\$\left(b_{+,R}(e),f_{+,R}(e)\right)$ and $\left(b_{-,R}(e),f_{-,R}(e)\right)$. Variables labeled with an $R$ stand at the start of an oriented edge whereas the label $L$ refers to variables at the end of an edge. The labels $(+,-)$ refer to the color of the variables/particles corresponding in the original model to the advanced and retarded variables. Finally, the $b$'s are usual complex variables and $f$-variables are Grassmann numbers.
Complex conjugates for bosonic/fermionic variables are denoted $b^*/f^*$. 
Integration over a bosonic/fermionic variable is defined with the following measure:
\begin{subequations}
\bea
\int\left[\diff b\right]\left(\ldots\right) &=& \frac{1}{\pi}\int\diff\left(\text{Re}\,b\right)\diff\left(\text{Im}\,b\right)\exp\left(-b^*b\right)\left(\ldots\right) \,, \\
\int\left[\diff f\right]\left(\ldots\right) &=& \int\diff f^*\diff f\exp\left(-f^*f\right)\left(\ldots\right)
\eea
\end{subequations}
Consider now a vertex $v$ with the orientation
\bea
\begin{tikzpicture}[baseline=8pt]
\draw[thick] (-1,-1) -- (1,1);
\draw[thick] (-1,1) -- (1,-1);
\draw (-1.2,-1.2) node {$1$};
\draw (1.2,1.2) node {$2$};
\draw (-1.2,1.2) node {$3$};
\draw (1.2,-1.2) node {$4$};
\arrowSW{0.5}{0.5}
\arrowNW{-0.5-0.1}{0.5+0.1}
\arrowNE{-0.5}{-0.5}
\arrowSE{0.5+0.1}{-0.5-0.1}
\end{tikzpicture}
\label{vertex1234}
\eea
The contribution of this vertex $v$ to the Boltzmann weight is given by
\bea
e^{S_v}&=&\exp\left(\sum_{j=1,2}\sum_{i=3,4}b^*_{+,R}(e_i)\mathcal{S}_{i,j}b_{+,L}(e_j)\right)\exp\left(\sum_{j=1,2}\sum_{i=3,4}f^*_{+,R}(e_i)\mathcal{S}_{i,j}f_{+,L}(e_j)\right)\nonumber\\
&&\times\exp\left(\sum_{j=1,2}\sum_{i=3,4}b^*_{-,L}(e_i)\mathcal{S}_{i,j}^*b_{-,R}(e_j)\right)\exp\left(\sum_{j=1,2}\sum_{i=3,4}f^*_{-,L}(e_i)\mathcal{S}_{i,j}^*f_{-,R}(e_j)\right) \,,
\eea
where scattering matrix $\mathcal{S} \in SU(2)$, taken real here, is parametrized as
\be
\mathcal{S}_{}  =\left( \begin{array}{cc} \alpha & \beta\\
          -\gamma & \delta\end{array} \right)     =\left( \begin{array}{cc} t & r \\
          -r & t\end{array} \right) \,, \label{Smatrix}
\ee
with $r=\sqrt{1-t^2}$.
The contribution of an edge $e$ to the Boltzmann weight  arises  from the $U(1)$ disorder in the underlying network model. It leads to  a term:

\bea
e^{S_e}=\sum_{m_e=0}^\infty\frac{(zz^*)^{m_e}}{(m_e!)^2}\left[\left(b^*_{+,L}(e)b_{+,R}(e)+f^*_{+,L}(e)f_{+,R}(e)\right)\left(b^*_{-,R}(e)b_{-,L}(e)+f^*_{-,R}(e)f_{-,L}(e)\right)\right]^{m_e}.
\eea

Essentially this term constrains the edge $e$ to carry the same number of particles $+$ and $-$. It also gives a weight ${|z|}^2$ to every pair of particles on $e$.
The partition function of the model is finally

\bea
Z=\int\prod_{e,\sigma}\left[\diff b_{\sigma,L}(e)\right]\left[\diff b_{\sigma,R}(e)\right]\left[\diff f_{\sigma,L}(e)\right]\left[\diff f_{\sigma,R}(e)\right]\left(\prod_ve^{S_v}\right)\left(\prod_ee^{S_e}\right)
\label{partitionfunction}
\eea

To reformulate the model in geometrical terms, we Taylor-expand the exponentials in the partition function, producing  terms of the form
\bea
\left(b^*_{+,R}(e_2)b_{+,L}(e_1)+f^*_{+,R}(e_2)f_{+,L}(e_1)\right)
\eea
or its complex conjugate in the case of retarded particles. Eq.~\eqref{partitionfunction} then leads to a sum of directed paths of advanced and retarded particles on the oriented lattice. This defines a two-color loop model with two important features. First, closed loops have a vanishing weight (this  is a direct consequence of supersymmetry). This implies in particular that in the partition function \eqref{partitionfunction} only the term with no path propagating contributes, so $Z=1$ as expected (of course, non-trivial path configurations will be generated by the expansion of physical observables). Second, an edge can carry an arbitrarily large number of strands, with the constraint that on any edge strands of different colors (representing advanced and retarded particles) should be of equal number.

\subsection{Truncation procedure}

The presence of  a infinite number of local degrees of freedom makes of course the model very hard to tackle analytically or numerically. It is a natural idea to investigate instead  truncations obtained by restricting the number of states on a given edge. Following Ikhlef et al. \cite{Ikhlef} we define the truncation at level $M$ by keeping only configurations such that the number of particles on an edge for each color is less or equal to $M$. In other words we truncate $e^{S_e}$ by keeping only the first $M+1$ terms,  while the interaction at the vertices $e^{S_v}$ is unchanged. We define $e^{S_e^M}$ such that

\bea
e^{S_e^M}=\sum_{m_e=0}^M\frac{(zz^*)^{m_e}}{(m_e!)^2}\left[\left(b^*_{+,L}(e)b_{+,R}(e)+f^*_{+,L}(e)f_{+,R}(e)\right)\left(b^*_{-,R}(e)b_{-,L}(e)+f^*_{-,R}(e)f_{-,L}(e)\right)\right]^{m_e}.
\label{truncatededge}
\eea

The original network model is known to be critical for $z=1$ but the truncated model for this value of $z$ is gapped, and criticality is restored only in the limit $M\rightarrow\infty$. However it is possible to increase the particle fugacity $z$ such that the model becomes critical for finite $M$. We assume in the rest of the paper that $z$ is real.

The original idea of studying such truncations comes from a work \cite{MarstonTsai} of Marston and Tsai (see also the earlier work \cite{MarstonKondev}). Using DMRG techniques, these authors  argued that the successive truncations of the $gl(2|2)$ Hamiltonian model defined by keeping $1+4M$ states at each site were not critical, but showed that the gap in the spectrum goes to $0$ as $M\rightarrow\infty$. However, no attempt was made to tune parameters of the Hamiltonian in order to study these models as a series of critical models. This idea was first discussed in the work of Ikhlef et al. \cite{Ikhlef} where they tuned the particle fugacity $z$ of the first truncation in order to restore criticality. 

The interpretation of the Chalker-Coddington model as a two-color loop model applies naturally to its truncations. We here turn to the first of these ($M=1$), already considered in \cite{Ikhlef}.  Since the truncation preserves the symmetry between bosons and fermions, each closed loop has weight $0$. 
The contribution of an edge $e$ is $e^{S_e^1}$ (given by \eqref{truncatededge}) and constrains $e$ to be either empty or to carry one strand of each color. 

The $R$-matrix encoding the local Boltzmann weights at each vertex is
\begin{eqnarray}
\check{R}&=& \Rt + u_1\left( \Rul + \Rur \right)+ u_2 \left( \Rud + \Ruu \right) \nonumber \\ & & + x\left( \Rei + \Rif \right) + w_1\Rii + w_2\Ref
\label{Rloopfirsttruncation}
\end{eqnarray}
where the weights are given below:
\begin{subequations}
\label{eq:loopweights}
\begin{eqnarray}
 u_1 &=&  z^2t^2 \\
  u_2 &=& z^2r^2  \\
    w_1 &=&  z^4t^4 \\
     w_2 &=& z^4r^4 \\
     x &=&  -z^4t^2r^2 \,.
\end{eqnarray}
\end{subequations}
The monomer fugacity $z$ has to be tuned in order for the model to be critical: an empty edge has therefore a weight $1$, and an edge with a pair of strands has a weight $z^2$. In the $M=1$ truncated model the critical value $z_C$ of $z$ is not known exactly but was estimated numerically to $z_C \simeq 1.03>1$ in \cite{Ikhlef}, with the natural interpretation that restricting the occupation number on each edge is compensated by
weighting occupied edges with a fugacity $z>1$.

\subsection{Integrable modification}

As discussed in \cite{Ikhlef}, it is  possible to slightly modify the truncated model to make it integrable. This makes the use of Bethe ansatz calculations possible, hence allowing a much more accurate and complete identification of the continuum limit. The question is however whether the required  ``slight'' modification matters, and in particular whether it changes the universality class. As shown in \cite{Ikhlef}, the integrable model involves the addition of two  tiles where the  strands can go straight,  with a corresponding integrable $\check{R}$ matrix 
\begin{eqnarray}
\check{R}(\varphi) &=& t(\varphi)
 \Rt + u_1(\varphi) \left( \Rul + \Rur \right)+ u_2(\varphi) \left( \Rud + \Ruu \right) \nonumber\\
 &+& v(\varphi) \left( \Rvlr + \Rvrl \right) + x(\varphi) \left( \Rei + \Rif \right) + w_1(\varphi) \Rii + w_2(\varphi) \Ref 
 \nonumber \\
\label{Rloop}
\end{eqnarray}
Here the weights are 
\begin{subequations}
\label{eq:loopweightsintegrable}
\begin{eqnarray}
t(\varphi) &=& -\cos\left(2 \varphi - 3\theta \right) - \cos 5 \theta + \cos 3\theta + \cos \theta \\
 u_1(\varphi) &=&  -2 \sin 2\theta \sin\left( \varphi- 3\theta \right) \\
  u_2(\varphi) &=& 2 \sin 2\theta \sin \varphi  \\
   v(\varphi) &=& -2 \sin 2\theta\varphi \sin\left( \varphi- 3\theta \right) \\
    w_1(\varphi) &=&  2 \sin\left( \varphi- 2\theta \right) \sin\left( \varphi- 3\theta \right) \\
     w_2(\varphi) &=& 2 \sin \varphi \sin\left( \varphi- \theta \right) \\
     x(\varphi) &=&  2 \sin \varphi \sin\left( \varphi- 3\theta \right) \\
     n &=& -2 \cos 2\theta \,.
\end{eqnarray}
\end{subequations}
The weights (\ref{eq:loopweightsintegrable}) contain an extra parameter $\theta$, that will permit us to adjust
the fugacity $n$ of closed loops to any value in the range $|n|\leq 2$. The model relevant for the quantum Hall effect is then retrieved in the end by setting $n=0$ (and, as we shall see, more precisely $\theta = \frac{3 \pi}{4}$).
 
The question of the equivalence between the first truncation of the Chalker-Coddington model and this integrable modification is not obvious.  This is due in large part to the fact that, in the integrable model, the loops can go straight, a fact which is incompatible with the underlying orientation of the original Chalker-Coddington model. It is easier to understand what this means---and how this might affect the universality class---by thinking in terms of symmetries. 

\subsection{Symmetries}

To proceed, it is convenient to introduce the formulation of the original model as a superspin chain, corresponding to the second quantization of the network model.
In the following, operators without a bar act on the states of even sites and operators with a bar act on the states of odd sites. The creation and annihilation operators verify the (anti)commutation relations 
\begin{subequations}
\bea
\left[b_\alpha,b_\beta^\dagger\right] &=& \left\{f_\alpha,f_\beta^\dagger\right\}=\delta_{\alpha,\beta} \,, \\
\left[\bar{b}_\alpha,\bar{b}_\beta^\dagger\right] &=& -\left\{\bar{f}_\alpha,\bar{f}_\beta^\dagger\right\}=\delta_{\alpha,\beta} \,,
\eea
\end{subequations}
where $\alpha,\beta=+,-$.
The space can be obtained from the vacuum---denoted $\ket{0}$ on even sites, and $\ket{\bar{0}}$ on odd sites---by the action of creation operators. The states for even sites are
\begin{subequations}
\bea 
\ket{4n+1}&=&\frac{1}{n!}\left(b_+^\dagger b_-^\dagger\right)^nf_+^\dagger f_-^\dagger\ket{0} \\
\ket{4n+2}&=&\frac{1}{\sqrt{n!(n+1)!}}\left(b_+^\dagger b_-^\dagger\right)^nb_+^\dagger f_-^\dagger\ket{0} \\
\ket{4n+3}&=&\frac{1}{\sqrt{n!(n+1)!}}\left(b_+^\dagger b_-^\dagger\right)^nb_-^\dagger f_+^\dagger\ket{0} \\
\ket{4n+4}&=&\frac{1}{(n+1)!}\left(b_+^\dagger b_-^\dagger\right)^{n+1}\ket{0}
\eea
\label{fockspaceeven}
\end{subequations}
and for odd sites
\begin{subequations}
\bea 
\ket{\overline{4n+1}}&=&\frac{1}{n!}\left(\bar{b}_+^\dagger \bar{b}_-^\dagger\right)^n\bar{f}_+^\dagger \bar{f}_-^\dagger\ket{\overline{0}} \\
\ket{\overline{4n+2}}&=&\frac{1}{\sqrt{n!(n+1)!}}\left(\bar{b}_+^\dagger \bar{b}_-^\dagger\right)^n\bar{b}_+^\dagger \bar{f}_-^\dagger\ket{\overline{0}} \\
\ket{\overline{4n+3}}&=&\frac{1}{\sqrt{n!(n+1)!}}\left(\bar{b}_+^\dagger \bar{b}_-^\dagger\right)^n\bar{b}_-^\dagger \bar{f}_+^\dagger\ket{\overline{0}} \\
\ket{\overline{4n+4}}&=&\frac{1}{(n+1)!}\left(\bar{b}_+^\dagger \bar{b}_-^\dagger\right)^{n+1}\ket{\overline{0}}
\eea
\label{fockspaceodd}
\end{subequations}
Note that  fermionic states on odd sites have norm square  $-1$, as a consequence of the anticommutation relation for barred fermionic generators . All states in (\ref{fockspaceeven}) and (\ref{fockspaceodd})  satisfy the constraint that comes from averaging over the $U(1)$ disorder
\bea
b_+^\dagger b_++f_+^\dagger f_+ = b_-^\dagger b_-+f_-^\dagger f_-,\quad\text{or}\quad\bar{b}_+^\dagger \bar{b}_+-\bar{f}_+^\dagger \bar{f}_+ =\bar{b}_-^\dagger \bar{b}_--\bar{f}_-^\dagger \bar{f}_-\label{disorderconstraint}.
\eea 
The transfer matrix at a node reads
\bea
R&=&\left(\mathcal{P}\otimes\bar{\mathcal{P}}\right)e^{\beta\left(b_+^\dagger\bar{b}_+^\dagger+f_+^\dagger\bar{f}_+^\dagger\right)+\beta^*\left(b_-^\dagger\bar{b}_-^\dagger+f_-^\dagger\bar{f}_-^\dagger\right)}\alpha^{\left(b_+^\dagger b_++f_+^\dagger f_+\right)}{\alpha^*}^{\left(b_-^\dagger b_-+f_-^\dagger f_-\right)}\nonumber\\ &&\quad\times\delta^{\left(\bar{b}_+^\dagger \bar{b}_+-\bar{f}_+^\dagger \bar{f}_+\right)}{\delta^*}^{\left(\bar{b}_-^\dagger \bar{b}_--\bar{f}_-^\dagger \bar{f}_-\right)}e^{\gamma\left(\bar{b}_+b_++\bar{f}_+f_+\right)+\gamma^*\left(\bar{b}_-b_-+\bar{f}_-f_-\right)}\left(\mathcal{P}\otimes\bar{\mathcal{P}}\right)
\label{transfermat}
\eea
Here, $\alpha,\beta,\gamma,\delta$ are the same as the parameters in the $S$-matrix  \eqref{Smatrix}. The operators  $\mathcal{P}/\bar{\mathcal{P}}$ (in the original model) project out states that do not respect the constraints \eqref{disorderconstraint}.

The truncation at level $M$ only keeps the first $1+4M$ states of \eqref{fockspaceeven} and \eqref{fockspaceodd}. Note that, 
since after this truncation we need to tune the value of $z$ to make the model critical,  the operator $\mathcal{P}$ (resp. $\bar{\mathcal{P}}$) gets modified in such a way that  for every state in \eqref{fockspaceeven} (resp. \eqref{fockspaceodd}), it gives a weight $z^N$ where $N=n_B+n_F$ is the number of bosons and fermions in the state, $0\leq N\leq 2M$. 

\begin{figure}
\centering
    \includegraphics[scale=.125]{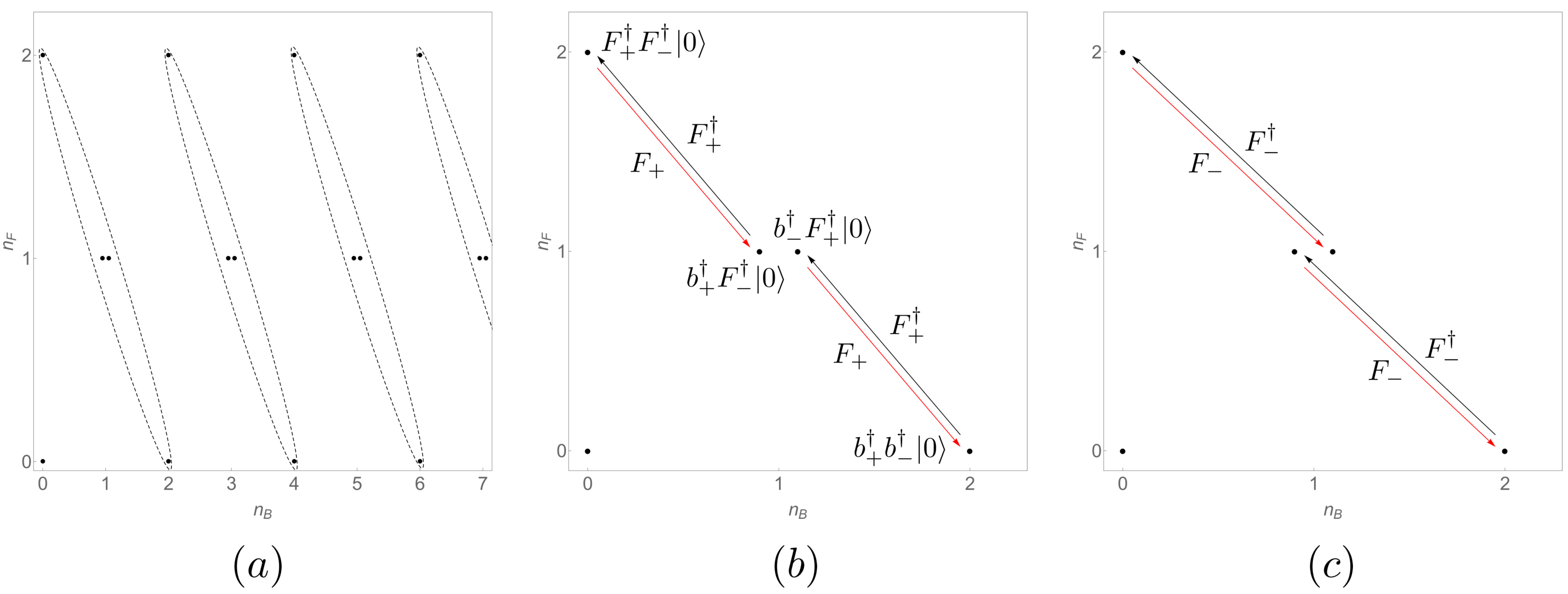}
    \caption{Structure of the Hilbert space on a given edge. The first panel $(a)$ represents the possible states corresponding to the quantum numbers $n_B$ and $n_F$. The vector space can be divided in quartet of states with an identical number of particles.The second and third panels ($(b)$ and $(c)$) shows the action of $F_+$ and $F_-$ within the first quartet (thus corresponding to the first truncation).}\label{Figu1}
\end{figure}

In the untruncated case (and thus  $z=1$), the transfer  matrix  (\ref{transfermat}) has $gl(2|2)$ symmetry. It is convenient in what follows to represent the Hilbert space on a given edge as in figure \ref{Figu1}. The state at the origin is the vacuum, while the ``quartets'' correspond to successive values of $n$ in (\ref{fockspaceeven}) or (\ref{fockspaceodd}), with the corresponding value of $N=n_B+n_F=2n+2$.  The four Cartans of $gl(2|2)$ correspond essentially to the numbers of bosons/fermions of type $\pm$: 
\begin{equation}
{\rm Cartan} = \left \lbrace b_+b_+^\dagger-{1\over 2}, f_+f_+^\dagger-{1\over 2},-b_-b_-^\dagger-{1\over 2},f_-f_-^\dagger-{1\over 2}\right \rbrace \,.
\end{equation}
The other  generators of this algebra can be conveniently encoded in matrix form as 
\begin{equation}
J=\left(\begin{array}{ccccc}
b_+b_+^\dagger-{1\over 2}&b_+f_+^\dagger&b_+b_-&b_+f_-\\
f_+b_+^\dagger& f_+f_+^\dagger-{1\over 2}&f_+b_-&f_+f_-\\
-b_-^\dagger b_+^\dagger&-b_-^\dagger f_+^\dagger&-b_-b_-^\dagger-{1\over 2}&-b_-^\dagger f_-\\
f_-^\dagger b_+^\dagger&f_-^\dagger f_+^\dagger&f_-^\dagger b_-& f_-f_-^\dagger-{1\over 2}\end{array}\right) \,. \label{J-gl22}
\end{equation}
We see that  some of these generators change the total number of particles: on the diagram they would move between different quartets, i.e. different values of  $n$. The first  non-trivial truncation is obtained by restricting to the first five states on a given edge, i.e., the vacuum and the first quartet ($n=1$).

In order to discuss the symmetries of the resulting model we introduce some notion about the $gl(1|1)$ symmetry.
It is easy to see that if we denote the particle numbers as 
\begin{subequations}
\begin{eqnarray}
b_+^\dagger b_++f_+^\dagger f_+ &=& N_+ \\
b_-^\dagger b_-+f_-^\dagger f_- &=& N_-
\end{eqnarray}
\end{subequations}
the new fermion operators defined as 
\begin{subequations}
\begin{eqnarray}
F_+ &\equiv& f_+ b_+^\dagger \,, \quad  F_+^\dagger= b_+ f_+^\dagger \,, \label{gl11+} \\
F_- &\equiv& b_-^\dagger f_- \,, \quad F_-^\dagger=f_-^\dagger b_- \label{gl11-}
\end{eqnarray}
\end{subequations}
obey
\begin{subequations}
\begin{eqnarray}
\{F_+,F_+^\dagger\} &=& N_+ \,, \\
\{F_-,F_-^\dagger\} &=& N_- \,.
\end{eqnarray}
\end{subequations}
These are well known to be $gl(1|1)$ commutation relations. And of course the $F_\pm ^{(\dagger)}$ generators are part of the $gl(2|2)$ symmetry: the matrix elements
$J_{1,2}$ and $J_{2,1}$ in \eqref{J-gl22} reproduce the fermionic generators \eqref{gl11+} of $gl(1|1)_+$, while $J_{3,4}$ and $J_{4,3}$ reproduce the generators
\eqref{gl11-} of $gl(1|1)_-$. Therefore we see that each quartet in figure \ref{Figu1} can be interpreted as the tensor  product of a pair of two-dimensional representations
of $gl(1|1)_+\times gl(1|1)_-$, where each of the $gl(1|1)$ acts on a given color ($\pm$). In our model moreover, $N_+=N_-$, so the two representations are isomorphic.
 
 This is all for one type of edge. Of course, for the other edges, we have the barred generators, with the conjugate action
\begin{eqnarray}
\{\bar{F}_+,\bar{F}^\dagger_+\}=-\bar{N}_+\nonumber\\
\{\bar{F}_-,\bar{F}^\dagger_-\}=-\bar{N}_-
\end{eqnarray}
which correspond to another (conjugate) pair of two-dimensional representations.

\subsection{Symmetries of the first truncation and its integrable modification}

Restricting to $N_+=N_-=1$ (i.e., $N=2$) gives the first non-trivial truncation. We see that it retains the $gl(1|1)_+\times gl(1|1)_-$ symmetry discussed above, here
acting on the tensor product $(V\otimes V^*)^{\otimes L}$, with the five-dimensional representations $V=1\oplus(\square_+\otimes \square_-)$ for even sites and $V^*=1\oplus(\bar{\square}_-\otimes \bar{\square}_+)$ for odd sites. We have here denoted the two-dimensional representations of $gl(1|1)$ simply by $\square$, $\bar{\square}$, without reference to the number $N$, as representations with different values of $N_\pm\neq 0$ are isomorphic.

For higher truncations, the symmetry is the same and acts again on a direct sum of the identity and products of pairs of two-dimensional representations.
Note that a key feature of the untruncated model is that the different quartets and the singlet are all related by action of $gl(2|2)$ generators:
this extra symmetry disappears in the truncations.

The presence of $V$ and $V^*$ modules is a consequence of the underlying orientation of the  Chalker-Coddington (CC) lattice. In geometrical terms, the arrows on the lattice edges define a unique orientation for all the loops.

A key feature of the integrable modification introduced in \cite{Ikhlef} is that it allows vertices which are not consistent with the CC orientation any more. This is a priori a dangerous manoeuver, since it breaks an underlying symmetry of the original (truncated as well as non-truncated) model.

\medskip

\begin{figure}[h]
\centering
\begin{tikzpicture}[scale=0.75]
\draw[red, line width=0.8mm, domain=-45:45] plot ({-1+0.707106*cos(\x)}, {0.707106*sin(\x)});
\draw[red, line width=0.8mm, domain=135:225] plot ({0.707106*cos(\x)}, {1+0.707106*sin(\x)});
\draw[red, line width=0.8mm, domain=45:135] plot ({0.707106*cos(\x)}, {1+0.707106*sin(\x)});
\draw[red, line width=0.8mm, domain=-135:-45] plot ({1+0.707106*cos(\x)}, {2+0.707106*sin(\x)});
\draw[red, line width=0.8mm, domain=-45:45] plot ({1+0.707106*cos(\x)}, {2+0.707106*sin(\x)});
\draw[red, line width=0.8mm, domain=45:135] plot ({1+0.707106*cos(\x)}, {2+0.707106*sin(\x)});
\draw[red, line width=0.8mm, domain=-135:-45] plot ({0.707106*cos(\x)}, {3+0.707106*sin(\x)});
\draw[red, line width=0.8mm, domain=-225:-135] plot ({0.707106*cos(\x)}, {3+0.707106*sin(\x)});
\draw[red, line width=0.8mm, domain=-45:45] plot ({-1+0.707106*cos(\x)}, {4+0.707106*sin(\x)});
\draw[red, line width=0.8mm, domain=-45:45] plot ({3+0.707106*cos(\x)}, {2+0.707106*sin(\x)});
\draw[red, line width=0.8mm, domain=-135:-45] plot ({3+0.707106*cos(\x)}, {2+0.707106*sin(\x)});
\draw[red, line width=0.8mm, domain=-225:-135] plot ({3+0.707106*cos(\x)}, {2+0.707106*sin(\x)});
\draw[red, line width=0.8mm, domain=-45:45] plot ({3+0.707106*cos(\x)}, {4+0.707106*sin(\x)});
\draw[red, line width=0.8mm, domain=45:135] plot ({3+0.707106*cos(\x)}, {4+0.707106*sin(\x)});
\draw[red, line width=0.8mm, domain=135:225] plot ({3+0.707106*cos(\x)}, {4+0.707106*sin(\x)});
\draw[red, line width=0.8mm, domain=-45:45] plot ({2+0.707106*cos(\x)}, {3+0.707106*sin(\x)});
\draw[red, line width=0.8mm, domain=135:225] plot ({4+0.707106*cos(\x)}, {3+0.707106*sin(\x)});
\foreach \x in {0,2,...,4}  
{ 
\foreach \y in {0,2,...,4}
{ 
\begin{scope}[very thick,decoration={
    markings,
    mark=at position 0.5 with {\arrow{stealth}}}
    ] 
    \draw[black,line width=1.0pt,postaction={decorate}] (\x-1,\y-1)--(\x,\y);
    \draw[black,line width=1.0pt,postaction={decorate}] (\x+1,\y+1)--(\x,\y);
    \draw[black,line width=1.0pt,postaction={decorate}] (\x,\y)--(\x-1,\y+1);
    \draw[black,line width=1.0pt,postaction={decorate}] (\x,\y)--(\x+1,\y-1);
\end{scope}
}}
\end{tikzpicture}  
  \caption{A configuration of the dilute oriented loop model.}
 \label{diluteconfiguration}
\end{figure}
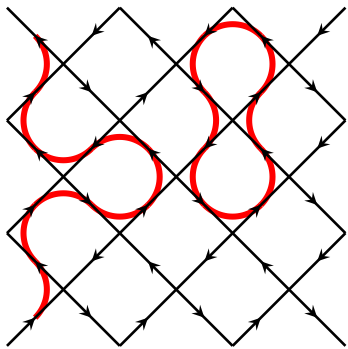

To understand more precisely the link between lattice orientation
and symmetry breaking, it is useful to consider a simpler version of the problem where only one color of loop would be allowed. The model consistent with the orientation of the CC lattice would then be a model of dilute self-avoiding loops on an oriented lattice. This model was studied in \cite{VJS15}, where it was called the {\em dilute oriented loop model}. A typical configuration of loops within this model is shown in figure \ref{diluteconfiguration}. 

The modified  model corresponding to this one-color toy model would then  be, by analogy, the ordinary dilute loop model on the square lattice \cite{WBN92}. This modification can be
made integrable by making an appropriate choice of the vertex weights.

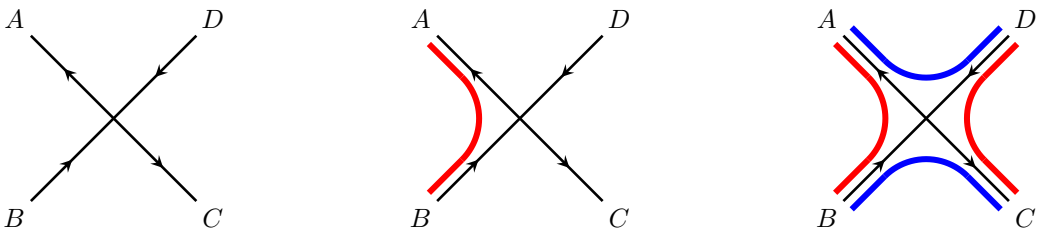
\begin{figure}[h]
\centering
\begin{tikzpicture}[scale=1.1]
\draw[black,line width=1.0pt,postaction={decorate,decoration={markings,mark=at position 0.5 with {\arrow{stealth}}}}] (-1,-1)--(0,0);
\draw[black,line width=1.0pt,postaction={decorate,decoration={markings,mark=at position 0.5 with {\arrow{stealth}}}}] (1,1)--(0,0);
\draw[black,line width=1.0pt,postaction={decorate,decoration={markings,mark=at position 0.6 with {\arrow{stealth}}}}] (0,0)--(0-1,1);
\draw[black,line width=1.0pt,postaction={decorate,decoration={markings,mark=at position 0.6 with {\arrow{stealth}}}}] (0,0)--(1,-1);
\draw (-1.2,1.2) node {$A$};
\draw (-1.2,-1.2) node {$B$};
\draw (1.2,-1.2) node {$C$};
\draw (1.2,1.2) node {$D$};
\end{tikzpicture}
\hspace{2cm}
\begin{tikzpicture}[scale=1.1]
\draw[red, line width=0.8mm, domain=-45:45] plot ({-1.2+0.707106*cos(\x)}, {0.707106*sin(\x)});
\draw[red, line width=0.8mm] (-0.7,-0.5)--(-1.1,-0.9);
\draw[red, line width=0.8mm] (-0.7,0.5)--(-1.1,0.9);
\draw[black,line width=1.0pt,postaction={decorate,decoration={markings,mark=at position 0.5 with {\arrow{stealth}}}}] (-1,-1)--(0,0);
\draw[black,line width=1.0pt,postaction={decorate,decoration={markings,mark=at position 0.5 with {\arrow{stealth}}}}] (1,1)--(0,0);
\draw[black,line width=1.0pt,postaction={decorate,decoration={markings,mark=at position 0.6 with {\arrow{stealth}}}}] (0,0)--(0-1,1);
\draw[black,line width=1.0pt,postaction={decorate,decoration={markings,mark=at position 0.6 with {\arrow{stealth}}}}] (0,0)--(1,-1);
\draw (-1.2,1.2) node {$A$};
\draw (-1.2,-1.2) node {$B$};
\draw (1.2,-1.2) node {$C$};
\draw (1.2,1.2) node {$D$};
\end{tikzpicture}
\hspace{2cm}
\begin{tikzpicture}[scale=1.1]
\draw[red, line width=0.8mm, domain=-45:45] plot ({-1.2+0.707106*cos(\x)}, {0.707106*sin(\x)});
\draw[red, line width=0.8mm] (-0.7,-0.5)--(-1.1,-0.9);
\draw[red, line width=0.8mm] (-0.7,0.5)--(-1.1,0.9);
\draw[red, line width=0.8mm, domain=-225:-135] plot ({1.2+0.707106*cos(\x)}, {0.707106*sin(\x)});
\draw[red, line width=0.8mm] (0.7,-0.5)--(1.1,-0.9);
\draw[red, line width=0.8mm] (0.7,0.5)--(1.1,0.9);
\draw[blue, line width=0.8mm, domain=45:135] plot ({0.707106*cos(\x)}, {-1.2+0.707106*sin(\x)});
\draw[blue, line width=0.8mm] (-0.5,-0.7)--(-0.9,-1.1);
\draw[blue, line width=0.8mm] (0.5,-0.7)--(0.9,-1.1);
\draw[blue, line width=0.8mm, domain=-135:-45] plot ({0.707106*cos(\x)}, {1.2+0.707106*sin(\x)});
\draw[blue, line width=0.8mm] (-0.5,0.7)--(-0.9,1.1);
\draw[blue, line width=0.8mm] (0.5,0.7)--(0.9,1.1);
\draw[black,line width=1.0pt,postaction={decorate,decoration={markings,mark=at position 0.5 with {\arrow{stealth}}}}] (-1,-1)--(0,0);
\draw[black,line width=1.0pt,postaction={decorate,decoration={markings,mark=at position 0.5 with {\arrow{stealth}}}}] (1,1)--(0,0);
\draw[black,line width=1.0pt,postaction={decorate,decoration={markings,mark=at position 0.6 with {\arrow{stealth}}}}] (0,0)--(0-1,1);
\draw[black,line width=1.0pt,postaction={decorate,decoration={markings,mark=at position 0.6 with {\arrow{stealth}}}}] (0,0)--(1,-1);
\draw (-1.2,1.2) node {$A$};
\draw (-1.2,-1.2) node {$B$};
\draw (1.2,-1.2) node {$C$};
\draw (1.2,1.2) node {$D$};
\end{tikzpicture}
 \caption{Examples of interactions at a vertex. The left panel shows an oriented vertex with no arc, corresponding to the identity in the action. The panel in the middle has an arc going from the edge $B$ to the edge $A$ corresponding in the action to the term $(\vec{z}_A^\dagger\cdot\vec{z}_B)$. The third panel on the right shows an other configuration with two colors coming from $(\vec{z}_A^\dagger \cdot \vec{z}_B)(\vec{z}_C^\dagger \cdot \vec{z}_D)(\vec{w}_A^\dagger \cdot \vec{w}_D)(\vec{w}_C^\dagger \cdot \vec{w}_B)$}
 \label{vectorcouplings}
\end{figure}

We now discuss (following \cite{VJS15}) a path-integral description of the one-color model for $n$ integer. In the oriented case, this is obtained by associating with every edge an $n$-dimensional complex vector $\vec{z}$ subject to $|\vec{z}|^2=1$.  Consider a vertex such as the one shown in the first diagram in figure \ref{vectorcouplings}. A  term $\vec{z}_A^\dagger \cdot \vec{z}_B$ will correspond to the arc drawn as in the second diagram in this figure. 

The convention is now that we go around a loop by following the arrows, and we take the dagger ($\dagger$) vector for the half edge entering a given vertex, and the non-dagger vector for the half-edge exiting the vertex. Each of $\vec{z}$ and $\vec{z}^\dagger$ appears once, due to the definition of loops. It is then possible to express the partition function of the
dilute oriented loop model as
\begin{equation}
Z\propto \int \prod_{\rm{edges}~ e} d\vec{z}_e \, e^{-S} \,.
\end{equation}
The corresponding action $S$ contains a total of seven terms: in addition to the identity it contains terms of the form $(\vec{z}_A^\dagger \cdot \vec{z}_B)(\vec{z}_C^\dagger \cdot \vec{z}_D)$ and $(\vec{z}_A^\dagger \cdot \vec{z}_D)(\vec{z}_C^\dagger \cdot \vec{z}_D)$, as well as 
$(\vec{z}_A^\dagger \cdot \vec{z}_B)$, $(\vec{z}_C^\dagger \cdot \vec{z}_D)$, $(\vec{z}_A^\dagger \cdot \vec{z}_D)$ and $(\vec{z}_C^\dagger \cdot \vec{z}_D)$. The loop diagrams
corresponding to two of these terms are shown in the first two panels of Figure~\ref{vectorcouplings}. This model has obviously $U(n)$ symmetry. Indeed, once the partition function is expanded in terms of loops, a local change of phase $\vec{z}_e\to e^{i\phi_e}\vec{z}_e$ does not change $Z$. As a matter of fact, the model is closely related to the sigma model on the complex projective space $CP^{n-1}=U(n)/U(n-1)\times U(1)$. 

Of course this definition works only for $n$ a positive integer. It can be extended to the case $n=0$ we are interested in after replacing $U(0)$ by the appropriate supergroup $U(p|p)$, with $p \ge 1$ integer. The simplest choice is obviously $U(1|1)$. 

It is possible to use, instead of a Euclidian version involving complex vectors, a transfer matrix or Hamiltonian version or the oriented one-color model. In this case, one needs to put on every edge the direct sum of the trivial and the fundamental representation for one orientation of $gl(1|1)$, and the
 direct sum of the trivial and the conjugate fundamental for the other. In other words, the ``Hilbert space'' is $(V\otimes V^*)^{\otimes L}$, with  $V=1\oplus\square$ for even sites and $V^*=1\oplus\bar{\square}$ for odd sites.  

\bigskip

Like the unmodified model, the modified one admits a different path-integral description where every edge now carries  a real $n$-dimensional vector $\vec{u}$. This is because loop pieces can connect either bonds of the CC lattice with compatible orientation, or bonds with opposite orientations, like those going straight through a vertex. This is only possible if the vectors are self-conjugate. One can then write a model with a similar kind of action, involving instead products of the type $(\vec{u}_A \cdot \vec{u}_B)(\vec{u}_C \cdot \vec{u}_D)$.
The possibility of going straight through a vertex will lead to two new interactions, $(\vec{u}_A \cdot \vec{u}_C)$ and $(\vec{u}_B \cdot \vec{u}_D)$, so that the expansion
over loops now gives a total of nine diagrams, in accordance with \cite{WBN92}.
The model has now $O(n)$ symmetry. 
Note that it is also possible to add an interaction of the form $(\vec{u}_A \cdot \vec{u}_C)(\vec{u}_B \cdot \vec{u}_D)$ that allows crossing at vertices without breaking
this symmetry.
Once the partition function is expanded in terms of loops, a local change of sign $\vec{u}_e\to -\vec{u}_e$ does not change the result. When $n=0$, $O(0)$ can be given sense using an orthosymplectic supergroup, the smallest of which is $OSp(2|2)$. 

{\sl Going from the oriented to the non-oriented model therefore breaks the symmetry from $U(n)$ to $O(n)$, or, in the case $n=0$, from $U(2|2)$ down to $OSp(2|2)$}. In general, such explicit symmetry-breaking does change the universality class. It was shown for instance in \cite{VJS15}  that, while the non-oriented model is in the universality class of  $O(n)$ criticality, the oriented model is in fact in the universality class of $O(2n)$. While these two universality classes {\sl are} different for general $n$, they coincide---remarkably---for $n=0$. The corresponding ``self-avoiding walk'' universality class is indeed extremely robust: not only is it insensitive to the lattice orientation, it is also known to be unaffected by the introduction of crossing vertices (giving rise to what is called self-avoiding trails \cite{JensenGuttmann98}), which remains compatible with $O(n)$ symmetry. 

\medskip

The symmetry analysis of the first truncation of the Chalker-Coddington model is very similar. In the Euclidian version, what we need now are two independent  complex vectors, $\vec{z}$ and $\vec{w}$, on each edge. Interactions are simple, but laborious to write down. For instance a diagram such as the third one in figure \ref{vectorcouplings}
corresponds to $(\vec{z}_A^\dagger \cdot \vec{z}_B)(\vec{z}_C^\dagger \cdot \vec{z}_D)(\vec{w}_A^\dagger \cdot \vec{w}_D)(\vec{w}_C^\dagger \cdot \vec{w}_B)$. The symmetry is obviously $U(n)\times U(n)\times Z_2$, the $Z_2$ corresponding to the symmetry between the two types of vectors (colors). In terms of spin chain, the Hilbert space for $n=0$  is 
 \begin{equation}
 {\cal H}=\left((1\oplus (\square_+\otimes\square_-))\otimes(1\oplus(\overline{\square}_-\otimes\overline{\square}_+))\right)^{\otimes L}
 \end{equation}
where $\square_+,\square_-$ are now  fundamental ($2p$-dimensional) representations of $U(p|p)$, and the total symmetry is now $U(p|p)\times U(p|p)\times Z_2$. It is then clear that the modification of this two-color model has lower symmetry, $O(n)\times O(n)\times Z_2$. For $n=0$ this becomes $OSp(2|2)\times OSp(2|2)\times Z_2$. 

While we lose symmetry when we  modify the model, interestingly, at its critical point, the modified truncation exhibits another kind of symmetry which is expected in the genuine, untruncated model, but would disappear in the first truncation per se. This symmetry can be seen qualitatively as relating the state with empty edges to the state with occupied edges, and is part of a $U_qso(5)$ symmetry, with $q=i$ for the point $n=0$ (this will be explicited in section \ref{sec:b21}).

The guess that the modified and unmodified truncated models are in the same universality class can of course be investigated numerically. This will be done in section~\ref{sec:intvsnonint} below.

Interestingly, we will find later on that the  universality class of the modified truncated  model is formally the same as the one of another model which {\sl does} respect the symmetry of the CC lattice: the $a_3^{(2)}$ integrable model \cite{AuyangPerk92,MartinsNienhuis98,FendleyJacobsen,VJS:a32}, which admits a loop formulation with the  following interaction:
\begin{eqnarray}
\check{R}&=& \left(\Rii +\Ref\right)+\lambda_c\left( \Rei + \Rif \right)  
\label{RloopUnmodified}
\end{eqnarray}

The situation is however a bit subtle because
integrable models can in general admit several different {\em regimes}, corresponding to different ranges of the ``crossing parameter'', denoted  $\theta$ in \eqref{eq:loopweightsintegrable}).
The critical exponents depend analytically on $\theta$ within a given regime, but distinct regimes are in general not related by analytical continuation. Therefore each regime
gives rise to a different universality class.

The regime of the $a_3^{(2)}$ which is of interest for us is the so-called regime III  \cite{VJS:a32}. This regime, however, does not include the value $n=0$ of the loop fugacity. Formally, however, the continuation of regime III for this model up to $n=0$ would give the same universality class as the one we shall find for the modified truncated model. This justifies {\sl a posteriori} that the modification does not affect the continuum limit significantly. Indeed, we will also find later that the continuum limit of the  modified truncated model exhibits a $U(n|n)\times U(n|n)\times Z_2$ symmetry in the continuum limit. In many ways, the  truncated  loop model is a "dilution" of the model in \cite{VJS:a32}. Like in the case of ordinary (single) loops, the dilution may change the relationship between the bare parameters and the field theories describing their continuum limit, but the two regimes are qualitatively very similar. 

\medskip
 As a last remark, we note that, while  the original truncated model can, after a gauge transformation, be formulated purely in terms of positive Boltzmann weights, the modified truncated model  involves some negative Boltzmann weights. This feature is however shared by higher truncations. It is not clear to what extent it is ``unphysical''.

\subsection{Modified and unmodified first truncations: a numerical comparison}
\label{sec:intvsnonint}

We finally turn to a numerical comparison of the exponents in the modified and unmodified truncated model. 
Conformal dimensions and the critical parameter $z_C$ can be studied using the ground state and excited states of the transfer matrix. Denoting by $\Lambda_{\ell_1,\ell_2}^L$ the largest eigenvalue of the sector $(\ell_1,\ell_2)$ propagating $\ell_1$ red strands and $\ell_2$ blue strands, we define approximations of   the associated dimension $x_{\ell_1,\ell_2}(L,z)$  by%
\footnote{The relation between dimensions and transfer matrix eigenvalues used here follows from the usual conformal mapping \cite{YellowBook}.}
\be
x_{\ell_1,\ell_2}(L,z)=-\frac{L}{2\pi}\log\frac{\Lambda_{\ell_1,\ell_2}(L,z)}{\Lambda_0(L,z)} \,,
\ee 
where $\Lambda_0(L,z)$ is simply the largest eigenvalue. The critical parameter $z_C$ can be obtained  using ``phenomenological renormalization'' \cite{Ikhlef} as the limit $L\rightarrow\infty$ of the series $z_C(L,L+2)$ solution of $x_{2,2}(L,z)=x_{2,2}(L+2,z)$. An estimate  of $x_{\ell_1,\ell_2}$  finally follows, either using for any $L$ the extrapolated value of $z_C$ found from phenomenological renormalization for the $x_{2,2}$ exponent, or using phenomenological renormalization size by size for the given exponent $x_{\ell_1,\ell_2}$. In the case of the integrable model things are simpler since the critical point is exactly known. 
The truncated model is not integrable, and due to the large Hilbert space (of dimension $5^L$ in the supersymmetric representation) only small sizes are reached using exact diagonalization. 

\begin{figure}
\centering
    \includegraphics[scale=.5]{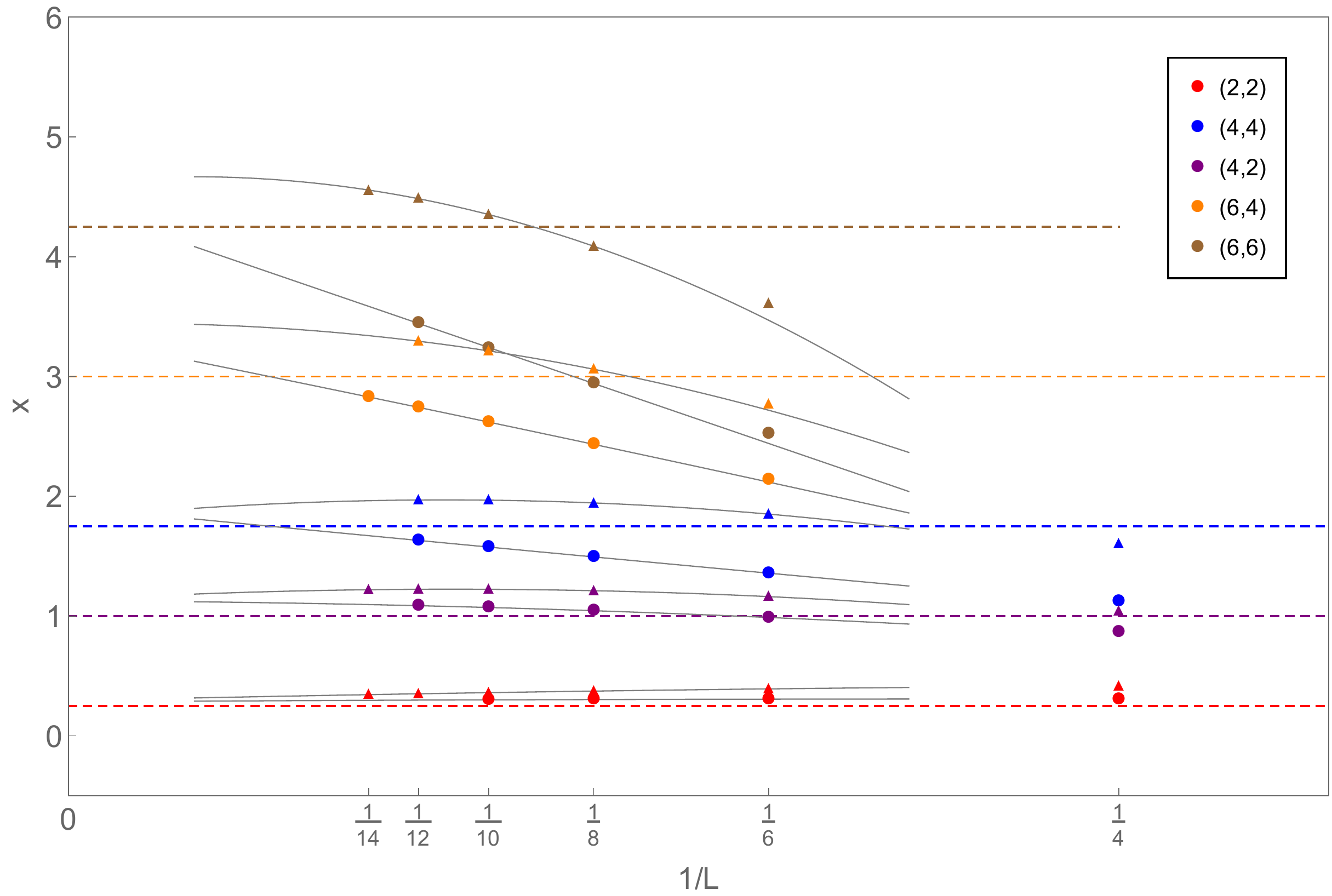}
   \caption{Estimate of dimensions $x_L=-\frac{L}{2\pi}\log{\lambda_i/\lambda_0}$ in even sectors for the truncated and integrable models. Each color refers to a different $(\ell_1,\ell_2)$ sector that propagates $\ell_1$ red strands and $\ell_2$ blue strands. Eigenvalues of the diagonal-to-diagonal transfer matrix are computed using exact diagonalization. The dots are estimates obtained from the integrable model. Triangles are obtained by considering the truncated unmodified model with a loop fugacity $z$ fixed to the estimated critical point $z_C=1.032$. Dashed colored lines are the exact  dimensions  corresponding to the watermelon exponents  $x_{\ell_1,\ell_2}$ which will be derived below in \eqref{eq:watermelons_s_n0}. Gray lines are polynomial fits of order $2$ over the $3$ or $4$ last data points for each exponent.  
}
\label{evensectors}
\end{figure}

Figure \ref{evensectors} shows numerical estimates in several sectors for the  unmodified truncated  model as well as  the modified integrable one, up to size $L=14$. It is clear  that despite strong finite-size corrections (expected to be logarithmic,  see below) the two models seem to have the same dimensions not only for the thermal state corresponding to the sector $(2,2)$ but also for higher excitations with even number of lines. 

\begin{figure}
\centering
    \includegraphics[scale=.5]{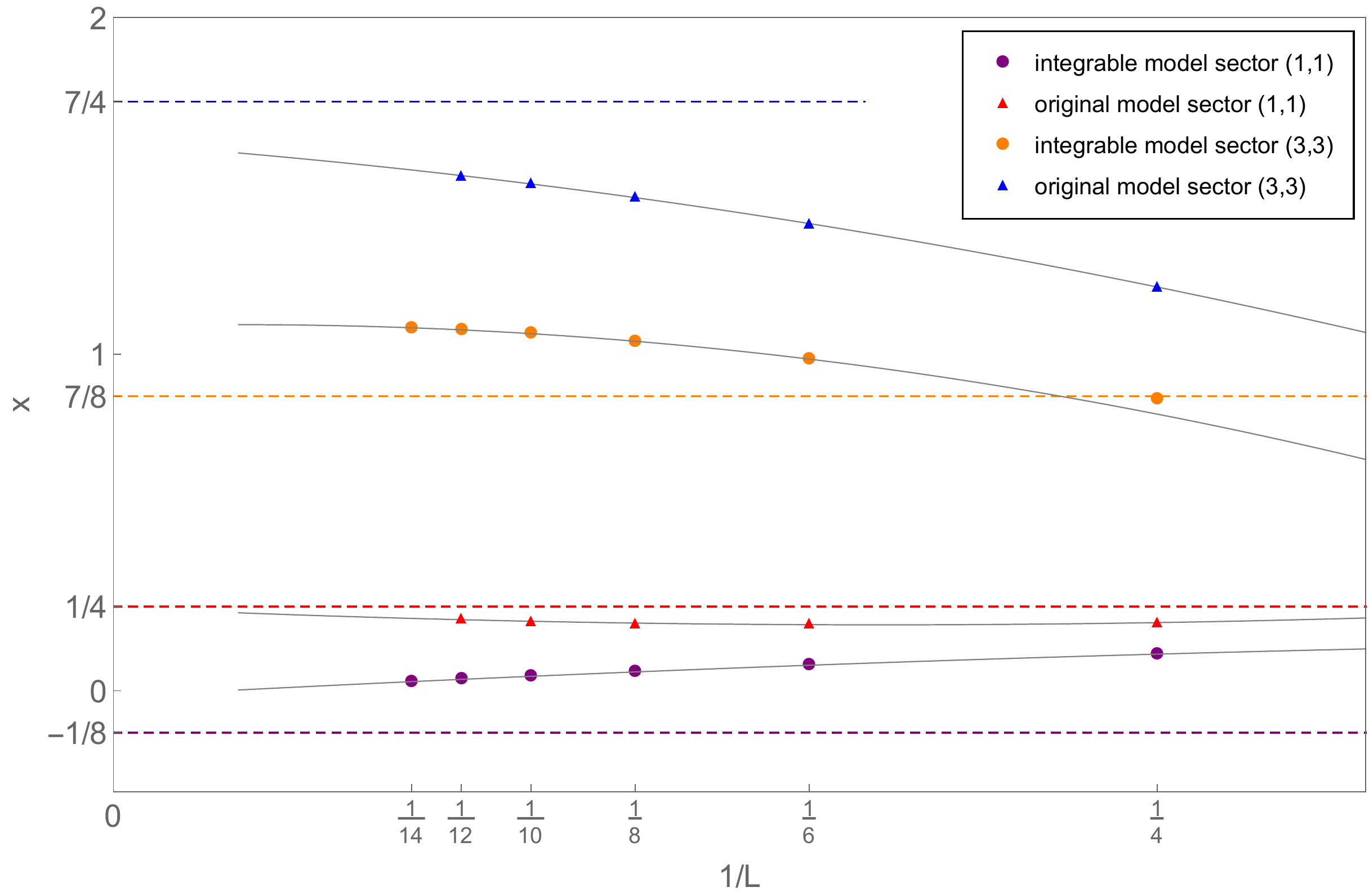}
    \caption{Estimate of  dimensions $x_L=-\frac{L}{2\pi}\log{\lambda_i/\lambda_0}$ in odd sectors for the truncated and integrable models. Eigenvalues of the diagonal-to-diagonal transfer matrix are computed using exact diagonalization. The dots are estimates obtained from the integrable model. Triangles are obtained by considering the truncated unmodified model with a loop fugacity $z$ fixed to the estimated critical point $z_C=1.032$. Dashed colored lines are the exact dimensions corresponding to the watermelon exponents. The propagation of $(1,1)$ lines corresponds to the purple color in the integrable model and red color in the truncated model. The propagation of $(3,3)$ lines corresponds to the orange color in the integrable model and blue color in the truncated model. Gray lines are polynomial fits of order $2$ over the $3$ or $4$ last data points for each exponent. It seems that the sectors $(1,1)$ and $(3,3)$ converge toward the exponent $(2,2)$ and $(4,4)$ in the truncated unmodified model---with values ${1\over 4}$ and ${7\over 4}$ respectively.  }

    \label{ComparisonModels}
\label{oddsectors}
\end{figure}

Meanwhile, we see that, while the integrable model has different dimensions for operators inserting an odd number of loops, these operators in the  unmodified truncated model seem to have the same dimensions as operators in the even sectors (cf Figure \ref{oddsectors}).  Similar features have been encountered sometimes in the past \cite{VJS15,Nahum}.

\section{The integrable dilute two-color model, and equivalence with $b_2^{(1)}$}
\label{sec:b21}

The integrable modification of the first truncation  was introduced in \cite{Ikhlef} by relating the two-color generators to those of a dilute Birman-Wenzl-Murakami algebra \cite{Grimm}. In the following, we will see how this modified model can also be related to the $b_{2}^{(1)}$ vertex model, which will allow us to use the well-known Bethe Ansatz equations of the latter.%
\footnote{
It is worth noting that the correspondence between dilutions of integrable models based on $D_n^{(1)}$ and integrable models based on $B_n^{(1)}$ had been already hinted at in \cite{Grimm2}. In the present case $n=2$, and $B_2^{(1)} = SO(4)^{(1)}$, which is built in terms of the generators of the $SO(4)$ BWM algebra.}

 To set the notations we define the quantum group deformation parameter $q=\mathrm{e}^{i \gamma}$, where $\gamma$ is related to the parameter $\theta$ used in \cite{Ikhlef} by 
\begin{equation}
\gamma = 2\pi - 2\theta \,.
\end{equation}
In particular the weight of closed loops is 
\be
n = -2 \cos \gamma \,,
\ee
and our main interest is the value $\gamma={\pi\over 2}$ for which $n=0$.

\subsection{The integrable $b_2^{(1)}$ $\check{R}$ matrix}

The $b_{2}^{(1)}$ model is defined by putting on each edge of a square lattice five-dimensional complex vector spaces (which correspond to the fundamental representation of the quantum group $U_q so(5)$), and specifying the local weights at each vertex in terms of the following  $\check{R}$ matrix \cite{Jimbo}, 
\begin{eqnarray}
\RR_{ab}(\lambda) &=& a(\lambda) \sum_{\stackrel{\alpha=1}{\alpha \neq \alpha'}}^{5} 
\hat{e}^{(a)}_{\alpha \alpha} \otimes \hat{e}^{(b)}_{\alpha \alpha}
+b (\lambda) \sum_{\stackrel{\alpha ,\beta=1}{\alpha \neq \beta,\alpha \neq \beta'}}^{5}
\hat{e}^{(a)}_{\beta \alpha} \otimes \hat{e}^{(b)}_{\alpha \beta}  \nonumber \\
&+& {\bar{c}} (\lambda) \sum_{\stackrel{\alpha ,\beta=1}{\alpha < \beta,\alpha \neq \beta'}}^{5} \hat{e}^{(a)}_{\alpha \alpha} \otimes \hat{e}^{(b)}_{\beta \beta}
+c (\lambda) \sum_{\stackrel{\alpha ,\beta=1}{\alpha > \beta,\alpha \neq \beta'}}^{5} \hat{e}^{(a)}_{\alpha \alpha} \otimes \hat{e}^{(b)}_{\beta \beta} \nonumber \\
&+& \sum_{\alpha ,\beta =1}^{5} d_{\alpha, \beta} (\lambda)
\hat{e}^{(a)}_{\alpha' \beta} \otimes \hat{e}^{(b)}_{\alpha \beta'} \,,
\label{RGM}
\end{eqnarray}
which is an endomorphism  $\mathbb{C}^5 \otimes \mathbb{C}^5 \to \mathbb{C}^5 \otimes \mathbb{C}^5$.
In the above formula to every index $\alpha=1,\ldots,5$ corresponds a conjugate index $\alpha' \equiv 6-\alpha$. The Boltzmann weights  $a(\lambda)$,
$b(\lambda)$, $c(\lambda)$ and ${\bar{c}} (\lambda)$ are determined by
\begin{subequations}
\label{bw1}
\begin{eqnarray}
a (\lambda) &=&(e^{2 \lambda} -\zeta)(e^{2 \lambda} -q^2) \,, \\
b (\lambda) &=&q(e^{2 \lambda} -1)(e^{2 \lambda} -\zeta) \,, \\
c (\lambda) &=&(1-q^2)(e^{2 \lambda} -\zeta) \,, \\
{\bar{c}} (\lambda) &=& e^{2 \lambda} c(\lambda) \,,
\end{eqnarray}
where $\zeta = q^3$, whilst $d_{\alpha\beta}(\lambda)$ has the form
\begin{equation}
d_{\alpha, \beta} (\lambda) = \left \lbrace
\begin{array}{ll}
 q(e^{2 \lambda} -1)(e^{2 \lambda} -\zeta) +e^{2\lambda}(q^2 -1)(\zeta -1) &
 \mbox{for } \alpha=\beta=\beta' \,, \\
 (e^{2 \lambda} -1)\left[ (e^{2 \lambda} -\zeta)q^{2} +e^{2\lambda}(q^2 -1) \right] &
 \mbox{for } \alpha=\beta \neq \beta' \,, \\
 (q^{2 }-1)\left[ \zeta(e^{2 \lambda} -1) q^{t_{\alpha}-t_{\beta}} -\delta_{\alpha ,\beta'} (e^{2\lambda} -\zeta) \right] &
 \mbox{for } \alpha < \beta \,, \\
 (q^{2 }-1) e^{2 \lambda} \left[ (e^{2 \lambda} -1) q^{t_{\alpha}-t_{\beta}} -\delta_{\alpha ,\beta'} (e^{2\lambda} -\zeta) \right] &
 \mbox{for } \alpha > \beta \,, \\
 \end{array} \right.
\end{equation}
\end{subequations}
where $t_{\alpha}=-{3 \over 2},-{1 \over 2},0,{1 \over 2},{3 \over 2}$ for $\alpha=1,2,3,4,5$ respectively.

\subsection{From the dilute two color model to $b_2^{(1)}$}
\label{sec:dilutetovertex}

We now show how the loop model (\ref{Rloop}) can me mapped onto the $b_2^{(1)}$ vertex model. 
In terms of the parameters $q$ and $z$ defined by\footnote{Note: in the following few equations, the parameter 
$z$ should not be confused with fugacity; $x(z)$ should not be confused with a  dimension.}

\begin{subequations}
\bea 
q &=& \mathrm{e}^{i \gamma} = \mathrm{e}^{i (2 \pi-2 \theta)} = \mathrm{e}^{- 2 i \theta} \\ 
z &=& \mathrm{e}^{- 2 i \varphi}
\eea
\end{subequations}
we first rewrite the weights (\ref{eq:loopweightsintegrable}) as
\begin{subequations}
\begin{eqnarray}
t(z) &=& \frac{q (z-1) \left(z-q^3\right)+\left(q^2-1\right) \left(q^3-1\right) z}{2 q^{5/2} z} \\
 u_1(z) &=& \frac{\left(1-q^2\right) \left(z-q^3\right)}{2 q^{5/2} z^{1/2}}\\
  u_2(z) &=&-\frac{\left(q^2-1\right) (z-1)}{2 q z^{1/2}}  \\
   v(z) &=& -\frac{(z-1) \left(z-q^3\right)}{2 q^{3/2} z} \\
    w_1(z) &=&  \frac{\left(z-q^2\right) \left(z-q^3\right)}{2 q^{5/2} z} \\
     w_2(z) &=& \frac{(z-1) \left(-q^3+\left(q^2-1\right) z+z\right)}{2 q^{5/2} z} \\
     x(z) &=& -\frac{(z-1) \left(q^3-z\right)}{2 q^{3/2} z}
\end{eqnarray}
\end{subequations}

Next, we assign orientations to loop segments, such that for each closed loop of either color we sum over the clockwise/anticlockwise orientations. 
The resulting loop weight is decomposed as 
\be
n = - 2 \cos \gamma = (-q) + (-q)^{-1} \,,
\ee
and we implement it in a local way by multiplying the weights of oriented vertices by the following angular contributions 
\be
\begin{tikzpicture}[scale=1,baseline={([yshift=-.5ex]current bounding box.center)}]
\draw[red,line width=0.3mm, rounded corners=7pt,-latex] (-0.4,-0.6) -- (0,0.) -- (0.4,-0.6);
\node at (0,-1) {$q^{1/2}$};
\end{tikzpicture}
\qquad 
\begin{tikzpicture}[scale=1,baseline={([yshift=-.5ex]current bounding box.center)}]
\draw[red,line width=0.3mm, rounded corners=7pt,latex-] (-0.4,-0.6) -- (0,0.) -- (0.4,-0.6);
\node at (0,-1) {$q^{-1/2}$};
\end{tikzpicture}
\qquad 
\begin{tikzpicture}[scale=1,baseline={([yshift=-.5ex]current bounding box.center)}]
\draw[blue,line width=0.3mm, rounded corners=7pt,-latex] (-0.4,-0.6) -- (0,0.) -- (0.4,-0.6);
\node at (0,-1) {$q^{-1/2}$};
\end{tikzpicture}
\qquad 
\begin{tikzpicture}[scale=1,baseline={([yshift=-.5ex]current bounding box.center)}]
\draw[blue,line width=0.3mm, rounded corners=7pt,latex-] (-0.4,-0.6) -- (0,0.) -- (0.4,-0.6);
\node at (0,-1) {$q^{1/2}$};
\end{tikzpicture}
\qquad 
\begin{tikzpicture}[scale=1,baseline={([yshift=-.5ex]current bounding box.center)}]
\draw[red,line width=0.3mm, rounded corners=7pt,-latex] (-0.4,0) -- (0,-0.6) -- (0.4,0);
\node at (0,-1) {$-q^{-1/2}$};
\end{tikzpicture}
\qquad 
\begin{tikzpicture}[scale=1,baseline={([yshift=-.5ex]current bounding box.center)}]
\draw[red,line width=0.3mm, rounded corners=7pt,latex-] (-0.4,0) -- (0,-0.6) -- (0.4,0);
\node at (0,-1) {$-q^{1/2}$};
\end{tikzpicture}
\qquad 
\begin{tikzpicture}[scale=1,baseline={([yshift=-.5ex]current bounding box.center)}]
\draw[blue,line width=0.3mm, rounded corners=7pt,-latex] (-0.4,0) -- (0,-0.6) -- (0.4,0);
\node at (0,-1) {$-q^{1/2}$};
\end{tikzpicture}
\qquad 
\begin{tikzpicture}[scale=1,baseline={([yshift=-.5ex]current bounding box.center)}]
\draw[blue,line width=0.3mm, rounded corners=7pt,latex-] (-0.4,0) -- (0,-0.6) -- (0.4,0);
\node at (0,-1) {$-q^{-1/2}$};
\end{tikzpicture}
\label{eq:angularcontribution}
\ee
Everything being now local, we can reinterpret the weights as those of a vertex model, every edge of the Chalker-Coddington lattice carrying a five-dimensional vector space of basis 
\be 
-2, -1, 0, 1, 2 =
\qquad 
\begin{tikzpicture}[scale=1,baseline={([yshift=-.5ex]current bounding box.center)}]
\draw[blue,line width=0.3mm, latex-] (-0.1,-0.6) -- (-0.1,0.);
\draw[red,line width=0.3mm, latex-] (-0.,-0.6) -- (0,0.);
\end{tikzpicture}
\qquad
\begin{tikzpicture}[scale=1,baseline={([yshift=-.5ex]current bounding box.center)}]
\draw[blue,line width=0.3mm, -latex] (-0.1,-0.6) -- (-0.1,0.);
\draw[red,line width=0.3mm, latex-] (0,-0.6) -- (0,0.);
\end{tikzpicture}
\qquad
\begin{tikzpicture}[scale=1,baseline={([yshift=-.5ex]current bounding box.center)}]
\draw[dashed,line width=0.3mm, ] (0,-0.6) -- (0,0.);
\end{tikzpicture}
\qquad
\begin{tikzpicture}[scale=1,baseline={([yshift=-.5ex]current bounding box.center)}]
\draw[blue,line width=0.3mm, latex-] (-0.1,-0.6) -- (-0.1,0.);
\draw[red,line width=0.3mm, -latex] (0,-0.6) -- (0,0.);
\end{tikzpicture}
\qquad
\begin{tikzpicture}[scale=1,baseline={([yshift=-.5ex]current bounding box.center)}]
\draw[blue,line width=0.3mm, -latex] (-0.1,-0.6) -- (-0.1,0.);
\draw[red,line width=0.3mm, -latex] (0,-0.6) -- (0,0.);
\end{tikzpicture}
\,,
\ee
respectively.
Each color is therefore associated with a spin $S_z^{(1,2)} = -1,0,1$, namely the states can be rewritten as 
\be 
-2, -1, 0, 1, 2 = |--\rangle, |-+\rangle, |00\rangle, |+-\rangle, |++\rangle \,,
 \ee
 such that both $S_z^{(1,2)}$ are individually conserved at a given vertex. 
Finally, we perform a gauge transformation on the vertex weights, namely  
 \be
 \begin{tikzpicture}[scale=1,baseline={([yshift=-.5ex]current bounding box.center)}]
\draw[black,line width=0.3mm] (-0.5,-0.5) node[below] {$a$} -- (0.5,0.5)node[above] {$d$};
\draw[black,line width=0.3mm] (0.5,-0.5) node[below] {$b$} -- (-0.5,0.5)node[above] {$c$} ;
\end{tikzpicture}
\qquad
\longrightarrow
\qquad
z^{ \frac{1}{4}\left( S_z^{(1)}(b)+ S_z^{(1)}(d)-S_z^{(1)}(a) -S_z^{(1)}(c)\right)}
(-1)^{\delta_{a,0}-\delta_{c,0}}
\qquad 
 \begin{tikzpicture}[scale=1,baseline={([yshift=-.5ex]current bounding box.center)}]
\draw[black,line width=0.3mm] (-0.5,-0.5) node[below] {$a$} -- (0.5,0.5)node[above] {$d$};
\draw[black,line width=0.3mm] (0.5,-0.5) node[below] {$b$} -- (-0.5,0.5)node[above] {$c$} ;
\end{tikzpicture} 
\,.
 \ee
 This results in a local $\check{R}$ matrix which is exactly proportional (namely by a factor $\frac{i}{2}$) to the integrable $b_2^{(1)}$ $\check{R}$-matrix (\ref{RGM}), where the states $1,2,3,4,5$ are in one-to-one correspondence with the $-2,-1,0,1,2$ used here, and where the spectral parameters are related through 
 \be 
 z = e^{2  \lambda} = e^{-2 i \varphi}
 \ee 
 
 We have argued in the first section that the modified truncated loop model could be formulated as a model with $Osp(2|2)$ symmetry. We see in fact that it is also possible to formulate it as a model with {\sl more symmetry}, namely $U_qso(5)$. As observed in other polymer problems (see, e.g., the dilute self-avoiding walks on the hexagonal lattice), the presence of this extra quantum group symmetry does not seem to be simply related with features of the continuum limit \cite{FSZ}.

\subsection{Transfer matrix formulation and boundary conditions}
\label{sec:twists}

To achieve a complete identification between the loop and vertex models, one also needs to properly choose the boundary conditions. 
In this paper we shall consider the loop model defined on a cylinder, namely with periodic boundary conditions in the horizontal direction.

It will be convenient, here and in the following, to work in the framework of transfer matrices, whose successive applications build up the various loops or vertex configurations by constructing the whole lattice row by row.
The vertex and loop transfer matrices are defined respectively as  
 \bea 
 T^{\rm vertex}(\lambda) &=& \mathrm{Tr}_0 \left( \check{R}^{\rm vertex}_{0,L}(\lambda)\check{R}^{\rm vertex}_{0,L-1}(\lambda) \ldots \check{R}^{\rm vertex}_{0,1}(\lambda) \right)  \,, \nonumber \\ 
  T^{\rm loops}(\lambda) &=& \mathrm{Tr}_0 \left( \check{R}^{\rm loop}_{0,L}(\lambda) \check{R}^{\rm loop}_{0,L-1}(\lambda) \ldots \check{R}^{\rm loop}_{0,1}(\lambda) \right) \,. 
 \eea  
They act on a Hilbert space constituted by the $L$ lattice sites in the horizontal direction, and are expressed as traces over an auxilliary space labeled $0$. As usual with quantum integrable models, transfer matrices corresponding to different choices of the spectral parameter $\lambda$ commute with one another.  
The essential physical information about the model is contained in the spectrum (and in particular the leading eigenvalues) of the transfer matrices, whose diagonalization will be presented in the following. 
To proceed, we first need to give some detail about the structure of the Hilbert space in each case.

\paragraph{Structure of the Hilbert space for the vertex model.}
For the vertex model, the Hilbert space on $L$ lattice sites is simply the tensor product $(\mathbb{C}^5)^{\otimes L}$. From the observations of section~\ref{sec:dilutetovertex}, it is easy to see that the total magnetizations 
\be
S_z^{(1)} = \sum_{i=1}^L (S_z^{(1)})_i \,, \qquad\qquad
S_z^{(2)} = \sum_{i=1}^L (S_z^{(2)})_i 
\ee
are individually conserved by the transfer matrix, so the latter can be diagonalized separately in each sector of specified $S_z^{(1)}, S_z^{(2)}$.

\paragraph{Structure of the Hilbert space for the loop model.}
The states of the loop model on $L$ sites are expressed as a set of connectivities, analogously to the case for simpler loop models  \cite{BloteNienhuis89}. For a given state some of the $L$ points can be empty, and the remaining points are occupied by pairs of blue and red loop segments, which either connect to other sites via an arc or remain single, which corresponds to ``legs'' propagating all the way from the bottom of the lattice.
It is easily seen that the loop transfer matrix has a block-triangular structure, as the number of legs of each color, called $\ell_1$ and $\ell_2$ respectively, can only be lowered (by contracting one leg with another) or kept fixed from one row to the next. In order to study the eigenspectrum we can therefore focus on the diagonal part, and hence study separately the different subsectors of fixed $\ell_1,\ell_2$. 
Up to subtletites to be discussed below, these sectors can be precisely related to the sectors of fixed magnetizations of the vertex model, with the correspondence 
\begin{equation}
S_z^{(1)} = \frac{\ell_1}{2} \,, \qquad S_z^{(2)} = \frac{\ell_2}{2} \,.
\label{conservedSz}
\end{equation}

In order to make the correspondence between loop and vertex transfer matrices complete, there is a last aspect that must be taken care of, namely the presence of non-contractible loops in the sectors where one or both of the $\ell_i$ is zero.   
 Because of the periodicity in the horizontal direction, there can exist in such sectors closed loops of color $i$ that wind horizontally around the periodic direction, which must have the same weight as the contractible ones. 
 In the vertex model formulation, this is guaranteed by inserting in the above definition of $ T^{\rm vertex}(u)$ a twist factor of the form 
 \begin{equation}
 \mathrm{e}^{ \mathrm{i} \left( \varphi_1 (S_z^{(1)})_0 +\varphi_2 (S_z^{(2)})_0 \right)  } \,,
\end{equation}
where $ (S_z^{(i)})_0 = \pm 1/2$ are the local magnetizations of the auxilliary space. The geometrical interpretation of such a twist is the following: a non-contractible red (resp.\ blue)
loop, closing around the axis of the cylinder does not get a factor of the type \eqref{eq:angularcontribution} because it does not turn through any angle, but instead comes with a factor $\mathrm{e}^{\pm i \varphi_{i}}$, depending on its orientation. Summing over both, we get for each non-contractible loop a fugacity 
\be
\tilde{n}_i = 2 \cos \varphi_{i} \,,
\ee
therefore this weight can be tuned to be the same as the weight $n$ or contractible loops by conveniently choosing $\varphi_1$ and $\varphi_2$. More precisely,
\begin{itemize}
 \item When $S_z^{(i)}=0$ there can exist non-contractible loops of color $i$. To recover $\tilde{n}_i = n = -2 \cos \gamma$, one must choose 
\be
\varphi_i  = \pi- \gamma \,. 
\ee
 \item When $S_i^{(z)} \neq 0$ the presence of through-lines (legs) forbids the presence of non-contractible loops of color $i$. The correct choice is then $\varphi_i = 0$, since otherwise the through-lines would pick up spurious phase factors when spiraling around the horizontal, periodic direction.
\end{itemize}

\subsection{Bethe ansatz equations}

The Bethe Ansatz construction for the $b_2^{(1)}$ vertex model is well known, see for instance \cite{Reshetikhin}. We use here the results \cite[eqs. (56) and (61)]{GalleasMartins04}.
The eigenvalues of the transfer matrix are expressed in terms of the set of $m_1$ roots $\{\lambda_i\}$ and $m_2$ roots $\{\mu_k\}$---each of which will henceforth be called a $\lambda$-root or a $\mu$-root---, related to the conserved magnetizations $S_z^{(1)},S_z^{(2)}$ (defined in \eqref{conservedSz}) by 
\begin{eqnarray}
m_1 &=& L-S_z^{(1)}-S_z^{(2)} \\
m_2 &=& L-2 S_z^{(1)} \,.
\label{eq:misi}
\end{eqnarray}
These are solution of a set of nested Bethe ansatz equations, which for purely periodic boundary conditions read
\begin{subequations}
\label{eq:untwisted_BAE}
\begin{align}
\left[  \frac{\sinh\left(\lambda_i-\mathrm{i}\frac{\gamma}{2}\right)} {\sinh\left(\lambda_i+\mathrm{i}\frac{\gamma}{2}\right)} \right]^L 
&= \prod_{j(\neq i)}^{m_{1}}{\frac{\sinh\left(\lambda_i-\lambda_j-\mathrm{i}\gamma\right)}{\sinh\left(\lambda_j-\lambda_i+\mathrm{i}\gamma\right)}}
\prod_{k=1}^{m_{2}}{\frac{\sinh\left(\lambda_i-\mu_k+\mathrm{i}\frac{\gamma}{2}\right)}{\sinh\left(\lambda_i-\mu_k-\mathrm{i}\frac{\gamma}{2}\right)}} \\
\prod_{i=1}^{m_{1}}{\frac{\sinh\left(\mu_k-\lambda_i-\mathrm{i}\frac{\gamma}{2}\right)}{\sinh\left(\mu_k-\lambda_i+\mathrm{i}\frac{\gamma}{2}\right)}}
&= \prod_{l(\neq k)}^{m_{2}}{\frac{\sinh\left(\mu_k-\mu_l-\mathrm{i}\frac{\gamma}{2}\right)}{\sinh\left(\mu_k-\mu_l+\mathrm{i}\frac{\gamma}{2}\right)}} \,.
\end{align}
\end{subequations}

The energy of the spin chain Hamiltonian, obtained as a completely anisotropic limit $\lambda\to 0$ of the transfer matrix, is
\begin{equation}
 E = -\epsilon \sum_{j=1}^{m_1} \frac{\sin \gamma}{\cosh2\lambda_j - \cos\gamma} \,,
 \label{Edef}
\end{equation}
where the two possible signs $\epsilon=\pm 1$ determine different regimes governed by a different physics, as will be detailed in the next paragraph. 
In particular the correspondence with the energy obtained from the loop Hamiltonian (given in eq. (3.16) of \cite{Ikhlef}) is 
\be
E_{\rm loops} = 4 \sin \gamma \sin \frac{3\gamma}{2} E \,.
\label{EloopsEBA}
\ee

In presence of the twist parameters introduced in paragraph (\ref{sec:twists}), the Bethe ansatz equations get modified as
\begin{subequations}
\label{BAEtwisted}
\begin{align}
\mathrm{e}^{2 \mathrm{i} \varphi_2} \left[  \frac{\sinh\left(\lambda_i-\mathrm{i}\frac{\gamma}{2}\right)} {\sinh\left(\lambda_i+\mathrm{i}\frac{\gamma}{2}\right)} \right]^L 
&= \prod_{j(\neq i)}^{m_{1}}{\frac{\sinh\left(\lambda_i-\lambda_j-\mathrm{i}\gamma\right)}{\sinh\left(\lambda_j-\lambda_i+\mathrm{i}\gamma\right)}}
\prod_{k=1}^{m_{2}}{\frac{\sinh\left(\lambda_i-\mu_k+\mathrm{i}\frac{\gamma}{2}\right)}{\sinh\left(\lambda_i-\mu_k-\mathrm{i}\frac{\gamma}{2}\right)}}
\label{BAEtwisted1} \\
\mathrm{e}^{\mathrm{i} (\varphi_1-\varphi_2)} \prod_{i=1}^{m_{1}}{\frac{\sinh\left(\mu_k-\lambda_i-\mathrm{i}\frac{\gamma}{2}\right)}{\sinh\left(\mu_k-\lambda_i+\mathrm{i}\frac{\gamma}{2}\right)}}
&= \prod_{l(\neq k)}^{m_{2}}{\frac{\sinh\left(\mu_k-\mu_l-\mathrm{i}\frac{\gamma}{2}\right)}{\sinh\left(\mu_k-\mu_l+\mathrm{i}\frac{\gamma}{2}\right)}}
\label{BAEtwisted2}
\end{align}
\end{subequations}
and the corresponding transfer matrix eigenvalues are given by 
\begin{eqnarray}
 \Lambda(\lambda) &=& \mathrm{e}^{- \mathrm{i}(\varphi_1+\varphi_2)} a(\lambda)^L \frac{Q_1\left(\lambda + \mathrm{i} {\gamma \over 2}\right)}{Q_1\left(\lambda - \mathrm{i} {\gamma \over 2}\right)} +  \mathrm{e}^{\mathrm{i}(\varphi_1+\varphi_2)} d_{5,5}(\lambda)^L \frac{Q_1\left(\lambda - 2 \mathrm{i} \gamma\right)}{Q_1\left(\lambda - \mathrm{i} \gamma\right)} 
 \nonumber \\ & & +
 b(\lambda)^L \left( \mathrm{e}^{\mathrm{i}(-\varphi_1+\varphi_2)} G_1(\lambda) + G_2(\lambda)   + \mathrm{e}^{\mathrm{i}(\varphi_1-\varphi_2)} G_3(\lambda)   \right) \,, 
 \end{eqnarray}
where the coefficients $a(\lambda), b(\lambda), d_{5,5}(\lambda)$ were defined in eqs. (\ref{bw1}), while 
\bea 
Q_1(\lambda) &=& \prod_{i=1}^{m_1} \sinh(\lambda-\lambda_i) \nonumber \\ 
Q_2(\lambda) &=& \prod_{k=1}^{m_2} \sinh(\lambda-\mu_k) \nonumber \\ 
G_1(\lambda) &=& \frac{Q_1\left(\lambda- \mathrm{i}\frac{3\gamma}{2}\right)}{Q_1\left(\lambda- \mathrm{i}\frac{ \gamma}{2}\right)}
\frac{Q_2\left(\lambda\right)}{Q_2\left(\lambda- \mathrm{i}\gamma\right)}
  \nonumber \\ 
G_2(\lambda) &=&
\frac{Q_2\left(\lambda- \mathrm{i}\frac{3 \gamma}{2}\right)}{Q_2\left(\lambda- \mathrm{i}\frac{\gamma}{2}\right)}
\frac{Q_2\left(\lambda\right)}{Q_2\left(\lambda- \mathrm{i}\gamma\right)}
  \nonumber \\ 
G_3(\lambda) &=&  \frac{Q_1\left(\lambda\right)}{Q_1\left(\lambda- \mathrm{i} \gamma\right)} \frac{Q_2\left(\lambda-  \mathrm{i}\frac{3 \gamma}{2}\right)}{Q_2\left(\lambda- \mathrm{i}\frac{\gamma}{2}\right)} \,. 
\eea 
(our notations are borrowed from \cite{GalleasMartins04}).

\subsection{Isotropic points and regimes}

First, a word on the range of $\gamma$ we need to  study. 
Notice that the model is $4\pi$ periodic in $\gamma$. It is also left unchanged by the transformation $\gamma \to - \gamma$.
We moreover see that the two regimes corresponding to the two different signs of $\epsilon$ in the expression of the energy are related by $\gamma \to 2\pi - \gamma$, therefore restricting  the range of values of $\gamma$ to be studied to the interval to $[0,\pi]$. In terms of the original parameter $\theta$ (recall, $\gamma=2\pi-2\theta$ and $n=-2\cos 2\theta$), this means $\theta \in \left[\frac{\pi}{2},\pi \right]$.

For a given value of $\gamma$, there are two isotropic values of the spectral parameter $\lambda$, corresponding to local maxima of the transfer matrix eigenvalues, and which are described by a different physics in the sense that they are not dominated by the same transfer matrix eigenstates. 
These are found to be 
\begin{equation}
 \lambda_+ = \mathrm{i}\frac{3\gamma}{4}  , \qquad  \lambda_- = \mathrm{i} \left(\frac{3\gamma}{4} + \frac{\pi}{2}\right) \,,
\end{equation}
and correspond, in the Hamiltonian point of view, to the two possible choices of sign $\epsilon = \pm 1$ in the definition of the energy (\ref{Edef}).  

As we will detail, this allows us to define  four  different regimes: 
\begin{center}
\begin{tabular}{c|c|c|c|c}
Regime~ &  ~Range of $\gamma$~  & ~Range of $\theta$~  & ~Isotropic point~ &  Sign of the Hamiltonian \\
\hline 
& & & &  \\ 
I &  $0 \leq \gamma \leq {2 \pi \over 3}$  &  ${2 \pi \over 3} \leq \theta \leq {\pi }$  & $\lambda =\mathrm{i}\frac{3\gamma}{4}$ &  $\epsilon = +1$ \\ 
& & & &  \\ 
II &  ${2 \pi \over 3} \leq \gamma \leq \pi$  &  ${ \pi \over 2} \leq \theta \leq {2 \pi  \over 3}$ & $\lambda =\mathrm{i}\frac{3\gamma}{4}$ &  $\epsilon = +1$ \\ 
& & & &  \\ 
III &  ${2 \pi \over 3} \leq \gamma \leq \pi$  &  ${ \pi \over 2} \leq \theta \leq {2 \pi  \over 3}$  & $\lambda =\mathrm{i} \left(\frac{3\gamma}{4} + \frac{\pi}{2}\right)$ &  $\epsilon = -1$ \\ 
& & & &  \\ 
IV &  $0 \leq \gamma \leq {2 \pi \over 3}$  &  ${2 \pi \over 3} \leq \theta \leq {\pi }$  & $\lambda =\mathrm{i} \left(\frac{3\gamma}{4} + \frac{\pi}{2}\right)$ &  $\epsilon = -1$ \\ 
\end{tabular} \,,
\end{center}
which are exactly those defined in table 2 of \cite{Ikhlef}.
Let us mention that the frontier between regimes I and II is a crossover. Its location is not precisely $\gamma = {2 \pi \over 3}$ at finite sizes, but goes precisely to this value in the thermodynamic limit. 

The point of interest for the study of the IQHE lies in the Regime IV, and corresponds to $n=0$, $\gamma = \frac{\pi}{2}$. In the following, this is the regime we will focus on. 
However, before doing so, it will  be useful to recall some results about the $a_2^{(2)}$ and $a_3^{(2)}$ models, which  exhibit, in the continuum limit, many features of interest to the present case.

\section{The $a_2^{(2)}$ and $a_3^{(2)}$ models, and the black hole CFT}
\label{sec:an2}

In \cite{AuyangPerk92,MartinsNienhuis98,FendleyJacobsen,VJS:a32}, a dense version of the two-color loop model was studied, which  shares important common features with the dilute one. For the dense model integrable weights were also derived, and were associated in \cite{VJS:a32} with those of the $a_3^{(2)}$ integrable vertex model. 
The $a_3^{(2)}$ model is part of a more general hierarchy of integrable vertex models, associated with the $a_r^{(2)}$ twisted affine algebras ($r\geq 2$) \cite{VJS:a22,VJS:an2}.

All of these models involve similarly a parameter $\gamma$, in terms of which the quantum group deformation parameter reads $q = e^{i \gamma}$, and whose range can be restricted, by using symmetries, and considering the two possible signs of the corresponding Hamiltonians, to $[0,\pi]$ for $r$ even and to $[0,\pi/2]$ for $r$ odd. In all cases, varying $\gamma$ and changing the sign of the Hamiltonian has been observed in \cite{VJS:an2} to lead to three different regimes, labeled I, II and III, the latter of which will be of particular interest to us here. In this section, we will therefore recall some details about the continuum limits of the $a_2^{(2)}$ and $a_3^{(2)}$ models in their regimes III.

\subsection{The $a_2^{(2)}$ model in regime III} 

The regime III of the $a_2^{(2)}$ model is associated with $\gamma \in \left[0, \frac{\pi}{4}  \right]$ and the antiferromagnetic sign of the Hamiltonian (equivalently, a given range of spectral parameters in the formulation as two-dimensional vertex model), and was studied in detail in \cite{VJS:a22}.
 
The model can be interpreted as a  spin-1 chain, whose global magnetization $m$ is conserved by the Hamiltonian and transfer matrix. Its eigenstates can therefore be parametrized by the corresponding eigenvalue of $m$, and, in the case of periodic boundary conditions in the space direction, by the value of the corresponding momentum. For simplicity we will focus on the zero-momentum, translationally invariant sector, in which the ground state and lowest excitations were found. 

As usual in gapless spin chains, the finite-size scaling of the low-energy levels can be used to obtain the corresponding central charge and conformal weights. More precisely, the energy of the ground state $E_0(L)$ and excitations $E_i(L)$  obey for $L$ large the following scaling relations \cite{YellowBook}
\begin{subequations}
\label{finitesizescaling}
\bea
E_0(L) &=&  L e_\infty  - \frac{\pi v_f}{6 L^2} c  + \ldots  \,, \\
E_i(L) - E_0(L) &=&  \frac{2 \pi v_f}{ L^2} (\Delta_i + \bar{\Delta}_i)  + \ldots \,,
\eea 
\end{subequations}
where $e_\infty$ is the non-universal extensive part to the energy, and $v_f$ is the Fermi velocity of the excitations, which can be computed for instance through Bethe ansatz. 
Here, and for all the models that we are concerned with in this paper, it is convenient to consider the effective central charges $c_i$ associated with the various excited levels, defined through 
\be
E_i(L) =  L e_\infty  - \frac{\pi v_f}{6 L^2} c_i  + \ldots \,,
\label{eq:cidef}
\ee 
that is, $c_i \equiv c - 12 (\Delta_i + \bar{\Delta}_i)$.  

In the case of periodic boundary conditions (thus with zero twist parameters, $\varphi_1 = \varphi_2 = 0$), it has been found in \cite{VJS:a22} that each sector of fixed momentum and magnetization contains a tower of eigenlevels parametrized by another index $j=0, 1,2 \ldots$, whose interpretation is best obtained  in terms of the corresponding Bethe roots structure. 
Restricting for simplicity to the zero-momentum sector, the effective central charges associated with different values of $m$ and $j$ were found after considerable analytical and numerical work to behave as 
 \begin{equation}
-\frac{c_{m,j}}{12} = -\frac{2}{12}+m^2\frac{\gamma}{4 \pi} + \left(N_{m,j}\right)^2 \frac{A(\gamma)}{\left[ B_{m,j}(\gamma) + \log L \right]^2}  \,,
\label{eq:a22:cmj}
\end{equation}
 where $ N_{m,j} = \frac{3 + (-1)^{m+1}}{2} + 2 j $ , that is, for $j$ large, $N_{m,j} \sim 2 j$, and
\begin{equation}
 A(\gamma) = \frac{5}{2} \frac{\gamma \left(\pi - \gamma \right)}{ \left(\pi - 3\gamma \right)^2} \,,
\end{equation}
whereas the functions $B_{m,j}(\gamma)$ (which extend to the more general family of functions $B_{m,j,w}$ when considering sectors of non-zero momentum) could not be accessed numerically, but are believed to be indeed universal functions with no $L$ dependence at this order (namely, the only further corrections to (\ref{eq:a22:cmj}) come from the `usual' terms which are positive powers of $1/L$). 
The ground state, in particular, corresponds to $m=j=0$, so the central charge is obtained in the $L \to \infty$ limit as $c=2$.
  
While the part depending quadratically on the magnetization $m$ in (\ref{eq:a22:cmj}) is usual, the $j$-depending part is much less familiar. In the continuum limit $L \to \infty$, we may replace the discrete index $j$ by some {\it continuous} index $s \sim \sqrt{\frac{5}{2}}\frac{\pi-\gamma}{\pi-3\gamma} {(2j) \over B_j+\log L} \sim  \frac{2j}{\log L} $ (the normalization is somewhat arbitrary at this stage), leading to conformal weights of the form 
\begin{equation}
\Delta_{m,s}+\bar{\Delta}_{m,s} = \frac{\gamma}{4 \pi} m^2  +  \frac{\gamma}{\pi - \gamma}s^2 \,.
\label{eq:a22:xms}
\end{equation}
Qualitatively, the spectrum of  dimensions is therefore made of a continuous part and a discrete part, as represented schematically in the following figure 
  \begin{center}
\begin{tikzpicture}[scale=1]
\draw[thick, opacity=1,->,>=latex] (-2,0) -- (-2,4);
\node[left] at (-2,4) {$\Delta, \bar{\Delta}$};

\draw[thick, opacity=1] (0,0) -- (1,0);
\shade[top color=white, bottom color=black] (0,0) rectangle +(1,2.);

\node[below, orange] at (0.5,-0.2) {$m=0$};
\draw[->] (1.2,0) -- (1.2,1.7);
\node[right] at (1.2,0.75) {$s$};

\begin{scope}[shift={(2.5,1)}]
\shade[top color=white, bottom color=black] (0,0) rectangle +(1,2.);
\node[below, orange] at (0.5,-0.2) {$m=1$};
\draw[->] (1.2,0) -- (1.2,1.7);
\node[right] at (1.2,0.75) {$s$};
\end{scope}

\begin{scope}[shift={(5,3)}]
\shade[top color=white, bottom color=black] (0,0) rectangle +(1,2.);
\node[below, orange] at (0.5,-0.2) {$m=2$};
\draw[->] (1.2,0) -- (1.2,1.7);
\node[right] at (1.2,0.75) {$s$};
\end{scope}
\end{tikzpicture} \,,
\end{center}
The corresponding CFT  is non-compact, and was identified \cite{VJS:a22} as the ``black hole sigma model''\cite{BlackHoleCFT}.

An important physical feature of the $a_2^{(2)}$ model in regime III, which was also instrumental in identifying its continuum limit, is the behaviour of the corresponding eigenlevels with  the twist $\varphi$ (analogous to the twists $\varphi_1, \varphi_2$ in the two-color model). Upon increasing $\varphi$ from its value $0$ in the periodic, untwisted case, the emergence of discrete states  popping up from the continuum  at special values of $\varphi$ was observed \cite{VJS:a22}. These can be  interpreted within  the CFT as states going from non-normalizable to normalizable. More explicitly, in a given sector of magnetization $m$ the level $j$ becomes discrete for $\varphi \geq (|m|+2 j+1)  \gamma$ (restricting to $0 \leq \varphi \leq \pi$)
\begin{equation}
  c_{m,j}(\varphi) = \begin{cases}  c_m^{*}  
- 12  {(N_{m,j})^2}\frac{A(\gamma)}{\left[ B_{m,j}(\gamma) + \log L \right]^2}
   & \mbox{for } \varphi\leq\ (|m|+2 j+1) \gamma 
\\  c_m^{*}  + {3 \over \gamma(\pi - \gamma)}\left[\varphi - (|m|+2 j+1)\gamma\right]^2 & \mbox{for } \varphi \geq (|m|+2 j+1)  \gamma  \,,
\end{cases}
\label{eq:a22:cmjtwisted}
\end{equation}
where 
\be
c^*_{m}(\varphi)   = 2 - 3 {\varphi^2 \over \pi \gamma} -\frac{3 \gamma m^2}{\pi}  \,.
\ee
The notations of equation (\ref{eq:a22:cmjtwisted}) suggest that the functions $A(\gamma)$ and $B_{m,j}(\gamma)$ have no dependence upon the twist $\varphi$. This could be wrong, as the system sizes $L$ for which the Bethe ansatz equations could be solved numerically were not enough to determine  $A(\gamma)$ and $B_{m,j}(\gamma)$  with sufficient precision, but since this aspect is not crucial to our analysis we will simply ignore it in the following.

\subsection{The $a_3^{(2)}$ model in regime III} 
 
The continuum limit of the $a_3^{(2)}$ model was studied in \cite{VJS:a32}. As the dilute version, it has two independent conserved magnetizations $m^{(1)}, m^{(2)}$ resulting from the corresponding two-color structure. We however point out that the relation between the loop weight $n$ and the parameter $\gamma$ in that case is different from that of the $b_2^{(1)}$ case, namely $n= 2 \cos \gamma$. 
The regime III for this model was identified as $\gamma \in \left[ 0 , \frac{\pi}{4}\right]$ for the antiferromagnetic sign of the Hamiltonian, and similarly to the $a_2^{(2)}$ case, in this regime there is in each sector of fixed momentum and magnetizations $m^{(1)}, m^{(2)}$ a tower of excited states $j=0, 1, 2, \ldots$, where once again the meaning of the index $j$ is well-defined in terms of the associated configurations of Bethe roots. The spectrum of effective central charges at zero twist is found to be
\begin{equation}
 -{c_{m^{(1)},m^{(2)},j} \over 12} = -\frac{3}{12}+  \frac{\gamma}{2\pi}\left( m^{(1)} + m^{(2)} \right)^2 + \left(N_{m^{(1)},m^{(2)},j}\right)^2 \frac{A(\gamma)}{\left[B_{m^{(1)},m^{(2)},j}(\gamma) + \log L \right]^2}  \,,
 \label{eq:a32:cmj}
\end{equation}
with $ N_{m^{(1)},m^{(2)},j} \sim 2j$ for $j$ large, and with reasonable numerical support for the following expression of $A(\lambda)$
\begin{eqnarray}
 A(\gamma) &=& 10 {\gamma (\pi - \gamma) \over (\pi - 4 \gamma)^2} \,,
  \label{a32:Agamma}
\end{eqnarray}
while that of $B(\lambda)$ remains unknown.%
\footnote{The factor of $10$ written here is only an indicative numerical value.}

In the light of the results for $a_2^{(2)}$, the spectrum (\ref{eq:a32:cmj}) can be interpreted in the continuum limit as giving rise to a continuum of exponents in each sector of fixed momentum and magnetizations $m^{(1)}, m^{(2)}$. 
The nature of the underlying CFT can in fact be best understood by studying the effect of the twists $\varphi_1, \varphi_2$ associated to the two loop colors. 
 Setting $m_\pm \equiv m^{(1)}\pm m^{(2)}$, $\varphi_\pm \equiv \varphi_1 \pm \varphi_2$, it is seen that the effective central charges can be decomposed as $c_{m^{(1)},m^{(2)},j} = c_{m_-} + c_{m_+,j}$, where
\begin{equation}
c_{m_-}(\varphi_-) = 1- 3 \frac{(\varphi_-)^2}{\pi \gamma}
- 3 \frac{ \gamma (m_-)^2}{\pi}  \,,
\label{a32:c-}
\end{equation}
and
\begin{equation}
c_{m_+,j}(\varphi_+) = 
\begin{cases}
c^*_{m_+} 
- 12  {(N_{m_+,j})^2}\frac{A(\gamma)}{\left[ B_{m_+,j}(\gamma) + \log L \right]^2}
\quad \mbox{ for } \varphi_+ \leq \left(m_+ +2j+1\right)\gamma
\\
c^*_{m_+} + 3 \frac{\left(\varphi_+ - (m_+ + 2j+1)\gamma\right)^2}{\pi(\pi-\gamma)}
\quad \mbox{ for } \varphi_+ \geq \left(m_+ +2j+1\right)\gamma \,.
\end{cases}
\label{a32:c+} \,,
\end{equation}
where 
\be
c^*_{m_+}(\varphi_+) =   2 - 3 \frac{(\varphi_+)^2}{\pi \gamma} - 3 \frac{\gamma (m_+)^2}{\pi} \,,
\ee
and where the same remark as that made in the $a_2^{(2)}$ case holds regarding the possible dependence of the functions $A(\gamma)$ and $B_{m_+,j}(\gamma)$ on the twist parameter. 

While (\ref{a32:c-}) can be interpreted as the spectrum of a free compact bosonic theory, the part (\ref{a32:c+}) has exactly the structure of the discrete levels found for the black hole CFT in the continuum limit of the $a_2^{(2)}$ model in regime III, see equation (\ref{eq:a22:cmjtwisted}). From there, the continuum limit of $a_3^{(2)}$ in regime III can be interpreted as the direct sum of a free compact bosonic CFT and of the black hole CFT.

\section{Bethe ansatz analysis of the $b_2^{(1)}$ model in regime IV}

We are now ready to turn to the study of regime IV of the $b_2^{(1)}$ model. First, we concentrate on the periodic, untwisted vertex model ($\varphi_1 = \varphi_2 = 0$), and only later will we study the effect of introducing a twist. In section \ref{sec:n0spectrum} below the focus will be put on the spectrum of the physical loop model, and in particular on the point $n=0$, associated with the truncated CC model.

\subsection{Spectrum at zero twist}

\subsubsection{The ground state}

From exact diagonalization of the quantum chain for finite system sizes $L=4,6,8$, we can draw conclusions about the configurations of Bethe roots describing the ground state and low-lying excitations. The ground state in regime IV at size $L$ corresponds, for $\gamma$ not too close to zero, to 
\begin{itemize}
 \item $L\over 2$ 2-strings of $\lambda$-roots with imaginary part close to (smaller than) ${\pi \over 2} - {\gamma \over 4}$, and
 \item $L\over 2$ 2-strings of $\mu$-roots with imaginary part close to (larger than) ${\pi \over 2} - {\gamma \over 4}$.
 \end{itemize}
 
\begin{figure}
\begin{center}
 \includegraphics[scale=0.7]{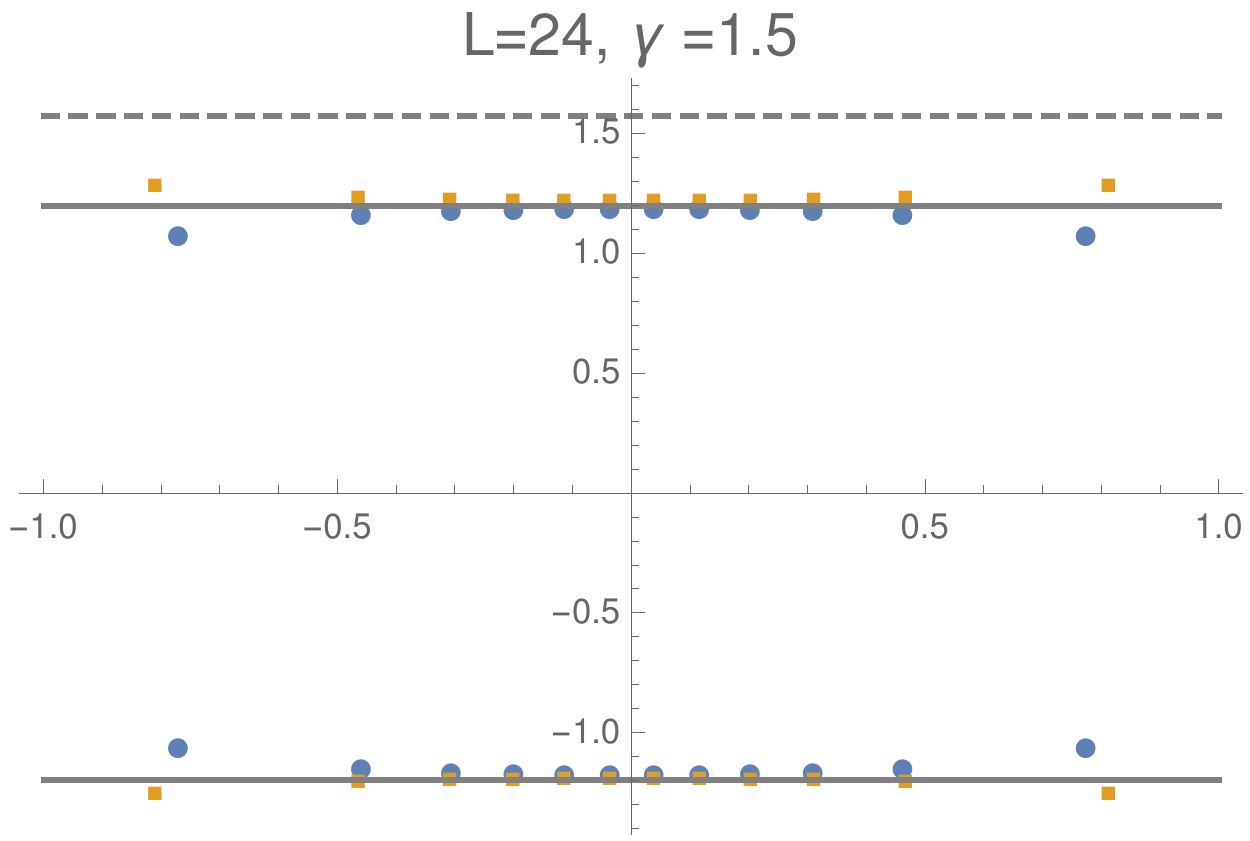}
 \\
 \vspace{1cm}
 \includegraphics[scale=0.7]{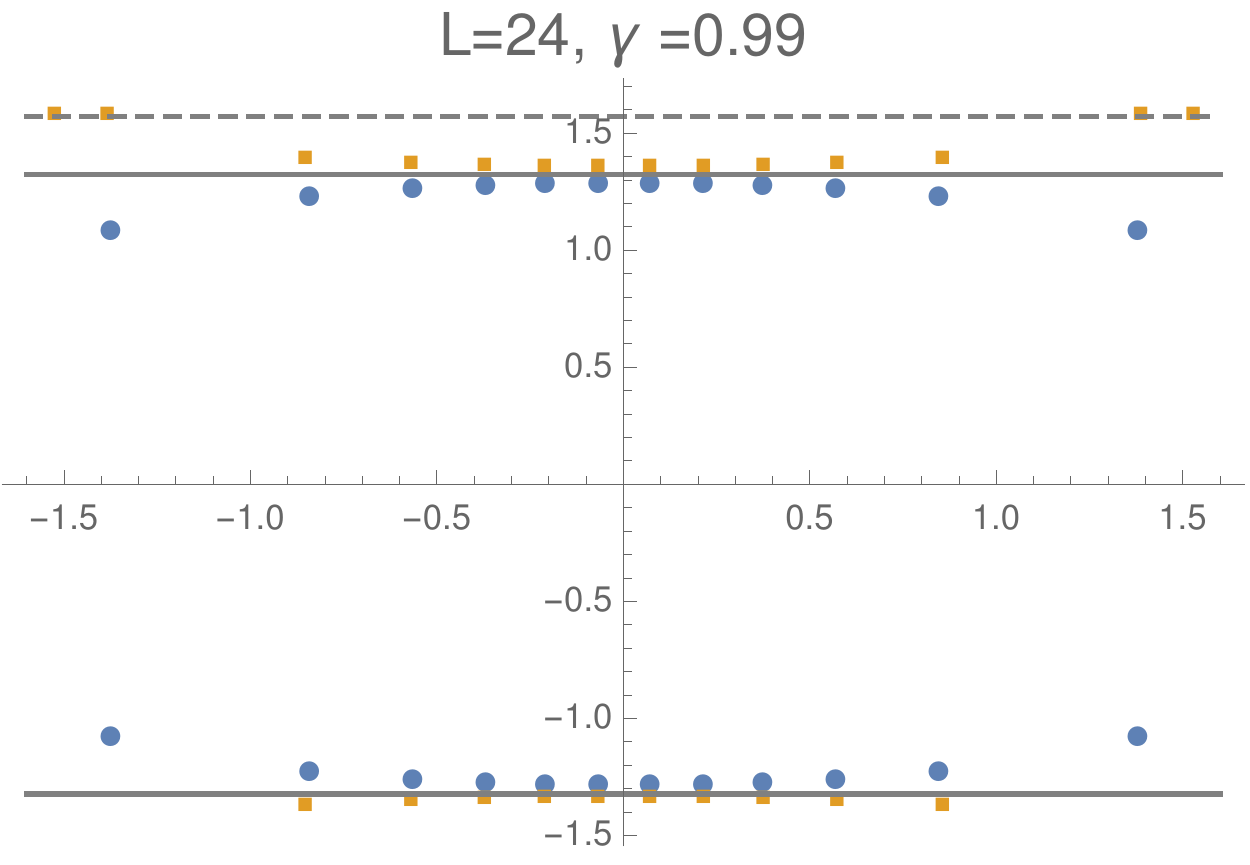}
 \end{center}
\caption{Roots configuration associated with the ground state of regime IV at $L=24$.  The $\lambda$ (resp $\mu$) roots are the blue (resp.\ orange) dots. We also plotted the lines of imaginary part $\pm \left({\pi \over 2} - {\gamma \over 4}\right)$ (solid lines) and $\pi \over 2$ (dashed line). For $\gamma$ large enough the ground state is described by $L\over 2$ 2-strings of each type. 
 At $\gamma \simeq 1.02$ the outer $\mu$ 2-strings' imaginary  part reaches $\pi \over 2$, and below this value these 2-strings have turned into roots of imaginary part precisely $\pi \over 2$. As $\gamma$ is lowered to zero, the same process is expected for all $\mu$ 2-strings.  
}                                 
\label{fig:E000_L24}                          
\end{figure}

Because of the $i \pi$-periodicity of the Bethe Ansatz equations \eqref{eq:untwisted_BAE}, it is clear from this that something must happen to the $\mu$-roots as $\gamma \to 0$, since the imaginary parts go to $\pi \over 2$ and then the two roots of each 2-string merge.
This is indeed the case: in finite size, as $\gamma$ is lowered, the $\mu$-roots with larger real parts, and hence larger imaginary parts, merge on the line of imaginary part $\pi \over 2$ for some finite value of $\gamma$, and below this value are turned into roots with different real parts and fixed imaginary part $\pi \over 2$. This mechanism is illustrated in figure \ref{fig:E000_L24}.
 
In the thermodynamic limit assume we can still take all the roots to have imaginary parts arbitrarily close to ${\pi \over 2} - {\gamma \over 4}$, and therefore need to consider only 2-strings over 2-strings. These are described by one single density of real parts $\rho$, for which the BAE translate into 
\begin{eqnarray}
   \rho + \rho^h  &=&   \frac{\sinh {\omega \gamma \over 4}  + \sinh {\omega {3 \gamma \over 4}}  }{\sinh {\omega \pi \over 2}}
  +  \left( \frac{ \sinh{\omega\left({\pi \over  2} - { \gamma}  \right)} +  \sinh{\omega\left({\pi \over  2} - {3 \gamma \over 2}  \right)}  - \sinh{\omega\left({\pi \over  2} - { \gamma \over 2}  \right)}  }{\sinh {\omega \pi \over 2}}  \right) \rho \,, 
  \label{RIVscatt}
 \end{eqnarray}
 that is
 \begin{equation}
 \rho + \rho^h = s + \Phi \star \rho
 \end{equation}
 where $s$ and $\Phi$ are the source term and scattering kernel respectively, whose Fourier transforms are directly read off (\ref{RIVscatt}).
The ground state density of string centers corresponds to a zero density of holes, and is therefore in Fourier space 
\begin{equation}
\rho(\omega) = \left(1-\Phi\right)^{-1} s =  \frac{1}{2 \cosh\left( \frac{\omega}{4}(2\pi - 3\gamma) \right)} \,,
\end{equation}
 that is, in real space, 
\begin{equation}
 \rho (t) = \frac{1}{2\pi - 3 \gamma} \frac{1}{\cosh \frac{2 \pi t}{3 \gamma - 2\pi}} \,.
\end{equation}

From there we can obtain the extensive part of the ground state energy in the thermodynamic limit 
\begin{equation}
 e_\infty = - \int_{-\infty}^{\infty}\mathrm{d}t \frac{1}{2\pi - 3 \gamma} \frac{1}{\cosh \frac{2 \pi t}{3 \gamma - 2\pi}}
 \frac{2 \sin \gamma \left( \cos {\gamma \over 2} \cosh 2 t  + \cos\gamma \right)}{\cos^2\gamma + 2 \cos\gamma \cos {\gamma \over 2} \cosh 2t + \frac{1}{2}(\cos\gamma + \cosh 2t)} \,.
\end{equation}

The Fermi velocity is also easily read from looking at the linear dispersion relation of low-lying hole excitations, and we find   
\begin{equation}
 v_f (\gamma) = \frac{\pi }{2\pi - 3 \gamma} \,.
\end{equation}
The associated Fermi velocity of the loop Hamiltonian is obtained from its relation with that of the vertex model, eq. (\ref{EloopsEBA}), and reads as a function of the parameter $\theta$
 \be
 v_f^{\rm loops} = \frac{2\pi \sin 2\theta \sin 3\theta}{2\pi-3\theta}  \,,
 \ee
which coincides with the expression found in \cite{Ikhlef} using discrete holomorphicity methods.

\begin{figure}
\begin{center}
 \includegraphics[scale=1.2]{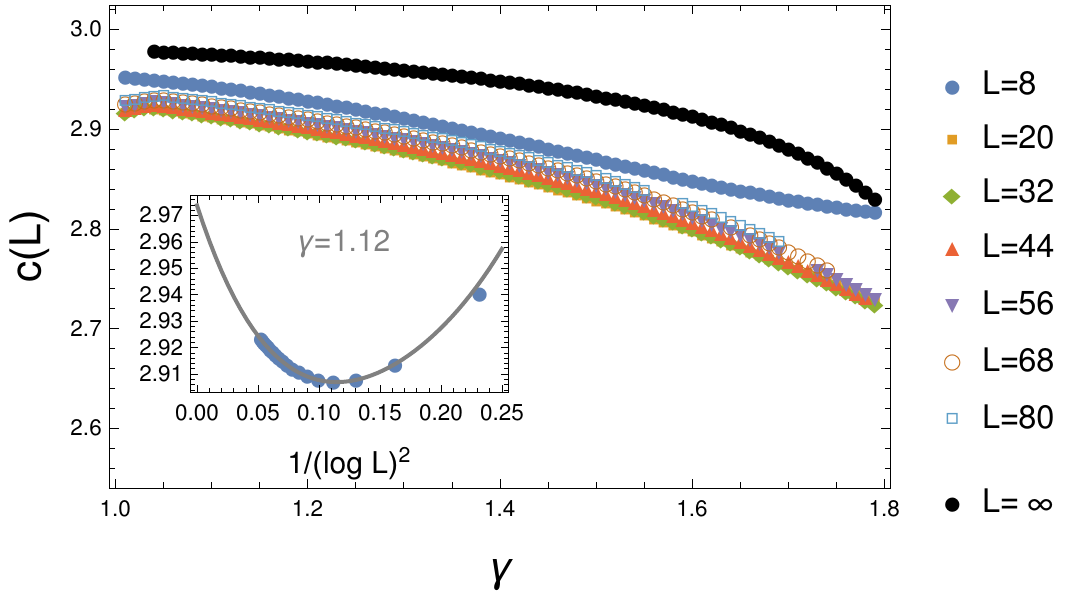}
 \end{center}
\caption{Central charge of the periodic $b_2^{(1)}$ chain in regime IV estimated from resolution of the Bethe ansatz equations at sizes $L = 8$ to $L=80$, plotted as a function of $\gamma$ (due to technical restrictions in the resolution of the Bethe ansatz equations, the represented range of $\gamma$ is not the full range of regime IV). 
The black dots are an $L = \infty$ extrapolation obtained from a quartic fit in $1 / (\log L)^2$, illustrated in the inset figure for a particular value of $\gamma$: notice the non-monotonic finite-size behaviour.}
                                
\label{fig:c000}                          
\end{figure}

From the expression of the extensive ground state energy and of the Fermi velocity, we can obtain finite-size estimations of the central charge at zero twist using the scaling formula (\ref{finitesizescaling}). The results are plotted in figure \ref{fig:c000}, and indicate a very slow convergence towards a conjectured value $c=3$. Using the experience gained with the $a_r^{(2)}$ models described in the previous section, this slow convergence hints at a possible non-compact direction. 
We therefore conjecture the possible scaling form 
\begin{equation}
c = 3 - 12\frac{A(\gamma)}{\left[B_{0,0,0}(\gamma) + \log L\right]^2} \,,
\label{eq:cLog}
\end{equation}
where the functions $A(\gamma)$ and $B_{0,0,0}(\gamma)$ can be estimated from data at finite $L$.

\subsubsection{Spin excitations}

In the following section we will identify two types of compact (and one type of non-compact) excitations over this ground state. However, it seems that the only direction tractable by the usual analytical means is the one corresponding to making holes of 2-strings over 2-strings in the Fermi sea.
Restricting for simplicity to states with no backscattering---i.e., those which are carrying zero momentum---, the conformal weights are obtained as  
\begin{equation}
 \Delta + \bar{\Delta} = \frac{1}{4}  (\delta m) \mathcal{Z}^{-1}  (\delta m) \,,
\end{equation}
where $\mathcal{Z}^{-1}$ is the $\omega \to 0$ limit of $1-\Phi$, that is $\mathcal{Z}^{-1} = \frac{4 \gamma}{\pi}$, and $\delta m$ is here the number of holes of strings over strings, that is each of the $\delta m$ holes corresponds to 2 $\lambda$ and 2 $\mu$-roots being removed. 
Using the identification (\ref{eq:misi}), this corresponds to $S_z^{(1)} = S_z^{(2)} = \delta m$. Using the symmetry between the two colors, 
\begin{equation}
 \Delta + \bar{\Delta} = \frac{\gamma}{2 \pi} \left[\left( S_z^{(1)} \right)^2 + \left( S_z^{(2)} \right)^2 \right] \,.
\label{Deltaspinsectors}
\end{equation}

This formula can be verified by numerically solving the associated Bethe equations.  
The roots patterns associated to the sectors of non-zero spins $S_z^{(1)}$ and $S_z^{(2)}$ are easily understood. Because of the model's symmetries (namely under exchange of the two colors, or under the reversal $S_z^{(1,2)} \to - S_z^{(1,2)}$ independently) we can restrict for instance to $S_z^{(1)} \geq S_z^{(2)} \geq 0$. These values are associated with the number of roots being respectively
\begin{equation}
m_1 = L- S_z^{(1)}- S_z^{(2)} \,, \qquad  m_2 = L-2 S_z^{(1)} 
\end{equation}

Now $S_z^{(1)}$ and $S_z^{(2)}$ are simultaneously either integer or half-integer. In the former case the ground state $(S_z^{(1)},S_z^{(2)},j=0)$ is obtained from the following root configuration: 
\begin{itemize}
\item $\frac{1}{2}(L-2 S_z^{(1)})$ 2-strings of $\mu$-roots over 2-strings of $\lambda$-roots, and
\item $S_z^{(1)}-S_z^{(2)}$ additional $\lambda$-roots with imaginary parts $\pi \over 2$.
\end{itemize}
In the case where $S_z^{(1)}$ and $S_z^{(2)}$ are half-integers, we have in addition one $\lambda$-root at $i \frac{\pi}{2}$ and one $\mu$-root equal to $0$.

This draws two independent ways of exciting the ground state Fermi sea of 2-strings over 2-strings, namely 
\begin{itemize}
\item hole excitations in the sea of strings over strings (excitations symmetric in the two colors);
\item hole excitations, together with the addition of as many pairs of $\lambda$-roots with imaginary part $\pi \over 2$ (excitations antisymmetric in the two colors) 
\end{itemize}
Solving numerically the associated Bethe equations for the first few values of $S_z^{(1)}$ and $S_z^{(2)}$, we indeed recover the expression (\ref{Deltaspinsectors}).
It is worth noting that this is the same formula as in the $a_3^{(2)}$ case. However, while for $a_3^{(2)}$ the charges $S_{z}^{(1)}$ and $S_{z}^{(2)}$ were restricted to integer (resp.\ half odd-integer) values for systems of even (resp.\ odd) size, in the present case $S_z^{(1)}$ and $S_{z}^{(2)}$ may be indifferently simultaneously integer or simultaneously half odd-integer for any system size.

\subsubsection{Non-compact excitations}

\begin{figure}
\begin{center}
 \includegraphics[scale=0.7]{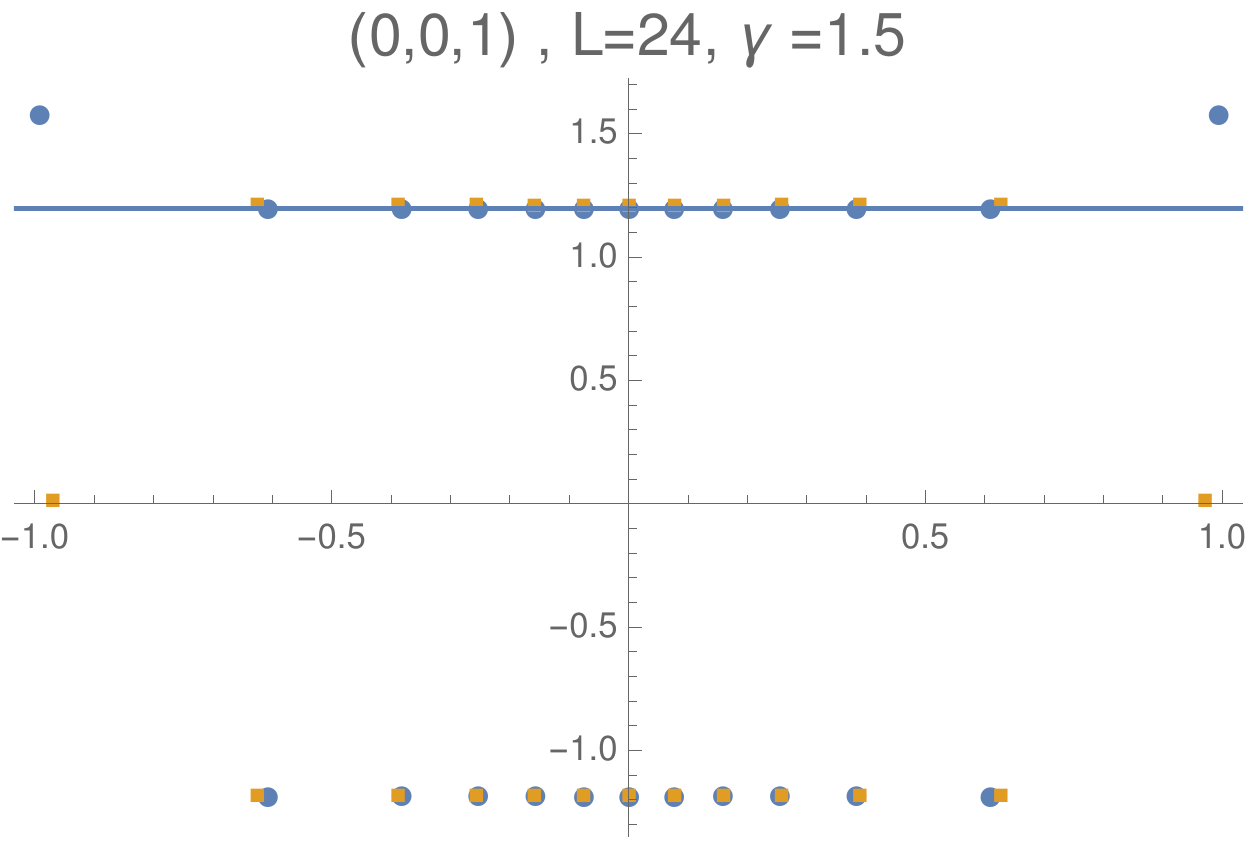}
 \end{center}
\caption{Roots configuration associated with the level $(0,0,1)$ for a system of size $L=24$.  The $\lambda$ (resp $\mu$) roots are the blue (resp red) dots. We also plotted the line of imaginary parts $\pm  \left({\pi \over 2} - {\gamma \over 4}\right)$ (gray line) and $\pi \over 2$ (black dotted line).
}                                 
\label{fig:E001_L24}                          
\end{figure}

The lowest-lying excitations with zero momentum in the $S_z^{(1)} = S_z^{(2)} =0$ sector can be indexed by an integer $j=0,1,2,\ldots$. They are obtained from the ground-state root configuration by simultaneously removing $j$ 2-strings over 2-strings, and replacing those by $j$ pairs of $\lambda$-roots with imaginary parts $\pi \over 2$ and $j$ pairs of real $\mu$-roots. The roots configuration for the first excitations of this type ($j=1$) for size $L=24$ is depicted in figure \ref{fig:E001_L24}.
We label these excitations $(0,0,j)$.

Taking in consideration the other magnetization sectors labeled by different values of  $S_z^{(1)}$ and $S_z^{(2)}$, we observe similarly excitations obtained by exciting the sea of 2-strings over 2-strings, which we generally label as 
\be
(S_z^{(1)},S_z^{(2)},j) \,.
 \nonumber
\ee
 For all these excitations, we introduce---as described for the $a_r^{(2)}$ models in eq.~\eqref{eq:cidef}---the effective central charges  $c_{S_z^{(1)},S_z^{(2)},j}$, which can be estimated numerically from resolution of the associated Bethe ansatz equations. 

It is quite difficult to solve the BAE for these excitations in a wide range of $\gamma$ and for large sizes. However the measure of the central charge $c_{0,0,1}$ for sizes up to $L \sim 30$ seems in accordance with a slow convergence towards $c=3$, which would correspond, in the $L\to \infty$ limit, to a non-compact direction.  

The lesson learned from our previous studies \cite{VJS:a22,VJS:a32,VJS:an2} of non-compact models indicates that the best way to properly identify the continuum limit associated with these excitations is to now turn on the twists $\varphi_1, \varphi_2$, in order to excite discrete states from the non-compact continuum.

\subsection{Spectrum at non-zero twist and discrete states}

We therefore now introduce the twists $\varphi_1, \varphi_2$ defined in section \ref{sec:twists}. The BAE become (\ref{BAEtwisted}), and their numerical solutions can be followed by  continuity as one  turns on the twists from their initial values $\varphi_1 = \varphi_2 = 0$.

Starting from the ground state at $\varphi_1 = \varphi_2 = 0$, in the region of $\gamma$ where it is made only of 2-strings, something happens again to the roots as the twists are increased: the rightmost 2-string of $\mu$-roots has its imaginary part increasing, until it reaches $\pi \over 2$ and the 2-string turns into a pair of roots with imaginary part $\pi \over 2$. The same is expected to happen to the following 2-strings as the twist is  increased further; however the numerical resolution of the Bethe ansatz equation is very unstable in such regions and we did not observe this phenomenon directly.

\begin{figure}
\begin{center}
 \includegraphics[scale=1.25]{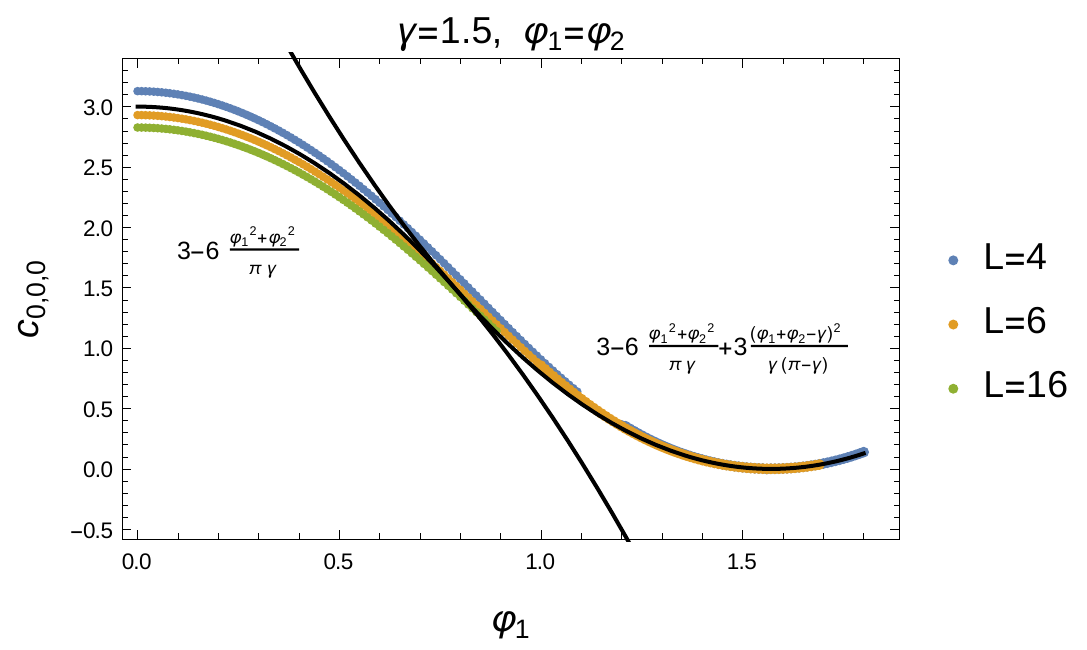}
 \end{center}
\caption{Discrete level structure of the level (0,0,0) in regime IV as a function of the twist $\varphi_1=\varphi_2$. The central charge was measured at $\gamma=1.5$ for sizes $L=4,6,16$.  The black lines show for comparison the two analytic expressions given by (\ref{c000conj}).
}                                 
\label{fig:c000_f1f2}                          
\end{figure}

Figure \ref{fig:c000_f1f2} shows numerical estimations of the central charge $c_{0,0,0}(\varphi_1,\varphi_2)$ at $\gamma=1.5$ for  $\varphi_1 = \varphi_2$, and gives good support for the following conjecture : 
\begin{equation}
c_{0,0,0} = \begin{cases}  3 - 6 \frac{\left(\varphi_1\right)^2 + \left(\varphi_2\right)^2}{\pi \gamma}
- 12   \frac{A(\gamma)}{\left[ B_{0,0,0}(\gamma) + \log L \right]^2}
  & \mbox{for } \varphi_1 + \varphi_2 \leq \gamma \,,
\\  3 - 6 \frac{\left(\varphi_1\right)^2 + \left(\varphi_2\right)^2}{\pi \gamma} + 3 \frac{\left(\varphi_1 + \varphi_2-\gamma\right)^2}{\gamma(\pi - \gamma)} & \mbox{for } \varphi_1 + \varphi_2 \geq \gamma \,,
\end{cases}
\label{c000conj}
\end{equation}
where, like in the $a_2^{(2)}$ and $a_3^{(2)}$ cases \cite{VJS:a22,VJS:a32}, we have written the logarithmic corrections in a form independent of the twists $\varphi_1, \varphi_2$.  As previsouly mentioned, we have not yet been able to provide fully conclusive evidence that the functions $A$ and $B$ do not depend on the twist parameters, but in the following we shall assume that they do not.

Turning to the excited states that appear as the twists $\varphi_1$ and $\varphi_2$ are turned on, the Bethe roots undergo qualitative changes remarkably similar to those that were already encountered in the $a_3^{(2)}$ case, and which are described in \cite{VJS:a32}.  
It turns out that the corresponding effective central charges in fact take exactly the same form as those found in the $a_3^{(2)}$ case---see eqs.~\eqref{a32:c-}--\eqref{a32:c+}---that we repeat here for convenience. Setting
\begin{subequations}
\label{eq:mphipm}
\bea
m_\pm &=&S_z^{(1)}\pm S_z^{(2)} \,, \\
\varphi_\pm&=&\varphi_1\pm \varphi_2 \,,
\eea
\end{subequations}
the effective central charge of the $b_2^{(1)}$ model can thus be decomposed as 
\begin{equation}
 c_{m^{(1)},m^{(2)},j} = c_{m_-} + c_{m_+,j} \,,
\end{equation}
where
\begin{subequations}
\label{b21:c}
\bea
 c_{m_-}(\varphi_-) &=& 1- 3 \frac{(\varphi_-)^2}{\pi \gamma}
- 3 \frac{ \gamma (m_-)^2}{\pi}  \,, \label{b21:c-} \\
c_{m_+,j}(\varphi_+)  &=& 
\begin{cases}
c^*_{m_+} 
- 12  {(N_{m_+,j})^2}\frac{A(\gamma)}{\left[ B_{m_+,j}(\gamma) + \log L \right]^2}
 & \mbox{ for } \varphi_+ \leq \left(m_+ +2j+1\right)\gamma \,,
\\
c^*_{m_+} + 3 \frac{\left(\varphi_+ - (m_+ + 2j+1)\gamma\right)^2}{\pi(\pi-\gamma)}
 & \mbox{ for } \varphi_+ \geq \left(m_+ +2j+1\right)\gamma \,,
\end{cases}
\label{b21:c+} \\
c^*_{m_+}(\varphi_+) &=&   2 - 3 \frac{(\varphi_+)^2}{\pi \gamma} - 3 \frac{\gamma (m_+)^2}{\pi} \,,
\label{b21:c+star}
\eea
\end{subequations}
and where the same remark as stated above holds regarding the possible dependence of the functions $A(\gamma)$ and $B_{m_+,j}(\gamma)$ on the twist parameter.

\subsection{Conclusion: The full continuum limit of regime IV}

\label{sec:RIVcontinuum}

With the common notation $q=e^{i\gamma}$, we find that the continuum limit of the $b_2^{(1)}$ and $a_3^{(2)}$ model are essentially identical. There are however two small differences:
\begin{itemize}
 \item The corresponding domains covered by $\gamma$ are different: 
the $b_2^{(1)}$ model is non-compact in regime IV (with $0 \le \gamma \le \frac{2 \pi}{3}$), while the $a_3^{(2)}$ model is non-compact in its regime III
(with $0 \le \gamma \le \frac{\pi}{4})$. We notice in particular that the $a_3^{(2)}$ model covers a smaller range of $\gamma$-values, which, in particular does {\em not} include the ``polymer point'' $\gamma = \frac{\pi}{2}$ that is essential to the understanding of the first truncation of the
CC model.
 \item  Whereas in the $a_3^{(2)}$ model the spins $S_z^{(1)}$ and $S_z^{(2)}$ are both integer (resp.\ half odd-integer) for even (resp.\ odd) size $L$, in the $b_2^{(1)}$ model both parities are possible for any value of $L$.
The expression of the conformal weights in terms of $S_z^{(1)}$ and $S_z^{(2)}$ is however formally exactly the same in the two models.
\end{itemize}
 
 We recall that the $a_3^{(2)}$ model can be formulated as a fully packed two-color loop model \cite{VJS:a32}, while the $b_2^{(1)}$ model allows for some ``dilution'' in
the form of empty edges. Our result is thus that the full packing of the $a_3^{(2)}$ does not change the universality class, except that it introduces a (natural) parity constraint on the
spins $S_z^{(1)}$ and $S_z^{(2)}$.%
\footnote{This phenomenon is familiar from simpler cases involving only one color of loops, e.g., when comparing the dense phase of the O($n$) model (expressed as a dilute loop model) and the Potts model (expressed as fully-packed Temperley-Lieb loop model).
}
Physically this means that the $b_2^{(1)}$ model in regime IV is in a {\em dense} universality class, and in particular the $n \to 0$ limit
of interest for the first CC truncation behaves as a certain type of dense two-colored polymers.%
\footnote{We note in passing that there are known cases \cite{KdGN96,JK98} where fully-packed loop models are in a different universality class than their dense counterparts.}

From the CFT perspective, the continuum limits of $b_2^{(1)}$ in regime IV and $a_3^{(2)}$ in regime III can both be described in terms of the tensor product of a $SL(2,\mathbb{R})_{K}/U(1)_{-K}$ coset CFT and a free boson  $U(1)_K$, where we have parametrized 
\begin{equation}
\gamma={2\pi\over K} \,.
\label{gammaK}
\end{equation}
The central charge is given by 
\begin{equation}
c=3+{6\over K-2} \,,
\label{eq:cc_intermsofK}
\end{equation}
and the exponents by
\begin{equation}
h(\bar{h})={(m_-\pm Kw_-)^2\over 4K}+{(m_+\pm Kw_+)^2\over 4K}-{J(J+1)\over K-2}
\end{equation}
where $J$ is an $SL(2,\mathbb{R})$ spin. The quantum numbers  $m_\pm$ have already been identified; the numbers $w_\pm$  are winding numbers, corresponding to global shifts of the Fermi seas. Normalizable states correspond to $J=-{1\over 2}+is$, with $s\in \mathbb{R}$. Note in particular that the ground state---which corresponds to $J=0$---is {\em not} normalizable, and that the lowest-lying normalizable state has conformal weight $h={1\over 4(K-2)}$, leading to the effective central charge $c_{\rm eff}=3$. In general, we shall denote by $h,\bar{h}$ the conformal weights (i.e., eigenvalues of the Virasoro generators $L_0,\bar{L}_0$) in the ``true'' CFT. We will denote by $\Delta,\bar{\Delta}$ the conformal weights measured as gaps over the normalizable, lowest-lying energy state. Setting $m_\pm=w_\pm=0$ for instance, we see that for the continuous spectrum we have $h+\bar{h}={s^2+{1\over 4}\over K-2}$ while $\Delta+\bar{\Delta}={s^2\over K-2}$. 

Without twist, there is no discrete state at zero winding number $w=0$. When a twist is imposed, discrete states do appear when $\varphi_+$ is large enough, leading to the structure of excitations described by \eqref{b21:c}. 

\section{The spectrum of the loop model}

\label{sec:n0spectrum}

We now go back to the loop model, whose spectrum is obtained from that of the vertex model by carefully adjusting the twist parameters $\varphi_1, \varphi_2$. Namely, recalling section \ref{sec:twists},  
\begin{itemize}
\item If for the color $i \in \{ 1,2\}$ there are no through-lines, that is $S_i^{(z)} =0$, one has to choose $\varphi_i = \pi - \gamma$. 
\item By contrast, if $S_i^{(z)} \neq 0$, that is in the presence of through-lines, one should choose $\varphi=0$. 

For completeness, let us mention that in such sectors there exist additional eigenvalues of the loop model associated to twists of the form $\varphi_i = 2 \pi p / S_i^{(z)}$, with $p \wedge S_i^{(z)} = 1$ and $p$ integer. 
Such eigenvalues correspond in the loop language to the fact that the $|S_i^{(z)}|$ through-lines can be taken around the vertical axis of the cylinder with each line picking up a $|S_i^{(z)}|$'th root of unity twist factor, without affecting the corresponding global weight. Such eigenvalues are important for studying for instance the winding angle distribution of the paths \cite{DS88}, but in the present work we will not consider them further. 
\end{itemize}

All the conformal weights can therefore be read off from \eqref{b21:c}, where we once again insist that for the non-compact part the dependence of the functions $A$ and $B_{m_+,j}$ in the twist $\varphi_+$ was not elucidated. In practice, in the $L\to \infty$ limit we will as in the untwisted case replace the discrete index $j$ by a continuous index $s$, but we do not exclude that the associated densities of states $\rho_{m_+}(s)$ might depend on $\varphi_+$.

\subsection{The central charge}

The central charge of the loop model is obtained from the state $j=0$ and $S_1^{(z)}=S_2^{(z)}=0$ of the vertex model, for twist parameters $\varphi_1 = \varphi_2 = \pi - \gamma$. 
Importantly, all through the regime IV this corresponds to a discrete state, for which there are therefore no logarithmic finite-size corrections to the central charge. We find analytically 
\be 
c = \frac{3 (\pi - 2 \gamma)^2}{\pi (\pi-\gamma)} \,, 
\ee
as already identified in \cite{Ikhlef}, and in good agreement with numerical results from exact diagonalization of the loop Hamiltonian.

\subsection{The thermal exponent $x_t$}

We discuss now the temperature exponent $x_t$, namely the exponent  associated with the first excited state in the sector with no legs. In the vertex-model language we expect that this state should correspond to $(0,0,1)$ at twists $\varphi_1 = \varphi_2 = \pi - \gamma$.
Using \eqref{b21:c}, the effective central charge for this state reads, as a function of the twist,
\begin{equation}
c_{0,0,1} = \begin{cases}  3 - 6 \frac{\left(\varphi_1\right)^2 + \left(\varphi_2\right)^2}{\pi \gamma} - O(\frac{1}{(\log L)^2}) & \mbox{for } \varphi_1 + \varphi_2 \leq 3\gamma \,,
\\  3 - 6 \frac{\left(\varphi_1\right)^2 + \left(\varphi_2\right)^2}{\pi \gamma} + 3 \frac{\left(\varphi_1 + \varphi_2-3\gamma\right)^2}{\gamma(\pi - \gamma)} & \mbox{for } \varphi_1 + \varphi_2 \geq 3\gamma \,.
\end{cases}
\label{c001conj}
\end{equation}
 Taking $\varphi_1 = \varphi_2 = \pi - \gamma$ we see that the relevant expression is the first one for $\frac{2 \pi}{5} \leq \gamma \leq \pi$, and the second one for $0\leq \gamma \leq \frac{2 \pi}{5}$.
We end up with 
\begin{equation}
x_t  =  -\frac{c_{0,0,1}(\pi - \gamma,\pi - \gamma)-c_{0,0,0}(\pi - \gamma,\pi - \gamma)}{12} = \begin{cases} -2 + \frac{\pi}{\gamma} + \frac{1}{4} \frac{\gamma}{\pi-\gamma} & \mbox{for } \frac{2\pi}{5}\leq \gamma \leq\frac{2\pi}{3}  \,,
\\ \frac{2(\pi-2 \gamma)}{\pi - \gamma} & \mbox{for } 0 \leq \gamma \leq \frac{2\pi}{5} \,.
\end{cases}
\label{eq:Xt}
\end{equation}

\begin{figure}
\begin{center}
 \includegraphics[scale=1.3]{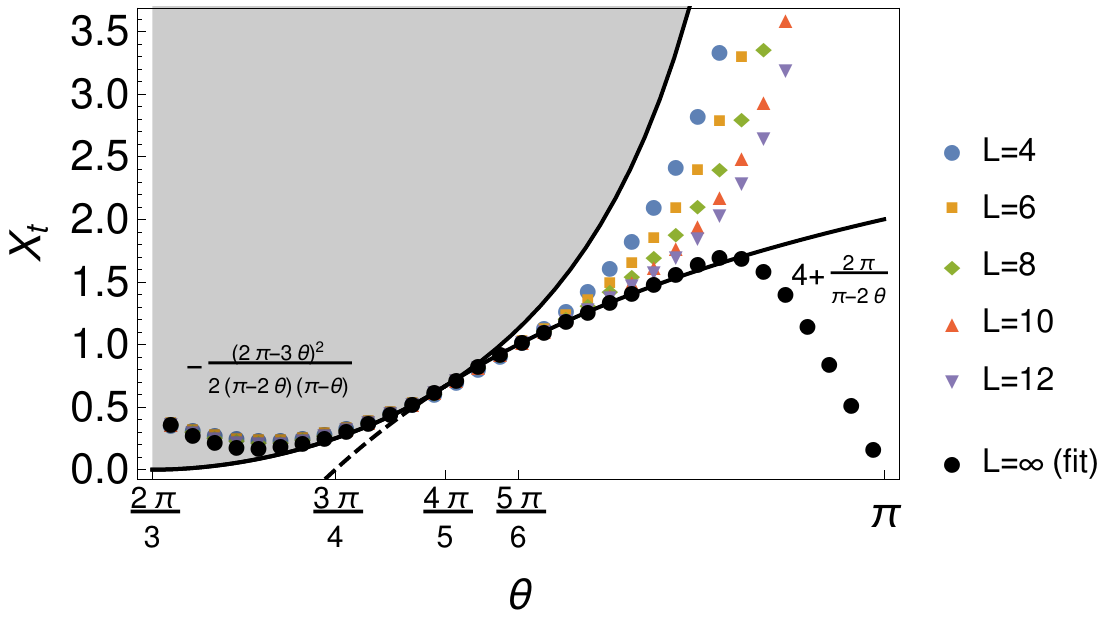}
 \end{center}
\caption{
Temperature exponent $x_t$ estimated from direct diagonalization of the loop hamiltonian at sizes $L=4, \ldots,12$, together with the corresponding extrapolation to the limit $L \to \infty$, plotted against the parameter $\theta$. This should be compared with \cite[Figure 8]{Ikhlef}. We plot for comparison the conjecture (\ref{eq:Xt}) (shown as dashed lines when the corresponding states are non-normalizable and therefore not observed in the physical spectrum), and also represent by a gray shading the continuum of states. 
}                                 
\label{fig:Xt}                          
\end{figure}

In order to facilitate comparison with the results presented in \cite{Ikhlef}, we recast it in terms of the parameter $\theta$ (recall the correspondence $\gamma=2\pi-2\theta$: $\gamma={\pi\over 2}$, that is $\theta={3\pi \over 4}$, for $n=0$). This yields 
\begin{equation}
x_t= \begin{cases} - \frac{(2\pi - 3\theta)^2}{2 (\pi - 2\theta)(\pi -\theta)} & \mbox{for }  \frac{2\pi}{3} \leq \theta \leq \frac{4 \pi}{5} \,,
\\ 4 + \frac{2\pi}{\pi-2\theta} & \mbox{for } \frac{4 \pi}{5} \leq \theta \leq \pi \,,
\end{cases}
\label{eq:Xttheta}
\end{equation}
In figure \ref{fig:Xt} we compare the result \eqref{eq:Xttheta} with numerical estimations obtained from the direct diagonalization of the Hamiltonian. These results should also be confronted with those of \cite[Figure 8]{Ikhlef}, where $x_t$ was calculated by direct diagonalization of the transfer matrix. 

The discrete-state expression: $x_t = 4 + \frac{2\pi}{\pi-2\theta}$, that holds in the region $\frac{4 \pi}{5}\leq \theta \leq \pi$---was already proposed in \cite{Ikhlef}.
It was supported there by the exact result $x_t = 1$ at $\theta= \frac{5 \pi}{6}$, which was derived from a mapping to a problem of free fermions. Note that both in our figure \ref{fig:Xt} and in \cite[Figure 8]{Ikhlef}, the finite-size estimations do not converge well in the region $\theta \lesssim \pi$, that is $\gamma \gtrsim 0$. The reason for this is that this corresponds to twists $\varphi_i \lesssim \pi$, for which the intersection between two different electric modes results in  a cusp in the central charge which is badly accommodated in finite size.

The ``continuous spectrum'' expression: $x_t =- \frac{(2\pi - 3\theta)^2}{2 (\pi - 2\theta)(\pi -\theta)}$, that holds for $\frac{4 \pi}{5} \leq \theta \leq \pi$---was, by contrast, not found in \cite{Ikhlef}. In the absence of a detailed analysis of the Bethe Ansatz equations, it would have been hard to infer this result from numerical diagonalizations for modest values of $L$. Indeed, the convergence of the finite-size estimations in this region (see the left part of figure \ref{fig:Xt}) is extremely slow---an effect which, as we have seen, can be traced back to logarithmic finite-size corrections.

\subsection{Exponents for non-zero winding}

States associated with non-zero winding numbers in the zero spin sector are obtained for a discrete set of values of $\varphi_\pm$ (as defined in eq. (\ref{eq:mphipm})) associated with the integer winding numbers $w_{\pm}$ as 
\begin{subequations}
\begin{align}
\varphi_+ &= 2(\pi-\gamma) + 2 \pi w_+ \\
\varphi_- &=  2 \pi w_-
\end{align}
\end{subequations}

In terms of the parameter $K = 2\pi /\gamma$ introduced in equation (\ref{gammaK}), the corresponding effective central charge coming from the $U(1)_K$ compact bosonic part is 
\begin{equation}
c_{w_-}= 1 -  \frac{12 \pi}{\gamma}(w_-)^2 = 1 - 6 K {w_-}^2
\end{equation}

For the $SL(2,\mathbb{R})_{K}/U(1)_K$ part, the effective central charge for $w_+ \neq 0$ is given by the first line in equation (\ref{b21:c+}):
\begin{equation}
c_{w_+}= 2 -  \frac{12 \pi}{\gamma}( w_+ + {\pi - \gamma  \over  \pi})^2 = 3 - {6 \over K} (K w_+ + K-2)^2 \,.
\end{equation}

Comparing with the central charge of the loop model, we find therefore the corresponding set of exponents 
\begin{equation}
\Delta + \bar{\Delta} 
= 
\frac{1}{4 K} \left[ 
\frac{(K-4)^2}{K-2} - K + 2 \left( (K w_-)^2 + (K w_+ + K-2)^2  \right)
\right]
\end{equation}

The case $w_+ = w_- =0$ is a bit particular, as it does not corresponds to the true ground state of the model. Rather, it can be interpreted as the bottom of the continuum at twists $\varphi_1 = \varphi_2 = \pi - \gamma$, which is practically the state $(0,0,1)$. 

On top of all such winding states, there is a continuum of states which we still parametrize by the continuous index $s$, even though we cannot exclude that the corresponding densities might depend on the winding number $w_+$.
The associated exponents read 
\begin{equation}
\left( \Delta + \bar{\Delta} \right)_{w_+, w_-, s}
= 
\frac{1}{4 K} \left[ 
\frac{(K-4)^2}{K-2} - K + 2 \left( (K w_-)^2 + (K w_+ + K-2)^2  \right)
\right] + \frac{2}{K-2} s^2 \,,
\label{eq:windexps}
\end{equation}
where the normalization of the index $s$ as a function of $\gamma$ has been chosen such as to match $a_2^{(2)}$.

\subsection{Watermelon exponents}

We now turn to the watermelon exponents $x_{\ell_1,\ell_2}$, which are associated to sectors of the loop model with $\ell_1, \ell_2$ through-lines of each color. 
As described previously, the lowest-lying levels of such sectors are related in the vertex model to the lowest levels in the sectors 
\begin{equation}
S_z^{(1)} = \frac{\ell_1}{2} \,, \qquad S_z^{(2)} = \frac{\ell_2}{2} \,,
\end{equation}
at twists $\varphi_1 = \varphi_2 =0$. 

We will in fact restrict the present study to sectors where both $\ell_1$ and $\ell_2$ are nonzero. Indeed, experience with the $a_3^{(2)}$ model \cite{VJS:a32} has shown us that the sectors $(\ell_1, 0)$ or $(0,\ell_2)$ may involve an intricate structure of crossovers, whose detailed investigation goes beyond the scope of this work. 

In sectors $\ell_1,\ell_2 \neq 0$, the effective central charge $c_{\ell_1, \ell_2}$ of the loop model is given by the effective central charge $c_{{\ell_1 \over 2}, {\ell_2\over 2},0}$ of the vertex model at twists $\varphi_1 = \varphi_2 =0$, that is, 
\begin{subequations}
\begin{eqnarray}
c_{\ell_1,\ell_2} &=& 3 - \frac{6 \gamma}{\pi} \left[ \left(\ell_1 \over 2\right)^2  +  \left(\ell_2 \over 2\right)^2 \right]  \,, \\
&=& 3 - \frac{12}{K}\left[ \left(\ell_1 \over 2\right)^2  +  \left(\ell_2 \over 2\right)^2 \right] \,,
\end{eqnarray}
\end{subequations}
and therefore 
\begin{subequations}
\label{eq:watermelons}
\begin{eqnarray}
x_{\ell_1,\ell_2} &=& \frac{1}{2}\frac{1}{K-2} +\frac{\left(\ell_1 \over 2\right)^2  +  \left(\ell_2 \over 2\right)^2-2}{K}  \,,
\\ 
&=& \frac{\gamma}{4(\pi-\gamma)} +\frac{2\pi}{\gamma}\left(\left(\ell_1 \over 2\right)^2  +  \left(\ell_2 \over 2\right)^2-2\right) \qquad \mbox{for } \ell_1, \ell_2  \neq 0 \,. 
\end{eqnarray}
\end{subequations}

Note that, in contrast with the sector of the ground state, in the sectors with non-zero numbers of legs, the spectrum of excitations over the watermelon exponents is continuous. This means in fact that in the $L \to \infty$ limit we can define in each sector the following continuum of exponents 
\begin{eqnarray}
x_{\ell_1, \ell_2, s} &=& \frac{1}{2}\frac{1}{K-2} +\frac{\left(\ell_1 \over 2\right)^2  +  \left(\ell_2 \over 2\right)^2-2}{K} +   \frac{2}{K-2}s^2 \nonumber
\\ 
&=&\frac{\gamma}{4(\pi-\gamma)} +\frac{2\pi}{\gamma}\left(\left(\ell_1 \over 2\right)^2  +  \left(\ell_2 \over 2\right)^2-2\right) 
+   \frac{\gamma}{\pi - \gamma}s^2
 \qquad \mbox{$\ell_1, \ell_2\neq 0$, $s>0$}   \,.
\label{eq:watermelons_s}
\end{eqnarray}

We have normalized the continuum index as in the $a_2^{(2)}$ case (see eq.~(\ref{eq:a22:xms})), and insist that since here the magnetic sectors have zero twist, the associated densities of states are the same as in the untwisted model.

\subsection{The special point $n=0$}

At the end of this long discussion, it is probably useful to summarize the results for the case  $n=0$ ($\gamma = \frac{\pi}{2}$, $K=4$,  $\theta = \frac{3 \pi}{4}$) which is relevant for the modified first truncation of the Chalker-Coddington model introduced at the beginning of this paper.

\smallskip

The \underline{central charge} follows from \eqref{eq:cc_intermsofK} and is of course 
\begin{equation}
c=0 \,.
\end{equation}
It  is associated with a {\sl discrete state}, and the corresponding ground state is separated by a finite gap from the continuum. This gap defines the 
 thermal exponent  given by \eqref{eq:Xttheta}:
\begin{equation}
x_t = \frac{1}{4} \,,
\end{equation}
with logarithmic corrections. This exponent is the lowest of a continuum parametrized by the continuous quantum number $s$. 

\smallskip

\underline{Winding exponents} (see \eqref{eq:windexps}):
\begin{equation}
x=\Delta + \bar{\Delta} 
= 
 - \frac{1}{4} + 2 \left( ( w_-)^2 + ( w_+ + 1/2)^2  \right)+s^2 \,.
\end{equation}
Recall that the thermal exponent corresponds to $w_+=w_-=0$.

For a non-vanishing number of legs we have, for any values of $\ell_1,\ell_2$, a continuum starting at the watermelon exponents:

\smallskip

\underline{Watermelon exponents} for $\ell_1\ell_2 \neq 0$ (see \eqref{eq:watermelons_s}):
\begin{eqnarray}
x_{\ell_1, \ell_2, s} &=&\frac{\left(\ell_1 \over 2\right)^2  +  \left(\ell_2 \over 2\right)^2-1}{4} +   s^2 
 \qquad \mbox{$\ell_1, \ell_2 \neq 0$, $s>0$}   \,.
\label{eq:watermelons_s_n0}
\end{eqnarray}

In particular, $x_{2,2}=\frac{1}{4}$ and therefore coincides with $x_t$ at this point, which agrees with the observation of \cite{Ikhlef} in the discussion preceding their eq.~(5.5).
Moreover, $x_{1,1}=-\frac{1}{8}$, a value which is close to but different from zero, contrarily to what was proposed by \cite{Ikhlef} in the beginning of their section 5.3.

Notice also that \cite{Ikhlef} gave the numerical result $d_{\rm f} = 2 - x_{2,2} \simeq 1.71$ for the fractal dimension
of a path (that is, $x_{2,2} \simeq 0.29$ for the exponent). This is well in line with previous observations \cite{VJSpolymer} that, in the presence of an (unexpected) continuous spectrum, numerical results have a tendency to overestimate
the critical exponents, somewhat as if the finite-size results were polluted by an effective non-zero value of $s$ in finite-size.

Now, we have to recall that while the original, non-integrable  trucated model and its integrable deformation were argued to be in the same universality class, some subtleties arise in sectors with an odd number of legs. Indeed, as discussed in section \ref{sec:intvsnonint}, the watermelon exponent of the non-integrable model with odd number of legs should correspond to the watermelons of the integrable models with the nearest larger even number of legs. This means in particular that the corresponding exponent $\tilde{x}_{1,1}$, which we indicate by a tilde in order to avoid confusion, should in fact be equal to $x_{2,2}$, that is 
\be 
\tilde{x}_{1,1} = \frac{1}{4} \,.
\ee

\section{Physical observations}

\subsection{The nature of the critical theory for the modified and unmodified truncated model}

We first discuss the nature of the conformal field theory describing the modified truncated model. This theory is made of two (possible twisted, depending on the sector) compact bosons of identical radius, and one non-compact boson. The fact that the central charge $c=0$ is however non-trivial, and due to the existence of a discrete state, which can only be explained by a more thorough description of the theory in terms of a ${SL(2,\mathbb{R})\over U(1)}\times U(1)$ model. Indeed (we restrict to the point $K=4,\gamma={\pi\over 2}$ in what follows), the ground state is obtained by choosing the twists $\varphi_1=\varphi_2={\pi\over 2}$. For a simple system of two twisted compact and one non-compact bosons, we would have the corresponding central charge
from (\ref{c001conj}) 
\begin{equation}
c_{\rm naive}=3-6{(\varphi_1)^2+(\varphi_2)^2\over \pi\gamma}=-3 \,. \label{cnaive}
\end{equation}
The fact that the true central charge is $c=0$ is possible only because there is a discrete state that appears in the spectrum. This means also that there is a gap between the ground state and the continuum in this sector. 

The value (\ref{cnaive}) is no accident. It can be interpreted at $-3=-4+1=2\times -2+1$. While the $+1$ can be ascribed to the non-compact boson, the $2\times -2=-4$ can be interpreted as the central charge $c^D=-4$  of a system of two independent dense polymers.  This can also be seen at the level of the exponents. Observe indeed that, from (\ref{eq:watermelons_s_n0})
\begin{equation}
x_{\ell_1,\ell_2}={\ell_1^2+\ell_2^2\over 16}-{1\over 4} \,,
\end{equation}
 Meanwhile, for a system of two decoupled dense polymers we would have (recall the watermelon exponents for polymers are $x_\ell^D={\ell^2\over 16}-{1\over 4}$)
\begin{equation}
x_{\ell_1,\ell_2}^D-{c^D\over 12}={\ell_1^2+\ell_2^2\over 16}-{1\over 2}+{4\over 12}={\ell_1^2+\ell_2^2\over 16}-{1\over 6}
\end{equation}
so finally
\begin{equation}
x_{\ell_1,\ell_2}=x_{\ell_1,\ell_2}^D-{c^D\over 12}-{1\over 12}
\end{equation}
Note that,   if the non-compact boson were to become massive, this would leave a simple theory of two dense polymers with $c^D=-4$. In fact, this is exactly what happens when one moves in the phase where the monomer fugacity is greater than the critical value, as we discuss below.  Dense polymers are known to have $gl(1|1)$ symmetry, so this is compatible with the presence (unbroken by increasing the fugacity) of the $gl(1|1)_+\times gl(1|1)_-$ symmetry in our model.

\begin{figure}
\centering
    \includegraphics[scale=.5]{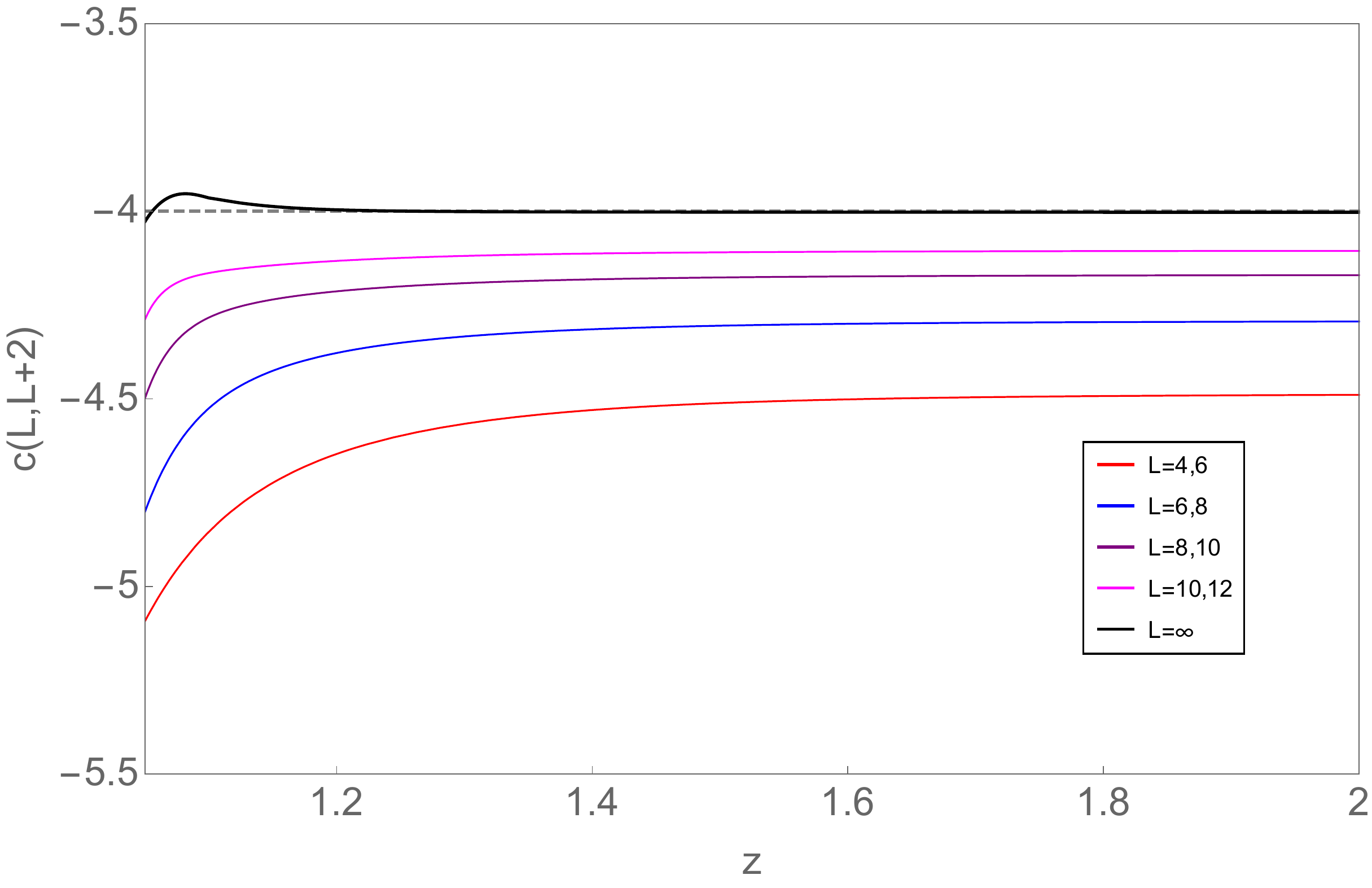}
    \caption{Estimate of the central charge for successive sizes as a function of the fugacity $z$. Using two sizes $L$ and $L+2$ at fixed $z$ we obtain an estimate of $c(L,L+2)$ that converges as $L$ increases. The model seems to have $c=-4$ for all $z$ higher than the critical value $z_C$. The ground state that we use comes from the sector with zero magnetization. The eigenvalue is exactly the same as for the sector propagating $(2,2)$ loops.
    }\label{centralchargedense}
\end{figure}

When $z$ is increased beyond $z_C\sim1.032$, the system  stays critical but the physics---like in the ordinary self-avoiding walk case \cite{DupSaldense}---is very different and the model is in a new universality class. The ground state is no longer a state of dilute polymers but becomes dense. The largest eigenvalue of the vacuum sector is no longer $1$ and does not contain the vacuum. The new ground state corresponds to the energy of the sector $(2,2)$. Numerical data show that as soon as we go higher than $z_C$, the system has a unique phase with central charge $c=-4$ (Figure \ref{centralchargedense}). This can be understood in the limit $z\rightarrow\infty$. The $R$-matrix in this limit is greatly simplified and corresponds to two decoupled dense polymers (one for each color). In particular, it reads
\bea
R=
\left(
\begin{tikzpicture}[scale=.4,baseline={([yshift=-.5ex]current bounding box.center)}]
\draw [black, line width=0.2pt] (0,-1)  -- (1,0); 
\draw [black, line width=0.2pt] (0,1) -- (1,0); 
\draw [black, line width=0.2pt] (-1,0) -- (0,-1); 
\draw [black, line width=0.2pt] (-1,0) -- (0,1);  
\draw[red, line width=0.3mm, domain=-45:45] plot ({-1+0.6*cos(\x)}, {0.6*sin(\x)});
\draw[red, line width=0.3mm, domain=135:225] plot ({1+0.6*cos(\x)}, {0.6*sin(\x)});
\end{tikzpicture}-\begin{tikzpicture}[scale=.4,baseline={([yshift=-.5ex]current bounding box.center)}]
\draw [black, line width=0.2pt] (0,-1)  -- (1,0); 
\draw [black, line width=0.2pt] (0,1) -- (1,0); 
\draw [black, line width=0.2pt] (-1,0) -- (0,-1); 
\draw [black, line width=0.2pt] (-1,0) -- (0,1);  
\draw[red, line width=0.3mm, domain=-135:-45] plot ({0.6*cos(\x)}, {1+0.6*sin(\x)});
\draw[red, line width=0.3mm, domain=45:135] plot ({0.6*cos(\x)}, {-1+0.6*sin(\x)});
\end{tikzpicture}\right)\otimes
\left(
\begin{tikzpicture}[scale=.4,baseline={([yshift=-.5ex]current bounding box.center)}]
\draw [black, line width=0.2pt] (0,-1)  -- (1,0); 
\draw [black, line width=0.2pt] (0,1) -- (1,0); 
\draw [black, line width=0.2pt] (-1,0) -- (0,-1); 
\draw [black, line width=0.2pt] (-1,0) -- (0,1);  
\draw[blue, line width=0.3mm, domain=-45:45] plot ({-1+0.6*cos(\x)}, {0.6*sin(\x)});
\draw[blue, line width=0.3mm, domain=135:225] plot ({1+0.6*cos(\x)}, {0.6*sin(\x)});
\end{tikzpicture}-\begin{tikzpicture}[scale=.4,baseline={([yshift=-.5ex]current bounding box.center)}]
\draw [black, line width=0.2pt] (0,-1)  -- (1,0); 
\draw [black, line width=0.2pt] (0,1) -- (1,0); 
\draw [black, line width=0.2pt] (-1,0) -- (0,-1); 
\draw [black, line width=0.2pt] (-1,0) -- (0,1);  
\draw[blue, line width=0.3mm, domain=-135:-45] plot ({0.6*cos(\x)}, {1+0.6*sin(\x)});
\draw[blue, line width=0.3mm, domain=45:135] plot ({0.6*cos(\x)}, {-1+0.6*sin(\x)});
\end{tikzpicture}\right)
\eea
at the isotropic point, where we kept separated the two colors for clarity and to emphasize that the system is decoupled. Note that the minus sign can be switched (indeed in the Temperley-Lieb algebra, the relations between the generators $e_i$ are also verified by $-e_i$ if the loop weight is vanishing.)

\begin{figure}
\centering
    \includegraphics[scale=.5]{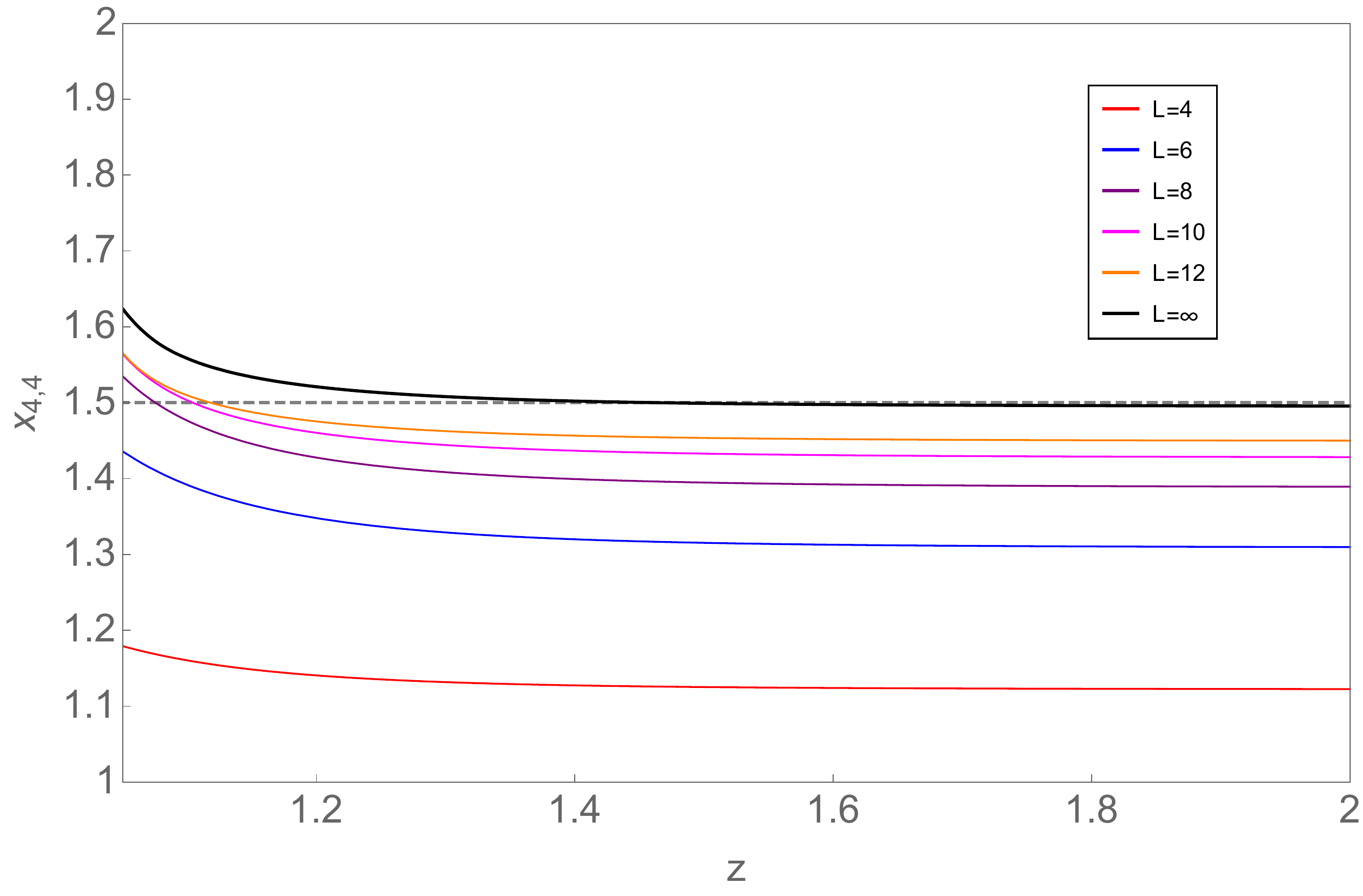}
    \caption{Estimate of the exponent $x_{4,4}$ corresponding to the propagation of $(4,4)$ through-lines. It converges toward the value $3/2$. The  dimension is twice the value expected in the dense polymer model because of the two colors.
    }\label{exponentdense}
\end{figure}

The central charge is hence twice the central charge of one dense polymer, $2\times(-2)=-4$. Watermelon exponents for dense polymers were given earlier, $x_\ell^D={\ell^2\over 16}-{1\over 4}$. Since  in the $z\rightarrow\infty$ limit the blue and red degrees of freedom are independent, the exponents in this limit are obtained simply by summing the dense values for each of the two colors. So for instance for the $(\ell_1=4,\ell_2=4)$ operator we have $x_{44}=2~x_4^D={3\over 2}$, a result we have checked   numerically (Figure \ref{exponentdense}).
 
\subsection{Comparison with expected properties in the untruncated case}

If we now restrict to  $w_-=0$, $\ell_1=\ell_2  ~\equiv  \ell$ and set $J={1\over 2}+is$, we can write the  exponents in the truncated model as 
\begin{equation}
x_{w,s}=2(w_++1)w_++J(1-J)\label{firstF}
\end{equation}
and 
\begin{equation}
x_{\ell,s}={\ell^2-4\over 8}+J(1-J)\label{secondF}
\end{equation}
We note meanwhile that a subset of the exponents for the critical theory in the  untruncated case was conjectured  in \cite{BWZII} to be 
\begin{equation}
x^{\rm CC}=\Delta+\bar{\Delta}=Xp(p+1)+Xq(1-q);~ p=0,1,2\ldots;~q={1\over 2}+is\label{Roberto}
\end{equation}
where $X\approx {1\over 4}$. This formula corresponds to   exponents  proportional to the quadratic Casimir of $gl(2|2)$, the number $p$ corresponding to integer discrete spins $j_1=p$ of $su(2)$, the number $q$ to the principal continuous series of $sl(2,R)$ (in \cite{BWZII}  the fact that  $p$ is integer is a consequence of the Hilbert space of the spin chain being of the form $(V\otimes V^*)^{\otimes L}$. Other choices (such as an extra $V$ or $V^*$) would  allow $p$ half-integer as well \cite{ReadSaleur}). 

As explained earlier, our first goal is to  study the exponents of the truncated model and compare them with the expected exponents of the original, untruncated problem.  It is not clear to us at this stage for instance whether one can go further and really identify, within the truncated model, the ''truncated version'' of physical observables  of the original problem.  For instance, we are not sure whether the number $p$ in (\ref{Roberto}) should be identified with $\ell$ or $w_+$ in the truncated model: see further discussion of this point below and in the appendix.  In any case, we observe  that (\ref{firstF}) or (\ref{secondF})  are qualitatively close to (\ref{Roberto}): both formulas  contain  a continuous as well as  a discrete part, and  these parts are exactly quadratic. The main differences are that the coefficients of the discrete and continuous  terms are different in (\ref{firstF}) but not in (\ref{secondF}), and that they are both off the expected value of $X\approx {1\over 4}$. 
Another difference, if we try to identify $p$ in (\ref{Roberto}) with $w_+$ in (\ref{firstF})  is that, unlike $p$,  $w_+$ is always even (since $w_+=w_1+w_2$ with $w_-=w_1-w_2=0$). Setting ${\ell^2-4\over 8}=2 ({\ell\over 4}-{1\over 2})({\ell\over 4}+{1\over 2})$, one can also try to identify $p$ in (\ref{Roberto}) with ${\ell\over 4}-{1\over 2}$ in  (\ref{secondF}), but, unlike $p$,  this latter quantity can be a half-integer is $\ell$ is a multiple of $4$. So the matching can never be complete.

The difference between the coefficients  of the discrete and continuous contributions in (\ref{firstF}) or (\ref{secondF})  as compared with (\ref{Roberto})  arises because the $gl(2|2)$ symmetry is broken in the truncated model. The fact that these coefficients  are both so much off  the value of $X$ is of course disappointing. If we assume that the spectrum of the $gl(2|2)$ chain can be obtained indeed by considering higher and higher truncations, this means in particular that the coefficient ${1\over 2}$ for the continuous part must evolve towards ${1\over 4}$, as the $gl(2|2)$ symmetry is closer to being realized.

On the other hand, it is remarkable that the structure of our theory is qualitatively similar to the one in \cite{BWZII}. Like these authors, we find a normalizable ground state separated by a finite gap from a continuum, and, when restricting to pure winding or pure watermelon operators,  two families of basic observables corresponding to $su(2)$ spins and the continuous $sl(2,\mathbb{R})$ series. 

A lasting mystery is the value of the thermal exponent $\nu$. It has been suggested \cite{Tsvelik} that the corresponding dimension might be obtained by putting $p=2,s=0$ in (\ref{Roberto}), leading to 
\begin{equation}
x^{\rm CC}_t\equiv \Delta+\bar{\Delta}=6X+{X\over 4}={25 X\over 4}\approx {25\over 16}\label{nuexp}
\end{equation}
Using $\Delta+\bar{\Delta}=2-{1\over \nu}$ leads to $\nu={16\over 7}\sim 2.29$, a value which, considering the very high sensitivity of the result on the numerical value $X$,  is not that different from the most recent numerical results of $\nu\approx 2.6$.  In our case, the model is simple enough that we can assert equality of the dimension $x_t$ with the $(2,2)$ leg operator, i.e., we find $x_t={1\over 4}$, vastly different from  (\ref{nuexp}). The problem of relating $x_t$ to the correlation length exponent $\nu$ is however non-trivial, and will be discussed below.

\section{Higher truncations}
\label{sec:higher-truncations}

We now give more detail about the higher truncations, together with the results of some numerical studies.

In the following we use the following combination of variables 
\bea
\Gamma^\sigma_{e_2\leftarrow e_1}=\left(b^*_{\sigma}(e_2)b_{\sigma}(e_1)+f^*_{\sigma}(e_2)f_{\sigma}(e_1)\right)
\eea
where $e_2$ and $e_1$ are edges. This term geometrically represents a loop of color $\sigma$ going from $e_1$ to $e_2$. We dropped the indices $L$ and $R$ for simplicity; they are useful to distinguish between the interaction at the vertices and the edges, but later the contribution of the edges is going to be directly taken into account.

\subsection{The second truncation}
We start with the second truncation. The operator \eqref{truncatededge} lying at an edge $e$ is
\bea
e^{S^2_e}\equiv\sum_{m=0}^2\frac{z^{2m}}{(m!)^2}\prod_{\sigma=+,-}\left(b^*_{\sigma,L}(e)b_{\sigma,R}(e)+f^*_{\sigma,L}(e)f_{\sigma,R}(e)\right)^m
\eea
with $M=2$. It constrains the edge to carry at most $2$ strands of each color. We are again going to expand the partition function in terms of the functions $\Gamma$ representing connectivities. We drop the indices $L,R$ and only consider the vertices for one color.
The full interaction is still the tensor product of individual tiles respecting the constraint of having the same number of strands per color at every edge. The vertex interaction for one color is then
\bea
e^{S^+_v}&=&\exp\left(\sum_{j=1,2}\sum_{i=3,4}\mathcal{S}_{i,j}\Gamma^+_{e_i\leftarrow e_j}\right) \,,
\eea
where $i$ and $j$ refers to the number of the edges in the vertex \eqref{vertex}. We now can expand this expression such that we only keep terms with at most $2$ strands on each $e_i$ and $e_j$. We need to expand the exponential to fourth order in $S$ to get all contributions. We can represent all terms with the following diagrams. First we have the contribution of order $0$ and the four ones of order $1$ that also appear in the first truncation:
\bea
\begin{tikzpicture}[scale=0.7,baseline={([yshift=-.5ex]current bounding box.center)}]
\draw [black, line width=0.2pt] (0,-1)  -- (1,0); 
\draw [black, line width=0.2pt] (0,1) -- (1,0); 
\draw [black, line width=0.2pt] (-1,0) -- (0,-1); 
\draw [black, line width=0.2pt] (-1,0) -- (0,1);  
\draw (0,-2) node{$1$};
\end{tikzpicture}
\quad
\begin{tikzpicture}[scale=0.7,baseline={([yshift=-.5ex]current bounding box.center)}]
\draw [black, line width=0.2pt] (0,-1)  -- (1,0); 
\draw [black, line width=0.2pt] (0,1) -- (1,0); 
\draw [black, line width=0.2pt] (-1,0) -- (0,-1); 
\draw [black, line width=0.2pt] (-1,0) -- (0,1);  
\draw[red, line width=0.3mm, domain=-45:45] plot ({-1+0.6*cos(\x)}, {0.6*sin(\x)});
\draw (0,-2) node{$t$};
\end{tikzpicture}
\quad
\begin{tikzpicture}[scale=0.7,baseline={([yshift=-.5ex]current bounding box.center)}]
\draw [black, line width=0.2pt] (0,-1)  -- (1,0); 
\draw [black, line width=0.2pt] (0,1) -- (1,0); 
\draw [black, line width=0.2pt] (-1,0) -- (0,-1); 
\draw [black, line width=0.2pt] (-1,0) -- (0,1);  
\draw[red, line width=0.3mm, domain=-135:-45] plot ({0.6*cos(\x)}, {1+0.6*sin(\x)});
\draw (0,-2) node{$r$};
\end{tikzpicture}
\quad
\begin{tikzpicture}[scale=0.7,baseline={([yshift=-.5ex]current bounding box.center)}]
\draw [black, line width=0.2pt] (0,-1)  -- (1,0); 
\draw [black, line width=0.2pt] (0,1) -- (1,0); 
\draw [black, line width=0.2pt] (-1,0) -- (0,-1); 
\draw [black, line width=0.2pt] (-1,0) -- (0,1);  
\draw[red, line width=0.3mm, domain=135:225] plot ({1+0.6*cos(\x)}, {0.6*sin(\x)});
\draw (0,-2) node{$t$};
\end{tikzpicture}
\quad
\begin{tikzpicture}[scale=0.7,baseline={([yshift=-.5ex]current bounding box.center)}]
\draw [black, line width=0.2pt] (0,-1)  -- (1,0); 
\draw [black, line width=0.2pt] (0,1) -- (1,0); 
\draw [black, line width=0.2pt] (-1,0) -- (0,-1); 
\draw [black, line width=0.2pt] (-1,0) -- (0,1);  
\draw[red, line width=0.3mm, domain=45:135] plot ({0.6*cos(\x)}, {-1+0.6*sin(\x)});
\draw (0,-2) node{$-r$};
\end{tikzpicture}
\eea

At order $2$ we have not only the tiles where we picked twice the same path in the expansion:
\bea
\begin{tikzpicture}[scale=0.7,baseline={([yshift=-.5ex]current bounding box.center)}]
\draw [black, line width=0.2pt] (0,-1)  -- (1,0); 
\draw [black, line width=0.2pt] (0,1) -- (1,0); 
\draw [black, line width=0.2pt] (-1,0) -- (0,-1); 
\draw [black, line width=0.2pt] (-1,0) -- (0,1);  
\draw[red, line width=0.3mm, domain=-45:45] plot ({-1+0.6*cos(\x)}, {0.6*sin(\x)});
\draw[red, line width=0.3mm, domain=-45:45] plot ({-1+0.8*cos(\x)}, {0.8*sin(\x)});
\draw (0,-1.5) node{$\frac{t^2}{2}$};
\end{tikzpicture}
\quad
\begin{tikzpicture}[scale=0.7,baseline={([yshift=-.5ex]current bounding box.center)}]
\draw [black, line width=0.2pt] (0,-1)  -- (1,0); 
\draw [black, line width=0.2pt] (0,1) -- (1,0); 
\draw [black, line width=0.2pt] (-1,0) -- (0,-1); 
\draw [black, line width=0.2pt] (-1,0) -- (0,1);  
\draw[red, line width=0.3mm, domain=-135:-45] plot ({0.6*cos(\x)}, {1+0.6*sin(\x)});
\draw[red, line width=0.3mm, domain=-135:-45] plot ({0.8*cos(\x)}, {1+0.8*sin(\x)});
\draw (0,-1.5) node{$\frac{r^2}{2}$};
\end{tikzpicture}
\quad
\begin{tikzpicture}[scale=0.7,baseline={([yshift=-.5ex]current bounding box.center)}]
\draw [black, line width=0.2pt] (0,-1)  -- (1,0); 
\draw [black, line width=0.2pt] (0,1) -- (1,0); 
\draw [black, line width=0.2pt] (-1,0) -- (0,-1); 
\draw [black, line width=0.2pt] (-1,0) -- (0,1);  
\draw[red, line width=0.3mm, domain=135:225] plot ({1+0.6*cos(\x)}, {0.6*sin(\x)});
\draw[red, line width=0.3mm, domain=135:225] plot ({1+0.8*cos(\x)}, {0.8*sin(\x)});
\draw (0,-1.5) node{$\frac{t^2}{2}$};
\end{tikzpicture}
\quad
\begin{tikzpicture}[scale=0.7,baseline={([yshift=-.5ex]current bounding box.center)}]
\draw [black, line width=0.2pt] (0,-1)  -- (1,0); 
\draw [black, line width=0.2pt] (0,1) -- (1,0); 
\draw [black, line width=0.2pt] (-1,0) -- (0,-1); 
\draw [black, line width=0.2pt] (-1,0) -- (0,1);  
\draw[red, line width=0.3mm, domain=45:135] plot ({0.6*cos(\x)}, {-1+0.6*sin(\x)});
\draw[red, line width=0.3mm, domain=45:135] plot ({0.8*cos(\x)}, {-1+0.8*sin(\x)});
\draw (0,-1.5) node{$\frac{r^2}{2}$};
\end{tikzpicture}
\eea
but also the tiles that we can pick by taking two different paths. They come with an extra factor $2$ cancelling the one from the expansion of the exponential function:
\bea
\begin{tikzpicture}[scale=0.7,baseline={([yshift=-.5ex]current bounding box.center)}]
\draw [black, line width=0.2pt] (0,-1)  -- (1,0); 
\draw [black, line width=0.2pt] (0,1) -- (1,0); 
\draw [black, line width=0.2pt] (-1,0) -- (0,-1); 
\draw [black, line width=0.2pt] (-1,0) -- (0,1);  
\draw[red, line width=0.3mm, domain=-45:45] plot ({-1+0.6*cos(\x)}, {0.6*sin(\x)});
\draw[red, line width=0.3mm, domain=-135:-45] plot ({0.6*cos(\x)}, {1+0.6*sin(\x)});
\draw (0,-1.5) node{$t r$};
\end{tikzpicture}
\quad
\begin{tikzpicture}[scale=0.7,baseline={([yshift=-.5ex]current bounding box.center)}]
\draw [black, line width=0.2pt] (0,-1)  -- (1,0); 
\draw [black, line width=0.2pt] (0,1) -- (1,0); 
\draw [black, line width=0.2pt] (-1,0) -- (0,-1); 
\draw [black, line width=0.2pt] (-1,0) -- (0,1);  
\draw[red, line width=0.3mm, domain=-135:-45] plot ({0.6*cos(\x)}, {1+0.6*sin(\x)});
\draw[red, line width=0.3mm, domain=135:225] plot ({1+0.6*cos(\x)}, {0.6*sin(\x)});
\draw (0,-1.5) node{$t r$};
\end{tikzpicture}
\quad
\begin{tikzpicture}[scale=0.7,baseline={([yshift=-.5ex]current bounding box.center)}]
\draw [black, line width=0.2pt] (0,-1)  -- (1,0); 
\draw [black, line width=0.2pt] (0,1) -- (1,0); 
\draw [black, line width=0.2pt] (-1,0) -- (0,-1); 
\draw [black, line width=0.2pt] (-1,0) -- (0,1);  
\draw[red, line width=0.3mm, domain=135:225] plot ({1+0.6*cos(\x)}, {0.6*sin(\x)});
\draw[red, line width=0.3mm, domain=45:135] plot ({0.6*cos(\x)}, {-1+0.6*sin(\x)});
\draw (0,-1.5) node{$-t r$};
\end{tikzpicture}
\quad
\begin{tikzpicture}[scale=0.7,baseline={([yshift=-.5ex]current bounding box.center)}]
\draw [black, line width=0.2pt] (0,-1)  -- (1,0); 
\draw [black, line width=0.2pt] (0,1) -- (1,0); 
\draw [black, line width=0.2pt] (-1,0) -- (0,-1); 
\draw [black, line width=0.2pt] (-1,0) -- (0,1);  
\draw[red, line width=0.3mm, domain=45:135] plot ({0.6*cos(\x)}, {-1+0.6*sin(\x)});
\draw[red, line width=0.3mm, domain=-45:45] plot ({-1+0.6*cos(\x)}, {0.6*sin(\x)});
\draw (0,-1.5) node{$-t r$};
\end{tikzpicture}
\quad
\begin{tikzpicture}[scale=0.7,baseline={([yshift=-.5ex]current bounding box.center)}]
\draw [black, line width=0.2pt] (0,-1)  -- (1,0); 
\draw [black, line width=0.2pt] (0,1) -- (1,0); 
\draw [black, line width=0.2pt] (-1,0) -- (0,-1); 
\draw [black, line width=0.2pt] (-1,0) -- (0,1);  
\draw[red, line width=0.3mm, domain=-45:45] plot ({-1+0.6*cos(\x)}, {0.6*sin(\x)});
\draw[red, line width=0.3mm, domain=135:225] plot ({1+0.6*cos(\x)}, {0.6*sin(\x)});
\draw (0,-1.5) node{$t^2$};
\end{tikzpicture}
\quad
\begin{tikzpicture}[scale=0.7,baseline={([yshift=-.5ex]current bounding box.center)}]
\draw [black, line width=0.2pt] (0,-1)  -- (1,0); 
\draw [black, line width=0.2pt] (0,1) -- (1,0); 
\draw [black, line width=0.2pt] (-1,0) -- (0,-1); 
\draw [black, line width=0.2pt] (-1,0) -- (0,1);  
\draw[red, line width=0.3mm, domain=-135:-45] plot ({0.6*cos(\x)}, {1+0.6*sin(\x)});
\draw[red, line width=0.3mm, domain=45:135] plot ({0.6*cos(\x)}, {-1+0.6*sin(\x)});
\draw (0,-1.5) node{$-r^2$};
\end{tikzpicture}
\eea

At order $3$ we have four contributions where we pick twice the same path:
\bea
\begin{tikzpicture}[scale=0.7,baseline={([yshift=-.5ex]current bounding box.center)}]
\draw [black, line width=0.2pt] (0,-1)  -- (1,0); 
\draw [black, line width=0.2pt] (0,1) -- (1,0); 
\draw [black, line width=0.2pt] (-1,0) -- (0,-1); 
\draw [black, line width=0.2pt] (-1,0) -- (0,1);  
\draw[red, line width=0.3mm, domain=-45:45] plot ({-1+0.6*cos(\x)}, {0.6*sin(\x)});
\draw[red, line width=0.3mm, domain=-45:45] plot ({-1+0.8*cos(\x)}, {0.8*sin(\x)});
\draw[red, line width=0.3mm, domain=135:225] plot ({1+0.7*cos(\x)}, {0.7*sin(\x)});
\draw (0,-1.5) node{$\frac{t^3}{2}$};
\end{tikzpicture}
\quad
\begin{tikzpicture}[scale=0.7,baseline={([yshift=-.5ex]current bounding box.center)}]
\draw [black, line width=0.2pt] (0,-1)  -- (1,0); 
\draw [black, line width=0.2pt] (0,1) -- (1,0); 
\draw [black, line width=0.2pt] (-1,0) -- (0,-1); 
\draw [black, line width=0.2pt] (-1,0) -- (0,1);  
\draw[red, line width=0.3mm, domain=-135:-45] plot ({0.6*cos(\x)}, {1+0.6*sin(\x)});
\draw[red, line width=0.3mm, domain=-135:-45] plot ({0.8*cos(\x)}, {1+0.8*sin(\x)});
\draw[red, line width=0.3mm, domain=45:135] plot ({0.7*cos(\x)}, {-1+0.7*sin(\x)});
\draw (0,-1.5) node{$-\frac{r^3}{2}$};
\end{tikzpicture}
\quad
\begin{tikzpicture}[scale=0.7,baseline={([yshift=-.5ex]current bounding box.center)}]
\draw [black, line width=0.2pt] (0,-1)  -- (1,0); 
\draw [black, line width=0.2pt] (0,1) -- (1,0); 
\draw [black, line width=0.2pt] (-1,0) -- (0,-1); 
\draw [black, line width=0.2pt] (-1,0) -- (0,1);  
\draw[red, line width=0.3mm, domain=135:225] plot ({1+0.6*cos(\x)}, {0.6*sin(\x)});
\draw[red, line width=0.3mm, domain=135:225] plot ({1+0.8*cos(\x)}, {0.8*sin(\x)});
\draw[red, line width=0.3mm, domain=-45:45] plot ({-1+0.7*cos(\x)}, {0.7*sin(\x)});
\draw (0,-1.5) node{$\frac{t^3}{2}$};
\end{tikzpicture}
\quad
\begin{tikzpicture}[scale=0.7,baseline={([yshift=-.5ex]current bounding box.center)}]
\draw [black, line width=0.2pt] (0,-1)  -- (1,0); 
\draw [black, line width=0.2pt] (0,1) -- (1,0); 
\draw [black, line width=0.2pt] (-1,0) -- (0,-1); 
\draw [black, line width=0.2pt] (-1,0) -- (0,1);  
\draw[red, line width=0.3mm, domain=45:135] plot ({0.6*cos(\x)}, {-1+0.6*sin(\x)});
\draw[red, line width=0.3mm, domain=45:135] plot ({0.8*cos(\x)}, {-1+0.8*sin(\x)});
\draw[red, line width=0.3mm, domain=-135:-45] plot ({0.7*cos(\x)}, {1+0.7*sin(\x)});
\draw (0,-1.5) node{$\frac{r^3}{2}$};
\end{tikzpicture}
\eea
and four where we pick $3$ different paths:
\bea
\begin{tikzpicture}[scale=0.7,baseline={([yshift=-.5ex]current bounding box.center)}]
\draw [black, line width=0.2pt] (0,-1)  -- (1,0); 
\draw [black, line width=0.2pt] (0,1) -- (1,0); 
\draw [black, line width=0.2pt] (-1,0) -- (0,-1); 
\draw [black, line width=0.2pt] (-1,0) -- (0,1);  
\draw[red, line width=0.3mm, domain=-45:45] plot ({-1+0.6*cos(\x)}, {0.6*sin(\x)});
\draw[red, line width=0.3mm, domain=-135:-45] plot ({0.6*cos(\x)}, {1+0.6*sin(\x)});
\draw[red, line width=0.3mm, domain=135:225] plot ({1+0.6*cos(\x)}, {0.6*sin(\x)});
\draw (0,-1.5) node{$t^2 r$};
\end{tikzpicture}
\quad
\begin{tikzpicture}[scale=0.7,baseline={([yshift=-.5ex]current bounding box.center)}]
\draw [black, line width=0.2pt] (0,-1)  -- (1,0); 
\draw [black, line width=0.2pt] (0,1) -- (1,0); 
\draw [black, line width=0.2pt] (-1,0) -- (0,-1); 
\draw [black, line width=0.2pt] (-1,0) -- (0,1);  
\draw[red, line width=0.3mm, domain=-135:-45] plot ({0.6*cos(\x)}, {1+0.6*sin(\x)});
\draw[red, line width=0.3mm, domain=135:225] plot ({1+0.6*cos(\x)}, {0.6*sin(\x)});
\draw[red, line width=0.3mm, domain=45:135] plot ({0.6*cos(\x)}, {-1+0.6*sin(\x)});
\draw (0,-1.5) node{$-t r^2$};
\end{tikzpicture}
\quad
\begin{tikzpicture}[scale=0.7,baseline={([yshift=-.5ex]current bounding box.center)}]
\draw [black, line width=0.2pt] (0,-1)  -- (1,0); 
\draw [black, line width=0.2pt] (0,1) -- (1,0); 
\draw [black, line width=0.2pt] (-1,0) -- (0,-1); 
\draw [black, line width=0.2pt] (-1,0) -- (0,1);  
\draw[red, line width=0.3mm, domain=135:225] plot ({1+0.6*cos(\x)}, {0.6*sin(\x)});
\draw[red, line width=0.3mm, domain=45:135] plot ({0.6*cos(\x)}, {-1+0.6*sin(\x)});
\draw[red, line width=0.3mm, domain=-45:45] plot ({-1+0.6*cos(\x)}, {0.6*sin(\x)});
\draw (0,-1.5) node{$-t^2 r$};
\end{tikzpicture}
\quad
\begin{tikzpicture}[scale=0.7,baseline={([yshift=-.5ex]current bounding box.center)}]
\draw [black, line width=0.2pt] (0,-1)  -- (1,0); 
\draw [black, line width=0.2pt] (0,1) -- (1,0); 
\draw [black, line width=0.2pt] (-1,0) -- (0,-1); 
\draw [black, line width=0.2pt] (-1,0) -- (0,1);  
\draw[red, line width=0.3mm, domain=45:135] plot ({0.6*cos(\x)}, {-1+0.6*sin(\x)});
\draw[red, line width=0.3mm, domain=-45:45] plot ({-1+0.6*cos(\x)}, {0.6*sin(\x)});
\draw[red, line width=0.3mm, domain=-135:-45] plot ({0.6*cos(\x)}, {1+0.6*sin(\x)});
\draw (0,-1.5) node{$-t r^2$};
\end{tikzpicture}
\eea

Finally, the fourth order gives $3$ different contributions:
\bea
\begin{tikzpicture}[scale=0.7,baseline={([yshift=-.5ex]current bounding box.center)}]
\draw [black, line width=0.2pt] (0,-1)  -- (1,0); 
\draw [black, line width=0.2pt] (0,1) -- (1,0); 
\draw [black, line width=0.2pt] (-1,0) -- (0,-1); 
\draw [black, line width=0.2pt] (-1,0) -- (0,1);  
\draw[red, line width=0.3mm, domain=-45:45] plot ({-1+0.6*cos(\x)}, {0.6*sin(\x)});
\draw[red, line width=0.3mm, domain=-45:45] plot ({-1+0.8*cos(\x)}, {0.8*sin(\x)});
\draw[red, line width=0.3mm, domain=135:225] plot ({1+0.6*cos(\x)}, {0.6*sin(\x)});
\draw[red, line width=0.3mm, domain=135:225] plot ({1+0.8*cos(\x)}, {0.8*sin(\x)});
\draw (0,-1.5) node{$\frac{t^4}{4}$};
\end{tikzpicture}
\quad
\begin{tikzpicture}[scale=0.7,baseline={([yshift=-.5ex]current bounding box.center)}]
\draw [black, line width=0.2pt] (0,-1)  -- (1,0); 
\draw [black, line width=0.2pt] (0,1) -- (1,0); 
\draw [black, line width=0.2pt] (-1,0) -- (0,-1); 
\draw [black, line width=0.2pt] (-1,0) -- (0,1);  
\draw[red, line width=0.3mm, domain=-135:-45] plot ({0.6*cos(\x)}, {1+0.6*sin(\x)});
\draw[red, line width=0.3mm, domain=-135:-45] plot ({0.8*cos(\x)}, {1+0.8*sin(\x)});
\draw[red, line width=0.3mm, domain=45:135] plot ({0.6*cos(\x)}, {-1+0.6*sin(\x)});
\draw[red, line width=0.3mm, domain=45:135] plot ({0.8*cos(\x)}, {-1+0.8*sin(\x)});
\draw (0,-1.5) node{$\frac{r^4}{4}$};
\end{tikzpicture}
\quad
\begin{tikzpicture}[scale=0.7,baseline={([yshift=-.5ex]current bounding box.center)}]
\draw [black, line width=0.2pt] (0,-1)  -- (1,0); 
\draw [black, line width=0.2pt] (0,1) -- (1,0); 
\draw [black, line width=0.2pt] (-1,0) -- (0,-1); 
\draw [black, line width=0.2pt] (-1,0) -- (0,1);  
\draw[red, line width=0.3mm, domain=135:225] plot ({1+0.6*cos(\x)}, {0.6*sin(\x)});
\draw[red, line width=0.3mm, domain=45:135] plot ({0.6*cos(\x)}, {-1+0.6*sin(\x)});
\draw[red, line width=0.3mm, domain=-45:45] plot ({-1+0.6*cos(\x)}, {0.6*sin(\x)});
\draw[red, line width=0.3mm, domain=-135:-45] plot ({0.6*cos(\x)}, {1+0.6*sin(\x)});
\draw (0,-1.5) node{$-t^2r^2$};
\end{tikzpicture}
\eea
 Notice that in all cases the weight is divided by a symmetry factor of $p!$ each time where $p$ strands run in parallel.

We still to discuss the diagrammatic gluing of two tiles. While it is not ambiguous for the first truncation, we can have now several strands and need to specify the right way to form connectivities. We need to go back to the path integral formulation. Gluing two edges with one strand corresponds to performing the integration
\bea
\int\left[\diff b_{e_2}\right]\left[\diff f_{e_2}\right]\Gamma^\sigma_{e_3\leftarrow e_2}\Gamma^\sigma_{e_2\leftarrow e_1}=\Gamma^\sigma_{e_3\leftarrow e_1}
\eea
which is exactly what is expected in the original loop model. Gluing one strand going from $e_1$ to $e_2$ to one strand going from $e_2$ to $e_3$ gives one strand between $e_1$ and $e_3$. Now for a pair of strands we have
\bea
\int\left[\diff b_{e_2}\right]\left[\diff f_{e_2}\right]\Gamma_{e_{B_1}\leftarrow e_2}\Gamma_{e_{B_2}\leftarrow e_2}\Gamma_{e_2\leftarrow e_{A_1}}\Gamma_{e_{2}\leftarrow e_{A_2}}=\Gamma_{e_{B_1}\leftarrow e_{A_1}}\Gamma_{e_{B_2}\leftarrow e_{A_2}}+\Gamma_{e_{B_2}\leftarrow e_{A_1}}\Gamma_{e_{B_1}\leftarrow e_{A_2}}
\eea
In the loop model this corresponds to gluing a pair of double strands together in the two possible ways.
\bea
\begin{tikzpicture}[scale=.8,baseline={([yshift=-.5ex]current bounding box.center)}]
\draw [black, line width=0.2pt] (0,-1)  -- (1,-1); 
\draw [black, line width=0.2pt] (1,-1) -- (1,1); 
\draw [black, line width=0.2pt] (1,1) -- (0,1); 
\draw [black, line width=0.2pt] (0,1) -- (0,-1);  
\draw [red, line width=0.3mm] (-1.5,.5) -- (0,.5);  
\draw [red, line width=0.3mm] (1,.5) -- (2.5,.5);  
\draw [red, line width=0.3mm] (-1.5,-.5) -- (0,-.5);  
\draw [red, line width=0.3mm] (1,-.5) -- (2.5,-.5);  
\draw (-1.9,.5) node{$e_{A_1}$};
\draw (-1.9,-.5) node{$e_{A_2}$};
\draw (2.9,.5) node{$e_{B_1}$};
\draw (2.9,-.5) node{$e_{B_2}$};
\end{tikzpicture}
=
\begin{tikzpicture}[scale=.8,baseline={([yshift=-.5ex]current bounding box.center)}]
\draw [red, line width=0.3mm] (-1.5,.5) -- (1.5,.5);  
\draw [red, line width=0.3mm] (-1.5,-.5) -- (1.5,-.5);  
\draw (-1.9,.5) node{$e_{A_1}$};
\draw (-1.9,-.5) node{$e_{A_2}$};
\draw (1.9,.5) node{$e_{B_1}$};
\draw (1.9,-.5) node{$e_{B_2}$};
\end{tikzpicture}
+
\begin{tikzpicture}[scale=.8,baseline={([yshift=-.5ex]current bounding box.center)}]
\draw [red, line width=0.3mm] (-1.5,.5) -- (1.5,.5);  
\draw [red, line width=0.3mm] (-1.5,-.5) -- (1.5,-.5);  
\draw (-1.9,.5) node{$e_{A_1}$};
\draw (-1.9,-.5) node{$e_{A_2}$};
\draw (1.9,.5) node{$e_{B_2}$};
\draw (1.9,-.5) node{$e_{B_1}$};
\end{tikzpicture}
\eea
The loop model is then not strictly planar: two loops can intersect. Closed loops still have a vanishing weight. For instance the following gluing, represented diagrammatically, gives a non-zero trivial contribution:
\bea
\begin{tikzpicture}[scale=.5,baseline={([yshift=-.5ex]current bounding box.center)}]
\draw [black, line width=0.2pt] (0,-1)  -- (1,-1); 
\draw [black, line width=0.2pt] (1,-1) -- (1,1); 
\draw [black, line width=0.2pt] (1,1) -- (0,1); 
\draw [black, line width=0.2pt] (0,1) -- (0,-1);  
\draw [red, line width=0.3mm] (0,1.5) -- (1,1.5);  
\draw[red, line width=0.3mm, domain=90:270] plot ({0.5*cos(\x)}, {1+0.5*sin(\x)});
\draw[red, line width=0.3mm, domain=-90:90] plot ({1+0.5*cos(\x)}, {1+0.5*sin(\x)});
\draw [red, line width=0.3mm] (-1.5,-.5) -- (0,-.5);  
\draw [red, line width=0.3mm] (1,-.5) -- (2.5,-.5);  
\end{tikzpicture}
=
\begin{tikzpicture}[scale=.5,baseline={([yshift=-.5ex]current bounding box.center)}]
\draw [red, line width=0.3mm] (0,1.5) -- (1,1.5);  
\draw [red, line width=0.3mm] (0,.5) -- (1,.5);  
\draw [red, line width=0.3mm] (0,-.5) -- (1,-.5);  
\draw[red, line width=0.3mm, domain=90:270] plot ({0.5*cos(\x)}, {1+0.5*sin(\x)});
\draw[red, line width=0.3mm, domain=-90:90] plot ({1+0.5*cos(\x)}, {1+0.5*sin(\x)});
\draw [red, line width=0.3mm] (-1.5,-.5) -- (0,-.5);  
\draw [red, line width=0.3mm] (1,-.5) -- (2.5,-.5);  
\end{tikzpicture}
+
\begin{tikzpicture}[scale=.5,baseline={([yshift=-.5ex]current bounding box.center)}]
\draw [red, line width=0.3mm] (0,1.5) -- (1,1.5);  
\draw [red, line width=0.3mm] (0.6,-0.1) -- (1,-.5);  
\draw [red, line width=0.3mm] (0,.5) -- (0.4,0.1);  
\draw [red, line width=0.3mm] (0,-.5) -- (1,.5);  
\draw[red, line width=0.3mm, domain=90:270] plot ({0.5*cos(\x)}, {1+0.5*sin(\x)});
\draw[red, line width=0.3mm, domain=-90:90] plot ({1+0.5*cos(\x)}, {1+0.5*sin(\x)});
\draw [red, line width=0.3mm] (-1.5,-.5) -- (0,-.5);  
\draw [red, line width=0.3mm] (1,-.5) -- (2.5,-.5);  
\end{tikzpicture}
=
\quad
\begin{tikzpicture}[scale=.5,baseline={([yshift=-.5ex]current bounding box.center)}]
\draw [red, line width=0.3mm] (0,0) -- (2,0);  
\end{tikzpicture}
\eea

The full $R$-matrix is then given by
\bea
R &=& 
\begin{tikzpicture}[scale=.4,baseline={([yshift=-.5ex]current bounding box.center)}]
\draw [black, line width=0.2pt] (0,-1)  -- (1,0); 
\draw [black, line width=0.2pt] (0,1) -- (1,0); 
\draw [black, line width=0.2pt] (-1,0) -- (0,-1); 
\draw [black, line width=0.2pt] (-1,0) -- (0,1);  
\end{tikzpicture}
+t^2
\begin{tikzpicture}[scale=.4,baseline={([yshift=-.5ex]current bounding box.center)}]
\draw [black, line width=0.2pt] (0,-1)  -- (1,0); 
\draw [black, line width=0.2pt] (0,1) -- (1,0); 
\draw [black, line width=0.2pt] (-1,0) -- (0,-1); 
\draw [black, line width=0.2pt] (-1,0) -- (0,1);  
\draw[red, line width=0.3mm, domain=-45:45] plot ({-1+0.6*cos(\x)}, {0.6*sin(\x)});
\end{tikzpicture}
\otimes
\begin{tikzpicture}[scale=.4,baseline={([yshift=-.5ex]current bounding box.center)}]
\draw [black, line width=0.2pt] (0,-1)  -- (1,0); 
\draw [black, line width=0.2pt] (0,1) -- (1,0); 
\draw [black, line width=0.2pt] (-1,0) -- (0,-1); 
\draw [black, line width=0.2pt] (-1,0) -- (0,1);  
\draw[blue, line width=0.3mm, domain=-45:45] plot ({-1+0.6*cos(\x)}, {0.6*sin(\x)});
\end{tikzpicture}
+r^2
\begin{tikzpicture}[scale=.4,baseline={([yshift=-.5ex]current bounding box.center)}]
\draw [black, line width=0.2pt] (0,-1)  -- (1,0); 
\draw [black, line width=0.2pt] (0,1) -- (1,0); 
\draw [black, line width=0.2pt] (-1,0) -- (0,-1); 
\draw [black, line width=0.2pt] (-1,0) -- (0,1);  
\draw[red, line width=0.3mm, domain=-135:-45] plot ({0.6*cos(\x)}, {1+0.6*sin(\x)});
\end{tikzpicture}
\otimes
\begin{tikzpicture}[scale=.4,baseline={([yshift=-.5ex]current bounding box.center)}]
\draw [black, line width=0.2pt] (0,-1)  -- (1,0); 
\draw [black, line width=0.2pt] (0,1) -- (1,0); 
\draw [black, line width=0.2pt] (-1,0) -- (0,-1); 
\draw [black, line width=0.2pt] (-1,0) -- (0,1);  
\draw[blue, line width=0.3mm, domain=-135:-45] plot ({0.6*cos(\x)}, {1+0.6*sin(\x)});
\end{tikzpicture}
+t^2
\begin{tikzpicture}[scale=.4,baseline={([yshift=-.5ex]current bounding box.center)}]
\draw [black, line width=0.2pt] (0,-1)  -- (1,0); 
\draw [black, line width=0.2pt] (0,1) -- (1,0); 
\draw [black, line width=0.2pt] (-1,0) -- (0,-1); 
\draw [black, line width=0.2pt] (-1,0) -- (0,1);  
\draw[red, line width=0.3mm, domain=135:225] plot ({1+0.6*cos(\x)}, {0.6*sin(\x)});
\end{tikzpicture}\otimes
\begin{tikzpicture}[scale=.4,baseline={([yshift=-.5ex]current bounding box.center)}]
\draw [black, line width=0.2pt] (0,-1)  -- (1,0); 
\draw [black, line width=0.2pt] (0,1) -- (1,0); 
\draw [black, line width=0.2pt] (-1,0) -- (0,-1); 
\draw [black, line width=0.2pt] (-1,0) -- (0,1);  
\draw[blue, line width=0.3mm, domain=135:225] plot ({1+0.6*cos(\x)}, {0.6*sin(\x)});
\end{tikzpicture}
+r^2
\begin{tikzpicture}[scale=.4,baseline={([yshift=-.5ex]current bounding box.center)}]
\draw [black, line width=0.2pt] (0,-1)  -- (1,0); 
\draw [black, line width=0.2pt] (0,1) -- (1,0); 
\draw [black, line width=0.2pt] (-1,0) -- (0,-1); 
\draw [black, line width=0.2pt] (-1,0) -- (0,1);  
\draw[red, line width=0.3mm, domain=45:135] plot ({0.6*cos(\x)}, {-1+0.6*sin(\x)});
\end{tikzpicture}\otimes
\begin{tikzpicture}[scale=.4,baseline={([yshift=-.5ex]current bounding box.center)}]
\draw [black, line width=0.2pt] (0,-1)  -- (1,0); 
\draw [black, line width=0.2pt] (0,1) -- (1,0); 
\draw [black, line width=0.2pt] (-1,0) -- (0,-1); 
\draw [black, line width=0.2pt] (-1,0) -- (0,1);  
\draw[blue, line width=0.3mm, domain=45:135] plot ({0.6*cos(\x)}, {-1+0.6*sin(\x)});
\end{tikzpicture}+\frac{t^4}{4}\begin{tikzpicture}[scale=.4,baseline={([yshift=-.5ex]current bounding box.center)}]
\draw [black, line width=0.2pt] (0,-1)  -- (1,0); 
\draw [black, line width=0.2pt] (0,1) -- (1,0); 
\draw [black, line width=0.2pt] (-1,0) -- (0,-1); 
\draw [black, line width=0.2pt] (-1,0) -- (0,1);  
\draw[red, line width=0.3mm, domain=-45:45] plot ({-1+0.6*cos(\x)}, {0.6*sin(\x)});
\draw[red, line width=0.3mm, domain=-45:45] plot ({-1+0.8*cos(\x)}, {0.8*sin(\x)});
\end{tikzpicture}\otimes
\begin{tikzpicture}[scale=.4,baseline={([yshift=-.5ex]current bounding box.center)}]
\draw [black, line width=0.2pt] (0,-1)  -- (1,0); 
\draw [black, line width=0.2pt] (0,1) -- (1,0); 
\draw [black, line width=0.2pt] (-1,0) -- (0,-1); 
\draw [black, line width=0.2pt] (-1,0) -- (0,1);  
\draw[blue, line width=0.3mm, domain=-45:45] plot ({-1+0.6*cos(\x)}, {0.6*sin(\x)});
\draw[blue, line width=0.3mm, domain=-45:45] plot ({-1+0.8*cos(\x)}, {0.8*sin(\x)});
\end{tikzpicture}\nonumber\\
&&
+\frac{r^4}{4}\begin{tikzpicture}[scale=.4,baseline={([yshift=-.5ex]current bounding box.center)}]
\draw [black, line width=0.2pt] (0,-1)  -- (1,0); 
\draw [black, line width=0.2pt] (0,1) -- (1,0); 
\draw [black, line width=0.2pt] (-1,0) -- (0,-1); 
\draw [black, line width=0.2pt] (-1,0) -- (0,1);  
\draw[red, line width=0.3mm, domain=-135:-45] plot ({0.6*cos(\x)}, {1+0.6*sin(\x)});
\draw[red, line width=0.3mm, domain=-135:-45] plot ({0.8*cos(\x)}, {1+0.8*sin(\x)});
\end{tikzpicture}\otimes
\begin{tikzpicture}[scale=.4,baseline={([yshift=-.5ex]current bounding box.center)}]
\draw [black, line width=0.2pt] (0,-1)  -- (1,0); 
\draw [black, line width=0.2pt] (0,1) -- (1,0); 
\draw [black, line width=0.2pt] (-1,0) -- (0,-1); 
\draw [black, line width=0.2pt] (-1,0) -- (0,1);  
\draw[blue, line width=0.3mm, domain=-135:-45] plot ({0.6*cos(\x)}, {1+0.6*sin(\x)});
\draw[blue, line width=0.3mm, domain=-135:-45] plot ({0.8*cos(\x)}, {1+0.8*sin(\x)});
\end{tikzpicture}+
\frac{t^4}{4}
\begin{tikzpicture}[scale=.4,baseline={([yshift=-.5ex]current bounding box.center)}]
\draw [black, line width=0.2pt] (0,-1)  -- (1,0); 
\draw [black, line width=0.2pt] (0,1) -- (1,0); 
\draw [black, line width=0.2pt] (-1,0) -- (0,-1); 
\draw [black, line width=0.2pt] (-1,0) -- (0,1);  
\draw[red, line width=0.3mm, domain=135:225] plot ({1+0.6*cos(\x)}, {0.6*sin(\x)});
\draw[red, line width=0.3mm, domain=135:225] plot ({1+0.8*cos(\x)}, {0.8*sin(\x)});
\end{tikzpicture}\otimes
\begin{tikzpicture}[scale=.4,baseline={([yshift=-.5ex]current bounding box.center)}]
\draw [black, line width=0.2pt] (0,-1)  -- (1,0); 
\draw [black, line width=0.2pt] (0,1) -- (1,0); 
\draw [black, line width=0.2pt] (-1,0) -- (0,-1); 
\draw [black, line width=0.2pt] (-1,0) -- (0,1);  
\draw[blue, line width=0.3mm, domain=135:225] plot ({1+0.6*cos(\x)}, {0.6*sin(\x)});
\draw[blue, line width=0.3mm, domain=135:225] plot ({1+0.8*cos(\x)}, {0.8*sin(\x)});
\end{tikzpicture}
+\frac{r^4}{4}
\begin{tikzpicture}[scale=.4,baseline={([yshift=-.5ex]current bounding box.center)}]
\draw [black, line width=0.2pt] (0,-1)  -- (1,0); 
\draw [black, line width=0.2pt] (0,1) -- (1,0); 
\draw [black, line width=0.2pt] (-1,0) -- (0,-1); 
\draw [black, line width=0.2pt] (-1,0) -- (0,1);  
\draw[red, line width=0.3mm, domain=45:135] plot ({0.6*cos(\x)}, {-1+0.6*sin(\x)});
\draw[red, line width=0.3mm, domain=45:135] plot ({0.8*cos(\x)}, {-1+0.8*sin(\x)});
\end{tikzpicture}\otimes
\begin{tikzpicture}[scale=.4,baseline={([yshift=-.5ex]current bounding box.center)}]
\draw [black, line width=0.2pt] (0,-1)  -- (1,0); 
\draw [black, line width=0.2pt] (0,1) -- (1,0); 
\draw [black, line width=0.2pt] (-1,0) -- (0,-1); 
\draw [black, line width=0.2pt] (-1,0) -- (0,1);  
\draw[blue, line width=0.3mm, domain=45:135] plot ({0.6*cos(\x)}, {-1+0.6*sin(\x)});
\draw[blue, line width=0.3mm, domain=45:135] plot ({0.8*cos(\x)}, {-1+0.8*sin(\x)});
\end{tikzpicture}+
t^2r^2\begin{tikzpicture}[scale=.4,baseline={([yshift=-.5ex]current bounding box.center)}]
\draw [black, line width=0.2pt] (0,-1)  -- (1,0); 
\draw [black, line width=0.2pt] (0,1) -- (1,0); 
\draw [black, line width=0.2pt] (-1,0) -- (0,-1); 
\draw [black, line width=0.2pt] (-1,0) -- (0,1);  
\draw[red, line width=0.3mm, domain=-45:45] plot ({-1+0.6*cos(\x)}, {0.6*sin(\x)});
\draw[red, line width=0.3mm, domain=-135:-45] plot ({0.6*cos(\x)}, {1+0.6*sin(\x)});
\end{tikzpicture}\otimes
\begin{tikzpicture}[scale=.4,baseline={([yshift=-.5ex]current bounding box.center)}]
\draw [black, line width=0.2pt] (0,-1)  -- (1,0); 
\draw [black, line width=0.2pt] (0,1) -- (1,0); 
\draw [black, line width=0.2pt] (-1,0) -- (0,-1); 
\draw [black, line width=0.2pt] (-1,0) -- (0,1);  
\draw[blue, line width=0.3mm, domain=-45:45] plot ({-1+0.6*cos(\x)}, {0.6*sin(\x)});
\draw[blue, line width=0.3mm, domain=-135:-45] plot ({0.6*cos(\x)}, {1+0.6*sin(\x)});
\end{tikzpicture}
+t^2r^2
\begin{tikzpicture}[scale=.4,baseline={([yshift=-.5ex]current bounding box.center)}]
\draw [black, line width=0.2pt] (0,-1)  -- (1,0); 
\draw [black, line width=0.2pt] (0,1) -- (1,0); 
\draw [black, line width=0.2pt] (-1,0) -- (0,-1); 
\draw [black, line width=0.2pt] (-1,0) -- (0,1);  
\draw[red, line width=0.3mm, domain=-135:-45] plot ({0.6*cos(\x)}, {1+0.6*sin(\x)});
\draw[red, line width=0.3mm, domain=135:225] plot ({1+0.6*cos(\x)}, {0.6*sin(\x)});
\end{tikzpicture}\otimes
\begin{tikzpicture}[scale=.4,baseline={([yshift=-.5ex]current bounding box.center)}]
\draw [black, line width=0.2pt] (0,-1)  -- (1,0); 
\draw [black, line width=0.2pt] (0,1) -- (1,0); 
\draw [black, line width=0.2pt] (-1,0) -- (0,-1); 
\draw [black, line width=0.2pt] (-1,0) -- (0,1);  
\draw[blue, line width=0.3mm, domain=-135:-45] plot ({0.6*cos(\x)}, {1+0.6*sin(\x)});
\draw[blue, line width=0.3mm, domain=135:225] plot ({1+0.6*cos(\x)}, {0.6*sin(\x)});
\end{tikzpicture}\nonumber\\
&&
+t^2r^2
\begin{tikzpicture}[scale=.4,baseline={([yshift=-.5ex]current bounding box.center)}]
\draw [black, line width=0.2pt] (0,-1)  -- (1,0); 
\draw [black, line width=0.2pt] (0,1) -- (1,0); 
\draw [black, line width=0.2pt] (-1,0) -- (0,-1); 
\draw [black, line width=0.2pt] (-1,0) -- (0,1);  
\draw[red, line width=0.3mm, domain=135:225] plot ({1+0.6*cos(\x)}, {0.6*sin(\x)});
\draw[red, line width=0.3mm, domain=45:135] plot ({0.6*cos(\x)}, {-1+0.6*sin(\x)});
\end{tikzpicture}\otimes
\begin{tikzpicture}[scale=.4,baseline={([yshift=-.5ex]current bounding box.center)}]
\draw [black, line width=0.2pt] (0,-1)  -- (1,0); 
\draw [black, line width=0.2pt] (0,1) -- (1,0); 
\draw [black, line width=0.2pt] (-1,0) -- (0,-1); 
\draw [black, line width=0.2pt] (-1,0) -- (0,1);  
\draw[blue, line width=0.3mm, domain=135:225] plot ({1+0.6*cos(\x)}, {0.6*sin(\x)});
\draw[blue, line width=0.3mm, domain=45:135] plot ({0.6*cos(\x)}, {-1+0.6*sin(\x)});
\end{tikzpicture}
+t^2r^2
\begin{tikzpicture}[scale=.4,baseline={([yshift=-.5ex]current bounding box.center)}]
\draw [black, line width=0.2pt] (0,-1)  -- (1,0); 
\draw [black, line width=0.2pt] (0,1) -- (1,0); 
\draw [black, line width=0.2pt] (-1,0) -- (0,-1); 
\draw [black, line width=0.2pt] (-1,0) -- (0,1);  
\draw[red, line width=0.3mm, domain=45:135] plot ({0.6*cos(\x)}, {-1+0.6*sin(\x)});
\draw[red, line width=0.3mm, domain=-45:45] plot ({-1+0.6*cos(\x)}, {0.6*sin(\x)});
\end{tikzpicture}\otimes
\begin{tikzpicture}[scale=.4,baseline={([yshift=-.5ex]current bounding box.center)}]
\draw [black, line width=0.2pt] (0,-1)  -- (1,0); 
\draw [black, line width=0.2pt] (0,1) -- (1,0); 
\draw [black, line width=0.2pt] (-1,0) -- (0,-1); 
\draw [black, line width=0.2pt] (-1,0) -- (0,1);  
\draw[blue, line width=0.3mm, domain=45:135] plot ({0.6*cos(\x)}, {-1+0.6*sin(\x)});
\draw[blue, line width=0.3mm, domain=-45:45] plot ({-1+0.6*cos(\x)}, {0.6*sin(\x)});
\end{tikzpicture}+\left(t^2
\begin{tikzpicture}[scale=.4,baseline={([yshift=-.5ex]current bounding box.center)}]
\draw [black, line width=0.2pt] (0,-1)  -- (1,0); 
\draw [black, line width=0.2pt] (0,1) -- (1,0); 
\draw [black, line width=0.2pt] (-1,0) -- (0,-1); 
\draw [black, line width=0.2pt] (-1,0) -- (0,1);  
\draw[red, line width=0.3mm, domain=-45:45] plot ({-1+0.6*cos(\x)}, {0.6*sin(\x)});
\draw[red, line width=0.3mm, domain=135:225] plot ({1+0.6*cos(\x)}, {0.6*sin(\x)});
\end{tikzpicture}-r^2\begin{tikzpicture}[scale=.4,baseline={([yshift=-.5ex]current bounding box.center)}]
\draw [black, line width=0.2pt] (0,-1)  -- (1,0); 
\draw [black, line width=0.2pt] (0,1) -- (1,0); 
\draw [black, line width=0.2pt] (-1,0) -- (0,-1); 
\draw [black, line width=0.2pt] (-1,0) -- (0,1);  
\draw[red, line width=0.3mm, domain=-135:-45] plot ({0.6*cos(\x)}, {1+0.6*sin(\x)});
\draw[red, line width=0.3mm, domain=45:135] plot ({0.6*cos(\x)}, {-1+0.6*sin(\x)});
\end{tikzpicture}\right)\otimes
\left(t^2
\begin{tikzpicture}[scale=.4,baseline={([yshift=-.5ex]current bounding box.center)}]
\draw [black, line width=0.2pt] (0,-1)  -- (1,0); 
\draw [black, line width=0.2pt] (0,1) -- (1,0); 
\draw [black, line width=0.2pt] (-1,0) -- (0,-1); 
\draw [black, line width=0.2pt] (-1,0) -- (0,1);  
\draw[blue, line width=0.3mm, domain=-45:45] plot ({-1+0.6*cos(\x)}, {0.6*sin(\x)});
\draw[blue, line width=0.3mm, domain=135:225] plot ({1+0.6*cos(\x)}, {0.6*sin(\x)});
\end{tikzpicture}-r^2\begin{tikzpicture}[scale=.4,baseline={([yshift=-.5ex]current bounding box.center)}]
\draw [black, line width=0.2pt] (0,-1)  -- (1,0); 
\draw [black, line width=0.2pt] (0,1) -- (1,0); 
\draw [black, line width=0.2pt] (-1,0) -- (0,-1); 
\draw [black, line width=0.2pt] (-1,0) -- (0,1);  
\draw[blue, line width=0.3mm, domain=-135:-45] plot ({0.6*cos(\x)}, {1+0.6*sin(\x)});
\draw[blue, line width=0.3mm, domain=45:135] plot ({0.6*cos(\x)}, {-1+0.6*sin(\x)});
\end{tikzpicture}\right)\nonumber\\
&&
+\left(\frac{t^3}{2}\begin{tikzpicture}[scale=.4,baseline={([yshift=-.5ex]current bounding box.center)}]
\draw [black, line width=0.2pt] (0,-1)  -- (1,0); 
\draw [black, line width=0.2pt] (0,1) -- (1,0); 
\draw [black, line width=0.2pt] (-1,0) -- (0,-1); 
\draw [black, line width=0.2pt] (-1,0) -- (0,1);  
\draw[red, line width=0.3mm, domain=-45:45] plot ({-1+0.6*cos(\x)}, {0.6*sin(\x)});
\draw[red, line width=0.3mm, domain=-45:45] plot ({-1+0.8*cos(\x)}, {0.8*sin(\x)});
\draw[red, line width=0.3mm, domain=135:225] plot ({1+0.7*cos(\x)}, {0.7*sin(\x)});
\end{tikzpicture}-tr^2\begin{tikzpicture}[scale=.4,baseline={([yshift=-.5ex]current bounding box.center)}]
\draw [black, line width=0.2pt] (0,-1)  -- (1,0); 
\draw [black, line width=0.2pt] (0,1) -- (1,0); 
\draw [black, line width=0.2pt] (-1,0) -- (0,-1); 
\draw [black, line width=0.2pt] (-1,0) -- (0,1);  
\draw[red, line width=0.3mm, domain=45:135] plot ({0.6*cos(\x)}, {-1+0.6*sin(\x)});
\draw[red, line width=0.3mm, domain=-45:45] plot ({-1+0.6*cos(\x)}, {0.6*sin(\x)});
\draw[red, line width=0.3mm, domain=-135:-45] plot ({0.6*cos(\x)}, {1+0.6*sin(\x)});
\end{tikzpicture}\right)\otimes
\left(\frac{t^3}{2}
\begin{tikzpicture}[scale=.4,baseline={([yshift=-.5ex]current bounding box.center)}]
\draw [black, line width=0.2pt] (0,-1)  -- (1,0); 
\draw [black, line width=0.2pt] (0,1) -- (1,0); 
\draw [black, line width=0.2pt] (-1,0) -- (0,-1); 
\draw [black, line width=0.2pt] (-1,0) -- (0,1);  
\draw[blue, line width=0.3mm, domain=-45:45] plot ({-1+0.6*cos(\x)}, {0.6*sin(\x)});
\draw[blue, line width=0.3mm, domain=-45:45] plot ({-1+0.8*cos(\x)}, {0.8*sin(\x)});
\draw[blue, line width=0.3mm, domain=135:225] plot ({1+0.7*cos(\x)}, {0.7*sin(\x)});
\end{tikzpicture}-tr^2
\begin{tikzpicture}[scale=.4,baseline={([yshift=-.5ex]current bounding box.center)}]
\draw [black, line width=0.2pt] (0,-1)  -- (1,0); 
\draw [black, line width=0.2pt] (0,1) -- (1,0); 
\draw [black, line width=0.2pt] (-1,0) -- (0,-1); 
\draw [black, line width=0.2pt] (-1,0) -- (0,1);  
\draw[blue, line width=0.3mm, domain=45:135] plot ({0.6*cos(\x)}, {-1+0.6*sin(\x)});
\draw[blue, line width=0.3mm, domain=-45:45] plot ({-1+0.6*cos(\x)}, {0.6*sin(\x)});
\draw[blue, line width=0.3mm, domain=-135:-45] plot ({0.6*cos(\x)}, {1+0.6*sin(\x)});
\end{tikzpicture}\right)+\left(-\frac{r^3}{2}
\begin{tikzpicture}[scale=.4,baseline={([yshift=-.5ex]current bounding box.center)}]
\draw [black, line width=0.2pt] (0,-1)  -- (1,0); 
\draw [black, line width=0.2pt] (0,1) -- (1,0); 
\draw [black, line width=0.2pt] (-1,0) -- (0,-1); 
\draw [black, line width=0.2pt] (-1,0) -- (0,1);  
\draw[red, line width=0.3mm, domain=-135:-45] plot ({0.6*cos(\x)}, {1+0.6*sin(\x)});
\draw[red, line width=0.3mm, domain=-135:-45] plot ({0.8*cos(\x)}, {1+0.8*sin(\x)});
\draw[red, line width=0.3mm, domain=45:135] plot ({0.7*cos(\x)}, {-1+0.7*sin(\x)});
\end{tikzpicture}+t^2r
\begin{tikzpicture}[scale=.4,baseline={([yshift=-.5ex]current bounding box.center)}]
\draw [black, line width=0.2pt] (0,-1)  -- (1,0); 
\draw [black, line width=0.2pt] (0,1) -- (1,0); 
\draw [black, line width=0.2pt] (-1,0) -- (0,-1); 
\draw [black, line width=0.2pt] (-1,0) -- (0,1);  
\draw[red, line width=0.3mm, domain=-45:45] plot ({-1+0.6*cos(\x)}, {0.6*sin(\x)});
\draw[red, line width=0.3mm, domain=-135:-45] plot ({0.6*cos(\x)}, {1+0.6*sin(\x)});
\draw[red, line width=0.3mm, domain=135:225] plot ({1+0.6*cos(\x)}, {0.6*sin(\x)});
\end{tikzpicture}\right)\otimes
\left(-\frac{r^3}{2}
\begin{tikzpicture}[scale=.4,baseline={([yshift=-.5ex]current bounding box.center)}]
\draw [black, line width=0.2pt] (0,-1)  -- (1,0); 
\draw [black, line width=0.2pt] (0,1) -- (1,0); 
\draw [black, line width=0.2pt] (-1,0) -- (0,-1); 
\draw [black, line width=0.2pt] (-1,0) -- (0,1);  
\draw[blue, line width=0.3mm, domain=-135:-45] plot ({0.6*cos(\x)}, {1+0.6*sin(\x)});
\draw[blue, line width=0.3mm, domain=-135:-45] plot ({0.8*cos(\x)}, {1+0.8*sin(\x)});
\draw[blue, line width=0.3mm, domain=45:135] plot ({0.7*cos(\x)}, {-1+0.7*sin(\x)});
\end{tikzpicture}+t^2r
\begin{tikzpicture}[scale=.4,baseline={([yshift=-.5ex]current bounding box.center)}]
\draw [black, line width=0.2pt] (0,-1)  -- (1,0); 
\draw [black, line width=0.2pt] (0,1) -- (1,0); 
\draw [black, line width=0.2pt] (-1,0) -- (0,-1); 
\draw [black, line width=0.2pt] (-1,0) -- (0,1);  
\draw[blue, line width=0.3mm, domain=-45:45] plot ({-1+0.6*cos(\x)}, {0.6*sin(\x)});
\draw[blue, line width=0.3mm, domain=-135:-45] plot ({0.6*cos(\x)}, {1+0.6*sin(\x)});
\draw[blue, line width=0.3mm, domain=135:225] plot ({1+0.6*cos(\x)}, {0.6*sin(\x)});
\end{tikzpicture}\right)\nonumber\\
&&+
\left(\frac{t^3}{2}\begin{tikzpicture}[scale=.4,baseline={([yshift=-.5ex]current bounding box.center)}]
\draw [black, line width=0.2pt] (0,-1)  -- (1,0); 
\draw [black, line width=0.2pt] (0,1) -- (1,0); 
\draw [black, line width=0.2pt] (-1,0) -- (0,-1); 
\draw [black, line width=0.2pt] (-1,0) -- (0,1);  
\draw[red, line width=0.3mm, domain=135:225] plot ({1+0.6*cos(\x)}, {0.6*sin(\x)});
\draw[red, line width=0.3mm, domain=135:225] plot ({1+0.8*cos(\x)}, {0.8*sin(\x)});
\draw[red, line width=0.3mm, domain=-45:45] plot ({-1+0.7*cos(\x)}, {0.7*sin(\x)});
\end{tikzpicture}-tr^2
\begin{tikzpicture}[scale=.4,baseline={([yshift=-.5ex]current bounding box.center)}]
\draw [black, line width=0.2pt] (0,-1)  -- (1,0); 
\draw [black, line width=0.2pt] (0,1) -- (1,0); 
\draw [black, line width=0.2pt] (-1,0) -- (0,-1); 
\draw [black, line width=0.2pt] (-1,0) -- (0,1);  
\draw[red, line width=0.3mm, domain=-135:-45] plot ({0.6*cos(\x)}, {1+0.6*sin(\x)});
\draw[red, line width=0.3mm, domain=135:225] plot ({1+0.6*cos(\x)}, {0.6*sin(\x)});
\draw[red, line width=0.3mm, domain=45:135] plot ({0.6*cos(\x)}, {-1+0.6*sin(\x)});
\end{tikzpicture}\right)\otimes
\left(\frac{t^3}{2}\begin{tikzpicture}[scale=.4,baseline={([yshift=-.5ex]current bounding box.center)}]
\draw [black, line width=0.2pt] (0,-1)  -- (1,0); 
\draw [black, line width=0.2pt] (0,1) -- (1,0); 
\draw [black, line width=0.2pt] (-1,0) -- (0,-1); 
\draw [black, line width=0.2pt] (-1,0) -- (0,1);  
\draw[blue, line width=0.3mm, domain=135:225] plot ({1+0.6*cos(\x)}, {0.6*sin(\x)});
\draw[blue, line width=0.3mm, domain=135:225] plot ({1+0.8*cos(\x)}, {0.8*sin(\x)});
\draw[blue, line width=0.3mm, domain=-45:45] plot ({-1+0.7*cos(\x)}, {0.7*sin(\x)});
\end{tikzpicture}-tr^2
\begin{tikzpicture}[scale=.4,baseline={([yshift=-.5ex]current bounding box.center)}]
\draw [black, line width=0.2pt] (0,-1)  -- (1,0); 
\draw [black, line width=0.2pt] (0,1) -- (1,0); 
\draw [black, line width=0.2pt] (-1,0) -- (0,-1); 
\draw [black, line width=0.2pt] (-1,0) -- (0,1);  
\draw[blue, line width=0.3mm, domain=-135:-45] plot ({0.6*cos(\x)}, {1+0.6*sin(\x)});
\draw[blue, line width=0.3mm, domain=135:225] plot ({1+0.6*cos(\x)}, {0.6*sin(\x)});
\draw[blue, line width=0.3mm, domain=45:135] plot ({0.6*cos(\x)}, {-1+0.6*sin(\x)});
\end{tikzpicture}\right)
+\left(\frac{r^3}{2}\begin{tikzpicture}[scale=.4,baseline={([yshift=-.5ex]current bounding box.center)}]
\draw [black, line width=0.2pt] (0,-1)  -- (1,0); 
\draw [black, line width=0.2pt] (0,1) -- (1,0); 
\draw [black, line width=0.2pt] (-1,0) -- (0,-1); 
\draw [black, line width=0.2pt] (-1,0) -- (0,1);  
\draw[red, line width=0.3mm, domain=45:135] plot ({0.6*cos(\x)}, {-1+0.6*sin(\x)});
\draw[red, line width=0.3mm, domain=45:135] plot ({0.8*cos(\x)}, {-1+0.8*sin(\x)});
\draw[red, line width=0.3mm, domain=-135:-45] plot ({0.7*cos(\x)}, {1+0.7*sin(\x)});
\end{tikzpicture}-t^2r\begin{tikzpicture}[scale=.4,baseline={([yshift=-.5ex]current bounding box.center)}]
\draw [black, line width=0.2pt] (0,-1)  -- (1,0); 
\draw [black, line width=0.2pt] (0,1) -- (1,0); 
\draw [black, line width=0.2pt] (-1,0) -- (0,-1); 
\draw [black, line width=0.2pt] (-1,0) -- (0,1);  
\draw[red, line width=0.3mm, domain=135:225] plot ({1+0.6*cos(\x)}, {0.6*sin(\x)});
\draw[red, line width=0.3mm, domain=45:135] plot ({0.6*cos(\x)}, {-1+0.6*sin(\x)});
\draw[red, line width=0.3mm, domain=-45:45] plot ({-1+0.6*cos(\x)}, {0.6*sin(\x)});
\end{tikzpicture}\right)\otimes
\left(\frac{r^3}{2}\begin{tikzpicture}[scale=.4,baseline={([yshift=-.5ex]current bounding box.center)}]
\draw [black, line width=0.2pt] (0,-1)  -- (1,0); 
\draw [black, line width=0.2pt] (0,1) -- (1,0); 
\draw [black, line width=0.2pt] (-1,0) -- (0,-1); 
\draw [black, line width=0.2pt] (-1,0) -- (0,1);  
\draw[blue, line width=0.3mm, domain=45:135] plot ({0.6*cos(\x)}, {-1+0.6*sin(\x)});
\draw[blue, line width=0.3mm, domain=45:135] plot ({0.8*cos(\x)}, {-1+0.8*sin(\x)});
\draw[blue, line width=0.3mm, domain=-135:-45] plot ({0.7*cos(\x)}, {1+0.7*sin(\x)});
\end{tikzpicture}-t^2r\begin{tikzpicture}[scale=.4,baseline={([yshift=-.5ex]current bounding box.center)}]
\draw [black, line width=0.2pt] (0,-1)  -- (1,0); 
\draw [black, line width=0.2pt] (0,1) -- (1,0); 
\draw [black, line width=0.2pt] (-1,0) -- (0,-1); 
\draw [black, line width=0.2pt] (-1,0) -- (0,1);  
\draw[blue, line width=0.3mm, domain=135:225] plot ({1+0.6*cos(\x)}, {0.6*sin(\x)});
\draw[blue, line width=0.3mm, domain=45:135] plot ({0.6*cos(\x)}, {-1+0.6*sin(\x)});
\draw[blue, line width=0.3mm, domain=-45:45] plot ({-1+0.6*cos(\x)}, {0.6*sin(\x)});
\end{tikzpicture}\right)\nonumber\\
&&+\left(\frac{t^4}{4}
\begin{tikzpicture}[scale=.4,baseline={([yshift=-.5ex]current bounding box.center)}]
\draw [black, line width=0.2pt] (0,-1)  -- (1,0); 
\draw [black, line width=0.2pt] (0,1) -- (1,0); 
\draw [black, line width=0.2pt] (-1,0) -- (0,-1); 
\draw [black, line width=0.2pt] (-1,0) -- (0,1);  
\draw[red, line width=0.3mm, domain=-45:45] plot ({-1+0.6*cos(\x)}, {0.6*sin(\x)});
\draw[red, line width=0.3mm, domain=-45:45] plot ({-1+0.8*cos(\x)}, {0.8*sin(\x)});
\draw[red, line width=0.3mm, domain=135:225] plot ({1+0.6*cos(\x)}, {0.6*sin(\x)});
\draw[red, line width=0.3mm, domain=135:225] plot ({1+0.8*cos(\x)}, {0.8*sin(\x)});
\end{tikzpicture}+\frac{r^4}{4}
\begin{tikzpicture}[scale=.4,baseline={([yshift=-.5ex]current bounding box.center)}]
\draw [black, line width=0.2pt] (0,-1)  -- (1,0); 
\draw [black, line width=0.2pt] (0,1) -- (1,0); 
\draw [black, line width=0.2pt] (-1,0) -- (0,-1); 
\draw [black, line width=0.2pt] (-1,0) -- (0,1);  
\draw[red, line width=0.3mm, domain=-135:-45] plot ({0.6*cos(\x)}, {1+0.6*sin(\x)});
\draw[red, line width=0.3mm, domain=-135:-45] plot ({0.8*cos(\x)}, {1+0.8*sin(\x)});
\draw[red, line width=0.3mm, domain=45:135] plot ({0.6*cos(\x)}, {-1+0.6*sin(\x)});
\draw[red, line width=0.3mm, domain=45:135] plot ({0.8*cos(\x)}, {-1+0.8*sin(\x)});
\end{tikzpicture}
-r^2t^2
\begin{tikzpicture}[scale=.4,baseline={([yshift=-.5ex]current bounding box.center)}]
\draw [black, line width=0.2pt] (0,-1)  -- (1,0); 
\draw [black, line width=0.2pt] (0,1) -- (1,0); 
\draw [black, line width=0.2pt] (-1,0) -- (0,-1); 
\draw [black, line width=0.2pt] (-1,0) -- (0,1);  
\draw[red, line width=0.3mm, domain=135:225] plot ({1+0.6*cos(\x)}, {0.6*sin(\x)});
\draw[red, line width=0.3mm, domain=45:135] plot ({0.6*cos(\x)}, {-1+0.6*sin(\x)});
\draw[red, line width=0.3mm, domain=-45:45] plot ({-1+0.6*cos(\x)}, {0.6*sin(\x)});
\draw[red, line width=0.3mm, domain=-135:-45] plot ({0.6*cos(\x)}, {1+0.6*sin(\x)});
\end{tikzpicture}\right)\otimes
\left(\frac{t^4}{4}
\begin{tikzpicture}[scale=.4,baseline={([yshift=-.5ex]current bounding box.center)}]
\draw [black, line width=0.2pt] (0,-1)  -- (1,0); 
\draw [black, line width=0.2pt] (0,1) -- (1,0); 
\draw [black, line width=0.2pt] (-1,0) -- (0,-1); 
\draw [black, line width=0.2pt] (-1,0) -- (0,1);  
\draw[blue, line width=0.3mm, domain=-45:45] plot ({-1+0.6*cos(\x)}, {0.6*sin(\x)});
\draw[blue, line width=0.3mm, domain=-45:45] plot ({-1+0.8*cos(\x)}, {0.8*sin(\x)});
\draw[blue, line width=0.3mm, domain=135:225] plot ({1+0.6*cos(\x)}, {0.6*sin(\x)});
\draw[blue, line width=0.3mm, domain=135:225] plot ({1+0.8*cos(\x)}, {0.8*sin(\x)});
\end{tikzpicture}+\frac{r^4}{4}
\begin{tikzpicture}[scale=.4,baseline={([yshift=-.5ex]current bounding box.center)}]
\draw [black, line width=0.2pt] (0,-1)  -- (1,0); 
\draw [black, line width=0.2pt] (0,1) -- (1,0); 
\draw [black, line width=0.2pt] (-1,0) -- (0,-1); 
\draw [black, line width=0.2pt] (-1,0) -- (0,1);  
\draw[blue, line width=0.3mm, domain=-135:-45] plot ({0.6*cos(\x)}, {1+0.6*sin(\x)});
\draw[blue, line width=0.3mm, domain=-135:-45] plot ({0.8*cos(\x)}, {1+0.8*sin(\x)});
\draw[blue, line width=0.3mm, domain=45:135] plot ({0.6*cos(\x)}, {-1+0.6*sin(\x)});
\draw[blue, line width=0.3mm, domain=45:135] plot ({0.8*cos(\x)}, {-1+0.8*sin(\x)});
\end{tikzpicture}
-r^2t^2
\begin{tikzpicture}[scale=.4,baseline={([yshift=-.5ex]current bounding box.center)}]
\draw [black, line width=0.2pt] (0,-1)  -- (1,0); 
\draw [black, line width=0.2pt] (0,1) -- (1,0); 
\draw [black, line width=0.2pt] (-1,0) -- (0,-1); 
\draw [black, line width=0.2pt] (-1,0) -- (0,1);  
\draw[blue, line width=0.3mm, domain=135:225] plot ({1+0.6*cos(\x)}, {0.6*sin(\x)});
\draw[blue, line width=0.3mm, domain=45:135] plot ({0.6*cos(\x)}, {-1+0.6*sin(\x)});
\draw[blue, line width=0.3mm, domain=-45:45] plot ({-1+0.6*cos(\x)}, {0.6*sin(\x)});
\draw[blue, line width=0.3mm, domain=-135:-45] plot ({0.6*cos(\x)}, {1+0.6*sin(\x)});
\end{tikzpicture}\right) \,,
\eea
where we have kept separated the two colors for clarity. Note that we did not include the extra parameter $z$ corresponding to the loop fugacity. Is it however direct to generalize the $R$-matrix for $z\neq1$ by considering the transformation $r\rightarrow zr,\quad t\rightarrow zt$. 

\subsection{Generalization}

The generalization of the above procedure is now clear. The $M$'th truncation is obtained by expanding $e^{S_v}$ using the functions $\Gamma$ defined earlier to the order $2M$ in power of elements of the scattering matrix $\mathcal{S}$. Directly in the loop model, the weight of a tile can be computed in the following way. Denote by $k_{ij}$ the number of strands joining the edge $i$ and $j$ of the following vertex (here the arrows are irrelevant)
\bea
\begin{tikzpicture}[baseline=8pt]
\draw[thick] (-1,-1) -- (1,1);
\draw[thick] (-1,1) -- (1,-1);
\draw (-1.2,-1.2) node {$1$};
\draw (1.2,1.2) node {$2$};
\draw (-1.2,1.2) node {$3$};
\draw (1.2,-1.2) node {$4$};
\arrowSW{0.5}{0.5}
\arrowNW{-0.5-0.1}{0.5+0.1}
\arrowNE{-0.5}{-0.5}
\arrowSE{0.5+0.1}{-0.5-0.1}
\end{tikzpicture}
\label{vertex}
\eea

The weight $w$ of a loop tile (with one color) $\mathcal{T}$ is given by
\bea
w(\mathcal{T})=\frac{(-1)^{k_{14}}}{k_{13}!k_{14}!k_{23}!k_{24}!}t^{k_{13}+k_{24}}r^{k_{14}+k_{23}}.
\eea

Moreover, gluing two edges together with $n$ strands correspond to summing over the $n!$ permutations. The $R$-matrix of the model is obtained in the end by taking the tensor product of tiles from each color that have the same number of strands at each edge. 

\subsection{Preliminary numerical  results}

The full Chalker-Coddington model is critical if $z$ is equal to the special value $z_C=1$. The modification of the critical parameter to $z_C\sim1.032$ in the first truncation ($M=1$) changes the symmetries to $gl(1|1)\otimes gl(1|1)$ instead of the full $gl(2|2)$ algebra. This observation remains true in the higher truncations. For instance in the second truncation ($M=2$), the critical value of $z$ is approximatively $z_C\sim1.014$ which again breaks the symmetries between the vacuum and the bosons/fermions. We expect that for the full series of truncated models, the critical value of the parameter $z$ stays strictly greater than $1$. As a consequence, the symmetry of the Chalker-Coddington model is only recovered in the limit of $M\rightarrow\infty$.

\begin{figure}
\centering
    \includegraphics[scale=.5]{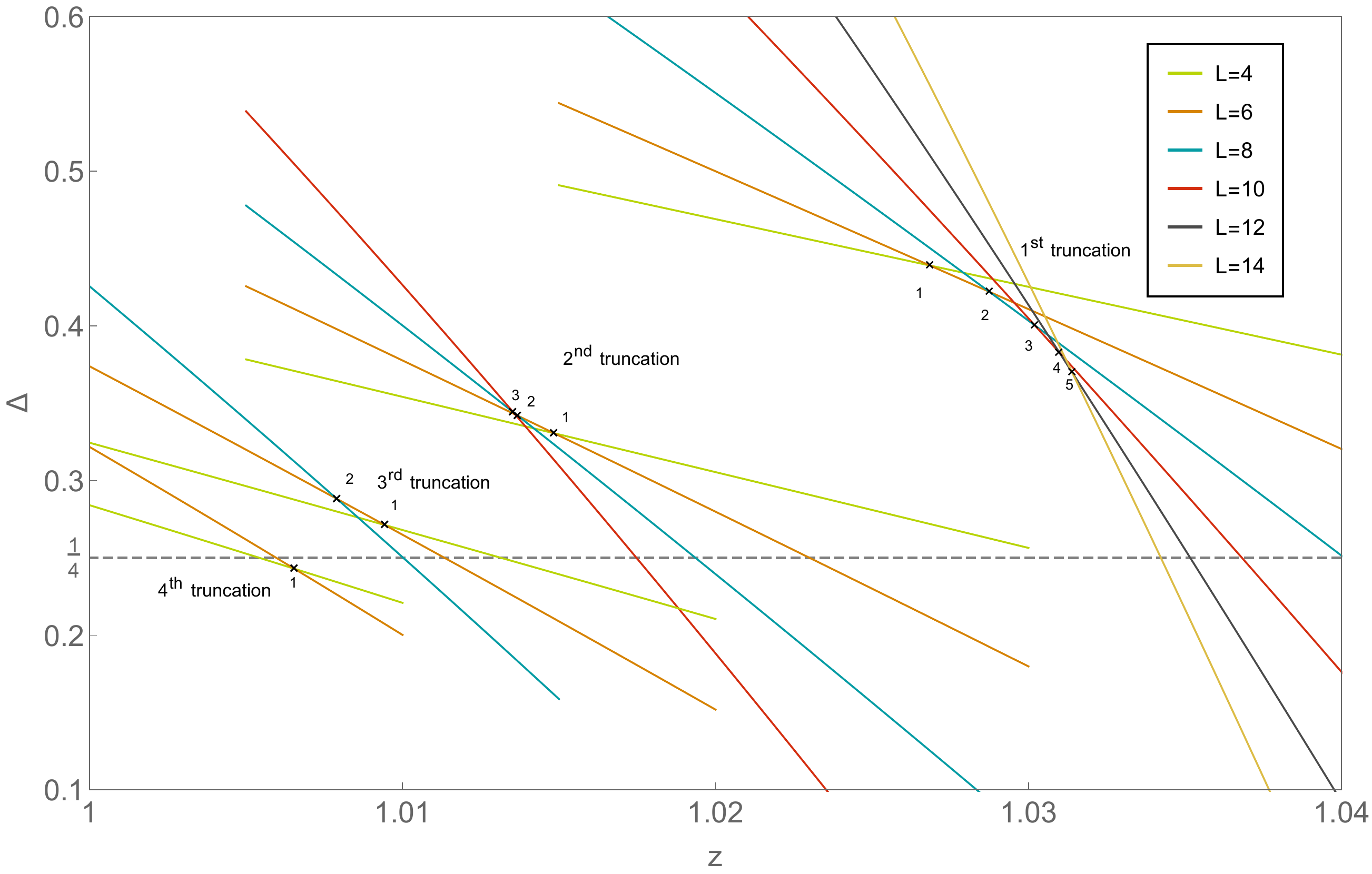}
    \caption{Estimate of  dimensions of $x_{2,2}$ in the first four truncations with periodic boundary conditions. The dotted line represent the scaling dimension $x_{2,2}=1/4$ in the first truncation.}\label{higherexp22}
\end{figure}

We can then investigate the value of the  dimension corresponding to the first excited state in the sector of the vacuum. As observed by Ikhlef et al.\ \cite{Ikhlef}, the energy of the first excited state corresponds to the one of the sector propagating two strands of each color. This property remains true for higher truncations. We can track the evolution of its associated critical dimension; see Figure~\ref{higherexp22}. While it is hard to carry out the numerics for higher truncations and large size, it looks like the  dimension decreases as the level of truncation increases.

\subsection{Exponent $\nu$}

An important quantity in our problem is the critical exponent $\nu$, which is believed to be around $2.6$ in the Chalker-Coddington model. On the lattice it is measured with a perturbation corresponding to a staggering of the weights that distinguishes between right/left turns at the vertices. If this perturbation is of order $\lambda$, the exponent $\nu$ is defined via the divergence of the correlation length,  $\xi\propto |\lambda|^{-\nu}$. 

In general, the exponent $\nu$ is related to the  dimension $x$ of the perturbing operator by
\bea 
x=2-\frac{1}{\nu}.
\eea

In the case of the first truncation, Ikhlef et al.\ concluded that the scaling relation between $\nu$ and the  dimension $x$ must be modified \cite[Section (5.2) and Appendix C]{Ikhlef} into
\bea
x_{2,2}=2-\frac{2}{\nu}
\eea
We know from our results that  $x=x_{2,2}=1/4$, leading to $\nu=8/7$. This  is close to the numerical value in the non-integrable model for which \cite[eq.\ (5.4)]{Ikhlef} report $\nu \simeq 1.1$. Their argument in favor of this modified scaling relation comes from an analytical result on the integrable model we discussed in section \ref{sec:b21} for the special value $\theta={5\pi\over 6}$.   At this point, they show that the continuum description of the lattice model with staggering lattice amplitude proportional to $\lambda$ is a massive Majorana fermion theory with mass  term  proportional to $\lambda^2$. The perturbation to the Hamiltonian is hence of order $2$ in the lattice perturbation parameter $\lambda$,
\bea 
H[\lambda]=H^*+\lambda^2\int\diff x\ \mathcal{O} \,,
\eea
where $H^*$ is the scale invariant Hamiltonian and $\mathcal{O}$ is the operator corresponding to the perturbation. The fact that the Hamiltonian is coupled to $\lambda^2$ (and not $\lambda$) changes the scaling relation:  if the modified   relation stays the same in the whole regime,  and in particular  at the point corresponding to the same universality class as the unmodified first truncation then indeed $\nu=8/7$, in agreement with numerical simulation. We first want to investigate numerically the unmodified truncated model to check directly if indeed this scenario holds. 

Generically, the partition function can be written as a path integral in the continuum limit as follows
\bea 
Z[\lambda]=\int\diff\phi \, e^{S[\phi]+\lambda\int\diff^2 x\ \mathcal{O}_\lambda+\lambda^2\int\diff^2 x\mathcal{O}_{\lambda^2}+\ldots} \,,
\eea
with $\mathcal{O}_{\lambda^n}$ the operator perturbing the system at order $n$ in $\lambda$. Using the transfer matrix we can study the perturbation on a finite lattice. We first expand the transfer matrix as a series in the order of the perturbation $\lambda$ around $\lambda=0$:
\bea
T[\lambda]=T[0]+\lambda\ \partial_\lambda T\bigr|_{\lambda=0}+\frac{\lambda^2}{2}\ \partial_\lambda^2 T\bigr|_{\lambda=0}+\ldots.
\eea
In the following we use the following notation $T^{(1)}\equiv \partial_\lambda T\bigr|_{\lambda=0}$ and $T^{(2)}\equiv \partial_\lambda^2 T\bigr|_{\lambda=0}$: for instance the operator $T^{(1)}$ contains some information about the nature of the operator $\mathcal{O}_\lambda$ if we study its action on the groundstate $\ket{\text{gs}}$: $\ket{\lambda}\equiv T^{(1)}\ket{\text{gs}}$. Indeed the state thus obtained should couple to the eigenstates of the transfer matrix corresponding to the operator $\mathcal{O}_\lambda$ in the continuum limit.

\subsubsection{The first truncation}
We start with a discussion of the first truncation in the unmodified model. Numerically we observe, for generic value of the fugacity, that the dominant eigenstate in the decomposition of $\ket{\lambda}$ (over the eigenstates of the transfer matrix) is $\ket{(2,2)}$ corresponding to the operator $(2,2)$ with scaling dimension $x_{2,2}=1/4$. However, in finite size, there exists a particular value of the fugacity near the critical point, such that $\ket{(2,2)}$ does not contribute to $\ket{\lambda}$. The values of this particular value of the fugacity $z$ are reported in the table below for small sizes and seem to converge toward the critical value $z_C$.
\bea
\begin{tabular}{|l|c|c|r|}
  \hline
  $L=4$ & $L=6$ & $L=8$ & $L=\infty$\\
  \hline
  1.04216 & 1.03811 & 1.03626 & 1.031...\\
  \hline
\end{tabular}
\eea
Moreover the eigenstate $\ket{(2,2)}$ always has a non-vanishing (in the vicinity of $z_C$) contribution to $\ket{\lambda^2}=T^{(2)}\ket{{\rm gs}}$ corresponding to the operator $\mathcal{O}_{\lambda^2}$ in the action. Therefore at criticality, the action has the following form:
\bea
S[\lambda]=S[0]+\lambda^2\int\diff^2 x\mathcal{O}_{x_{2,2}=1/4}
\eea
thus providing indeed the scaling relation
\bea
x_{2,2}=2-\frac{2}{\nu}.
\label{scalingfirsttruncation}
\eea
Note that it is possible for other operators to appear in the action at first order in $\lambda$ but they have to be coupled to a less relevant field (for instance the $(4,4)$ operator). We assume here that the scaling relation is determined by $x_{2,2}$.

\subsubsection{The second truncation}

To measure the value of the critical exponent, we assume  as usual (and as done in \cite{Ikhlef}) that the correlation length $\xi_L$ of a system of length $L$ with a perturbation $\lambda$ is determined by a universal function

\bea
\frac{1}{\xi_L}=\frac{1}{L}F(\lambda L^{1/\nu})
\eea
With the second trucation, a good collapse for the correlation lengths at different sizes is obtained for 
\bea\frac{1}{\nu}\sim0.6-0.7\eea
which provides $\nu\sim1.5$. This value is still very different from the one corresponding to the IQHE transition but nevertheless shows that higher truncations give a better approximation of the full Chalker-Coddington model where $\nu\sim2.6$ than the first truncation ($\nu\sim1.1$). 

It is then natural to ask whether the scaling relation (\ref{scalingfirsttruncation}) holds for higher truncations (and the full Chalker-Coddington model), or whether it might be modified in a way that depends on the order of the truncation. 
Like for  the first truncation, we find numerically that  the state $\ket{(2,2)}$ does not appear in the decomposition of $\ket{\lambda}$ near the critical point. As a consequence, we expect again that the usual scaling relation does not hold. However,  \eqref{scalingfirsttruncation} would give, based on our measures of $\nu$, a value  $x_{2,2}\sim0.6$, which is  very far from our direct numerical estimates ($x_{2,2}\sim0.2$ roughly): it seems that \eqref{scalingfirsttruncation} does not hold for the second truncation.   In fact, numerical data suggests that  $\ket{(2,2)}$ may not appear in the decomposition of $\ket{\lambda^2}$  either! We may thus expect that the scaling relation is modified to $x_{2,2}=2-p/\nu$ with $p>2$. Another possibility is that the operator involved in the scaling relation is no longer $\mathcal{O}_{x_{2,2}}$ but a less relevant one.

\subsection{About the universality class of higher truncations}

Like in the case $M=1$, all higher truncations are expected to give rise to non-compact CFTs, and to  exhibit very important finite-size effects.  Even though the measured  exponent $\nu$ in the second truncation is slightly higher than in the first one (and thus closer to the one of the Chalker-Coddington model), it could well be that the numerics is biased by the small system sizes available. In fact---as anticipated in the introduction---it is difficult to claim whether different values of $M$ correspond to different universality classes or not.  See the conclusion below for more discussion of this point.

\section{Conclusion}

It is clear that the first truncation of the Chalker-Coddington model presents many interesting features similar to those of the untruncated model, and as such provides a useful playground to understand, in particular, the role of the continuous spectrum and of normalizability issues. At the same time, it is also clear that the quantitative difference between the truncated and untruncated models remains substantial. 

Note that our strategy in most of this paper has been to study the spectrum of the transfer matrix/Hamiltonian for the various truncations, and try to infer from our results features of the corresponding spectrum for the untruncated model, that is, basically, the $gl(2|2)$ alternating spin chain. It is tempting of course to proceed a bit differently, and to wonder  whether  quantities which are natural in our approach---such as the watermelon operators---can be suitably interpreted as ``truncated versions" of physical  (in particular, transport) observables in the scaling limit of the quantum Hall critical point. While such physical observables have been the subject of a lot of attention recently  \cite{BWZ,BWZII}, their analysis  in the truncated models is fraught with ambiguities. 

To illustrate this point, we recall that  in the untruncated model, the calculation of 
$\overline{|\mathcal{G}(e_1,e_2)|^2}$   involves paths  that can  go an infinite number of times over every edge, including $e_1,e_2$. More detailed analysis of the problem has led to further discussion of the nature of the edges, and the realization that a very different role is played by {\sl observation points} (denoted below by $\mathbf{r}$) or  {\sl point contacts} (denoted by  ($\mathbf{c}$).  In terms of paths, they both correspond to a path insertion, but for the former they may be visited an arbitrary number of times, while for the latter they cannot be visited more than once. This in turn has led to the introduction  \cite{BWZ,BWZII} of  quantities  such as  the observables $Z_{\ell,q}$ 
\begin{equation}
Z_{1,q} = \mathbb{E}\left(  |\psi_{\mathbf{c}_1}(\mathbf{r}_2)|^{2q} \right) 
\end{equation}
 together with their ``cousins'' 
\begin{equation}
\tilde{Z}_{1,q} = \mathbb{E}\left(  |\psi_{\mathbf{c}_1}(\mathbf{c}_2)|^{2q} \right) 
\end{equation}
The point is, these observables are expected to have different scaling behaviors. For $q=1$ for instance, one expects a vanishing exponent for $Z_{1,1}$ but an exponent $x={X\over 4}$  for $\tilde{Z}_{1,1}$ (this observable coincides with the point contact conductance). 

Now the distinction between observation points and point contacts  disappears in the first truncation, since  one cannot go more than once over a given edge anyway! In this case,  the exponent for $Z_{1,1}$ and for $\tilde{Z}_{1,1}$ can be identified with the one leg operator exponent $x_{1,1}$. It is not clear whether one should consider this exponent as an approximation of zero or of ${X\over 4}$.  The authors  in \cite{Ikhlef} had found a  numerical estimate  $x_{1,1} \sim 0$ in the truncated integrable model, and thus suggested the first option. However, we have seen that this is not correct and that  in fact $x_{1,1} = -1/8$ in this model. Moreover,  for the unmodified  truncated model, we have also seen  that exponents for odd and even numbers of legs coincide, and thus $x_{1,1}=x_{2,2}={1\over 4}$. This is neither  close to zero nor to ${X\over 4}\approx {1\over 16}$. A more complete discussion is provided in appendix B.

In conclusion, we believe that the next step in this line of approach is to understand more generally how the physics of non-compact spin chains can be approached by the study of truncated versions, and whether the study of higher truncations might be able to give results closer to those of the untruncated model.  The Hall effect---with its complexity of retarded and advanced paths (which lead to the appearance of two different colors of loops)---is certainly not the best set-up to do this. A much simpler question could be, how can we possibly approach the physics of Brownian motion (realised by self-intersecting paths of a single color) by studying truncated versions where, say, the paths can go a large but finite number of times over a given edge? What are the corresponding spin chains?  This will be discussed in a forthcoming paper \cite{CJS}. 

Finally, we note that while this paper was being completed, a new proposal appeared for the CFT of the Chalker-Coddington model, which involves, in particular, a level-four $GL(1|1)$ Wess-Zumino-Witten model \cite{Zirnbauer18}.  More work is needed to see whether our results are more compatible with this proposal than with previous ones. Note that in \cite{Zirnbauer18},  the value for $\nu$ is expected to be $\nu=\infty$,  despite the large but certainly finite values obtained in  simulations on the CC model.

\bigskip

\noindent{\bf Acknowledgments:} We thank R.~Bondesan, P.~Fendley,  I.~Gruzberg, Y.~Ikhlef, T.~Quella, A.~Tsvelik and M.~Zirnbauer for stimulating discussions. This work was supported by the ERC Advanced Grant NuQFT.  EV also acknowledges support by the EPSRC under grant EP/N01930X. JLJ was supported by the Institut Universitaire de France.

\section*{Appendices}

\appendix

\section{The action for the first truncation}

Using the equivalence between the $SL(2,\mathbb{R})/U(1)$ model and the sine-Liouville theory \cite{HikiScho} we can write the action describing the continuum limit of the $b_2^{(1)}$ model  or the $a_3^{(2)}$ model as 
\begin{equation}
A=\int d^2x\left\{ {1\over 16\pi}\left[(\partial_\mu \varphi)^2+(\partial_\mu \phi_+)^2+(\partial_\mu \phi_-)^2\right]
-2\mu
e^{\alpha\varphi}\cos(\delta\phi_+)\right\}\label{modLiouapp} \,,
\end{equation}
with
\begin{equation}
\alpha^2={K-2\over 4} \,, \qquad \delta^2={K\over 4} \,.
\end{equation}
This can be seen as the UV limit of the perturbed CFT with action
\begin{equation}
A=\int d^2x\left\{ {1\over 16\pi}\left[(\partial_\mu \varphi)^2+(\partial_\mu \phi_+)^2+(\partial_\mu \phi_-)^2\right]
-2\mu
\left[e^{\alpha\varphi}\cos(\delta\phi_+)+ e^{-\alpha\varphi}\cos(\delta\phi_-)\right]\right\}\label{modLiou} \,,
\end{equation}
which is itself a particular case of the more general ``SS model'' \cite{Fateev}
\begin{eqnarray}
A=\int d^2x\left\{ {1\over 16\pi}\left[(\partial_\mu \phi)^2+(\partial_\mu \varphi_1)^2+(\partial_\mu \varphi_2)^2\right]\right.\left.-2\mu \left[
e^{\alpha\phi}\cos(\beta\varphi_1+\gamma\varphi_2)+
e^{-\alpha\phi}\cos(\beta\varphi_1-\gamma\varphi_2)\right]\right\}\label{SSM}
\end{eqnarray}
(with $\beta^2+\gamma^2-\alpha^2={1\over 2}$) for $\beta=\gamma$. Model (\ref{SSM}) when 
\begin{equation}
4\alpha^2=p_1+p_2-2,~4\beta^2=p_1,~4\gamma^2=p_2,~p_1,p_2\in N
\end{equation}
can be restricted to a perturbation of the coset theory $SU(2)_{p_1-2}\times SU(2)_{p_2-2}/SU(2)_{p_1+p_2-4}$. The particular case $p_1=p_2$ corresponds as well to the coset $SO(4)_{p-2}/SO(3)_{p-2}$. This is  in agreement with a general expectations explored in particular in \cite{VJS:an2} that $a_r^{(2)}$ spin chains possess a regime with continuum limit the diagonal coset $SO(r+1)/SO(r)$ . When $p_1=p_2$, the {\sl two compact bosons are identical}: this symmetry is inherited from the symmetry under exchange of the two $SU(2)$ in $SO(4)$. It is of course more surprising that the same continuum limit would appear in the context of a $b_2^{(1)}=SO(5)$ spin chain. We do not have enough examples at this stage to speculate to what extent this might be a general phenomenon.

It is well known, meanwhile, that the SS  model is dual to a non-linear sigma model \cite{Fateev}. It is interesting to study directly this model when $p_1=p_2$, in order to recover the black-hole theory  in the UV limit. We proceed as follows (closely following calculations in \cite{Bakas}. We adopt all notations from reference \cite{Fateev}
where we have, after setting $p_1=p_2$, 
\begin{subequations}
\begin{eqnarray}
a&=&{\pi\nu\over 2}\coth{\xi\over 2} \,, \\
b&=&{\pi\nu\over 2}\tanh{\xi\over 2} \,, \\
c&=&0 \,, \\
d&=&-b \,, \\
u&=&a+b=\pi\nu\coth\xi \,,
\end{eqnarray}
\end{subequations}
where $\nu={2\over p_1+p_2}$. The metric for the sigma model
\begin{eqnarray}
ds^2={u\over 4(1-z^2)(a^2-b^2z^2)}(dz)^2+{(1+z)(u+d(1+z))\over 2(a^2-b^2z^2)}(d\chi_1^2)\nonumber\\
+{(1-z)(u+d(1-z))\over 2(a^2-b^2z^2)}(d\chi_2^2)
\end{eqnarray}
reduces in this symmetric case to 
\begin{eqnarray}
ds^2={a+b\over 4(1-z^2)(a^2-b^2z^2)}(dz)^2+{1+z\over 2(a+bz)}(d\chi_1)^2+{1-z\over 2(a-bz)}(d\chi_2)^2 \,.
\end{eqnarray}
Here we have 
\begin{equation}
z=n_1^2+n_2^2-n_3^2-n_4^2\,, \quad  |z|\leq 1 \,; \qquad e^{2i\chi_1}={n_1+in_2\over n_1-in_2} \,, \qquad e^{2i\chi_2}={n_4+in_3\over n_4-in_3} \,.
\end{equation}
 We set $z\equiv \cos2\theta$ and introduce $y$ such that 
\begin{equation}
e^{y}=\coth\theta \,,
\end{equation}
so $y\in [-\infty,\infty]$. Replacing in the metric gives
\begin{eqnarray}
(ds)^2={\sinh\xi\over\pi\nu}\left[{\cosh\xi\over (1+e^{-2y}\cosh\xi)(1+e^{2y}\cosh\xi)}(dy)^2\right.\nonumber\\
\left.+{(d\chi_1)^2\over e^{-2y}+\cosh\xi}+{(d\chi_2)^2\over e^{2y}+\cosh\xi}\right] \,.
\end{eqnarray}
The point is now that this metric approaches the conformal metric of interest in the deep UV region, which is $\xi\to\infty$. In this limit indeed, setting 
\begin{equation}
y\equiv {\xi\over 2}-{\ln 2\over 2}-\ln\rho
\end{equation}
and letting $\xi\to\infty$ at fixed $\rho$ (in other words, keeping the product $e^{-\xi}e^{2y}$ finite), we find
\begin{equation}
(ds)^2\to {1\over\pi\nu} \left[{(d\rho)^2+\rho^2(d\chi_2)^2\over 1+\rho^2}+(d\chi_1)^2\right]\label{actTrMod}
\end{equation}
We thus see that the UV limit of the SS sigma model for  $p_1=p_2$ is indeed the product of a cigar theory and a circle theory (free boson), the latter having the same radius, asymptotically, as the cigar. Equation (\ref{actTrMod}) describes the continuum limit of the $b_2^{(1)}$ vertex model (and the modified truncated Chalker-Coddington model, up to boundary conditions), with $\chi_1$ and $\chi_2$ corresponding to the symmetric and antisymmetric combinations denoted by $\pm$ on the lattice.

\section{Network observables of Bondesan {\it et al.}}

In this appendix we recall some aspects of the network observables constructed in \cite{BWZ,BWZII}. We thank Roberto Bondesan for sharing with us some further details in a private communication.

Let $\mathcal{U}$ be the discrete-time evolution operator of a single-electronic wavefunction on the Chalker-Coddington network : 
\be
\mathcal{U} = \bigotimes_{\mbox{\small edge $e$}} \mathcal{U}_e
\bigotimes_{\mbox{\small vertex $v$}} \mathcal{U}_v \,.
\ee
The observables of \cite{BWZ,BWZII} are defined by introducing a set of {\it point-contacts} $\mathbf{c}_k$, $k=1, \ldots r$, which amount to cut edges open at which current can enter or exit the system. 
They are encoded by the projector 
\be 
\mathcal{Q} = \prod_k(1- |\mathbf{c}_k \rangle \langle\mathbf{c}_k | )=1- \sum_{k=1}^r |\mathbf{c}_k \rangle \langle \mathbf{c}_k | \,,
\ee 
such that the stationary wave function (``scattering state'') associated with current injected at the point contact $\mathbf{c}_{k}$ reads 
\be
| \psi_{\mathbf{c}_k}  \rangle = \mathcal{QU}(1 - \mathcal{Q} \mathcal{U}) | \mathbf{c}_k \rangle \,.
\ee
Let $\mathbf{r}_i$, $i=1, \ldots \ell$ be a set of {\it observation points} lying in the bulk of the system, that is, distinct from the point-contacts. In this framework the observables of Bondesan {\it et al.} are then defined as 
\bea 
Z_{1,q} &=& \mathbb{E}\left(  |\psi_{\mathbf{c}_k}(\mathbf{r}_i)|^{2q} \right) 
\nonumber \\ 
Z_{\ell,q} &=& 
 \mathbb{E}\left( 
\left|
\begin{array}{ccc}
\psi_{\mathbf{c}_1}(\mathbf{r}_1) & \ldots & \psi_{\mathbf{c}_1}(\mathbf{r}_\ell) \\
\vdots & &  \vdots \\
\psi_{\mathbf{c}_\ell}(\mathbf{r}_1) & \ldots & \psi_{\mathbf{c}_\ell}(\mathbf{r}_\ell) 
\end{array}
\right|^{2q}
\right) \,,
\label{eq:observablesBWZ}
\eea 
where $\mathbb{E}(\ldots)$ is used here to denote the average over disorder.

Following the construction of \cite{BWZ,BWZII}, the observables $Z_{\ell,q}$ are expected to become pure-scaling ones in a continuum limit of the network model which takes the contact and observation regions to single points, namely $\mathbf{c}_k \to \mathbf{c}$ and $\mathbf{r}_k \to \mathbf{r}$  while $\mathbf{r}$ and $\mathbf{c}$ remain distinct :
\be 
Z_{\ell,q} \sim |\mathbf{r} - \mathbf{c}|^{-2 X \ell q (\ell-q)} \,,
\ee 
with $X \simeq 1/4$. 
A particular case is that of $Z_{1,1}$, which is shown to be trivial: 
\be
Z_{1,1} = 1 \,.
\ee
It is easy to  relate these observables to the geometric formulation used in this paper. 
Proceeding as in section \ref{sec:pathintegral}, the amplitudes $ \psi_{\mathbf{c}_k}(\mathbf{r}_i) \equiv   \langle \mathbf{r}_i | \psi_{\mathbf{c}_k}  \rangle$ can be decomposed as a sum over (advanced) Feynman paths starting at  $\mathbf{c}_k$ and ending at $\mathbf{r}_k$ while avoiding all point-contacts,
\be
 \psi_{\mathbf{c}_k}(\mathbf{r}_i) = \sum_{\omega : \mathbf{c}_k \to \mathbf{r}_i } W(\omega) \,,
\ee
where the statistical weights $W(\omega)$ collects all phases accumulated by the path $\omega$ at the visited edges as well as the weight coming from scattering at vertices.
Similarly, the complex conjugate $ \psi_{\mathbf{c}_k}(\mathbf{r}_i)^*$ is written as a sum over retarded Feynman paths $\bar{\omega}$ starting at $\mathbf{r}_k$ and ending at $\mathbf{c}_k$ , with weight $W(\bar{\omega})^*$. From there, a geometrical interpretation can be given to the observables (\ref{eq:observablesBWZ}), when $q\equiv n$ is an integer. For instance, let us take the case of only one contact point $\mathbf{c}$ and one observation point $\mathbf{r}$, 
\bea 
Z_{1,n} &=&\mathbb{E}\left(  |\psi_\mathbf{c}|^{2(q=n)} \right) \nonumber \\ 
&=&
 \sum_{\omega_1, \ldots , \omega_n : \mathbf{c} \to \mathbf{r} }
~
 \sum_{\bar{\omega}_1, \ldots , \bar{\omega}_n : \mathbf{r} \to ~\mathbf{c} }  \prod_e \delta\left( \sum_{i=1}^n n_e(\omega_i)- \sum_{i=1}^n n_e(\bar{\omega}_i) \right)
\nonumber \\
& & 
\qquad\qquad\qquad\qquad\qquad
\times W'(\omega_1) \otimes W'(\omega_n) W'(\bar{\omega}_1)^* \ldots W'(\bar{\omega}_n)^*  \,,
\eea
where $n_e(\omega)$ is the number of times $\omega$ visits and the weights $W'$ correspond to $W$ with random phases removed.
The observable $Z_{1,1}$ for instance is given by a sum over configurations of one advanced (resp.\ one retarded) path going from $\mathbf{r}$ to $\mathbf{c}$ (resp.\ from $\mathbf{c}$ to $\mathbf{r}$) with the constraint that the path never visits the edge $\mathbf{c}$ except at the end (resp.\ at the beginning). Similarly, for higher values of $q=n$ an integer, we get $q$ paths (in advanced/retarded pairs) relating $\mathbf{r}$ to $\mathbf{c}$ with the constraint that each of these paths visits $\mathbf{c}$ only once. 

We note that instead of $Z_{1,q}$, one can also define another observable involving two contact points $\mathbf{c_1}$ and $\mathbf{c_2}$ 
\bea
\tilde{Z}_{1,q}=\mathbb{E}\left(  |\psi_\mathbf{c_1}(\mathbf{c_2})|^{2q} \right).
\eea
The geometric interpretation of $\tilde{Z}_{1,q}$ now involves $q$ paths (in advanced /retarded pairs)  relating points $c_1$ and $c_2$ with the constraint that these paths pass only once through $c_1$ and $c_2$. This is in contrast with $Z_{1,q}$ where $c_2$ is replaced by an observation point $r_2$, where  the paths can pass an arbitrary number of times. 

It was argued in \cite{JMZ}  that the behaviour of $\tilde{Z}_{1,q}$  for $|q|\geq1/2$ is
\bea
\tilde{Z}_{1,q}=\Gamma(q)^{-2}\int_0^\infty\lvert\Gamma\left(q-1/2-i\lambda/2\right)\rvert r^{-2x_\lambda}\mu(\lambda)\diff\lambda
\eea
with $\mu(\lambda)=\frac{\lambda}{2}\tanh\frac{\pi\lambda}{2}$. Unlike $Z_{1,q}$, this is not a pure-scaling variable. At large distances, for all $q\geq1/2$ the integral is dominated by the contribution of the same scaling dimension $x_{\lambda=0}=X/4$ corresponding to $p=0$ and $q=1/2$ in \eqref{Roberto}. 

In the first truncation, edges can be occupied at most once. Watermelon observables involve $\ell$ paths (coming in advanced/retarded pairs), and their two point functions are essentially truncated versions of $Z_{1,q}$ or $\tilde{Z}_{1,q}$ with $q=\ell$---it is not clear which.%
\footnote{The paths in the watermelon observables  start and finish at neighboring points instead of exactly at the same point, but this does not affect the scaling behavior.}
Since for watermelons $q\geq 1$, the interpretation in terms of $\tilde{Z}_{1,q}$ would lead to exponents at the bottom of the continuum, while we find the watermelon exponents in the discrete part of the spectrum, the interpretation in terms of $Z_{1,q}$ seems more natural.%
\footnote{Note however that the value $x_{1,1}={1\over 4}$ for $Z_{11}$ {\sl is} precisely at the bottom of the continuum! But this is not the case for higher values of $q$.}
For higher truncations, a difference between contact points and observation points can be introduced in the lattice model: for instance for $Z_{11}$ it is possible, for the second and higher truncations, to impose that the paths go only once on the contact-edges, while they can go once or more over the observation edges. This, however, does not seem to make much difference: in all cases, the correlation function is dominated by the $x_{11}$ dimension, and we do not know how to build objects corresponding to  different $Z_{11}$ and $\tilde{Z}_{11}$.

\end{document}